\documentclass[11pt]{article}
\pdfoutput=1

\usepackage{jheppub}
\usepackage[utf8]{inputenc}
\usepackage{booktabs}
\usepackage{graphicx}
\usepackage{bbold}
\usepackage{subcaption}
\usepackage{mathtools}
\usepackage{marvosym}

\usepackage{xkeyval}

\usepackage{tikz,fp}
\usetikzlibrary{external,fixedpointarithmetic,decorations.pathmorphing,positioning,calc,fadings,decorations.markings,decorations.pathreplacing,patterns,arrows,arrows.meta}

\tikzset{
  label distance=-3pt,
  inner sep=1.5pt,
  >=stealth,
  boson/.style={
    decoration={snake, segment length=2mm, amplitude=0.5mm},
    decorate,
  },
  graviton/.style={
  	line width=0.1pt,
    decoration={snake, segment length=2mm, amplitude=0.5mm},
    decorate,
  },
  quark/.style={
    postaction={decorate},
    decoration={markings,mark=at position .5 with {\draw[->] (-0.01,0) -- (0.07,0);}}
  },
  aquark/.style={
    postaction={decorate},
    decoration={markings,mark=at position .5 with {\draw[->] (0.01,0) -- (-0.07,0);}}
  },
  gluonOld/.style={
    line width=0.7pt,
    decorate,
    decoration={coil,aspect=-0.5,amplitude=-2pt,segment length=2.2pt}
  },
  gluon/.style={
    line width=0.5pt,
    decorate,
    decoration={coil,aspect=-0.75,amplitude=-1.5pt,segment length=2pt}
  },
  agluon/.style={
    line width=0.5pt,
    decorate,
    decoration={coil,aspect=0.75,amplitude=1.5pt,segment length=2pt}
  },
  scalar/.style={
    dashed
  },
  dscalar/.style={
    dashed,
    postaction={decorate},
    decoration={markings,mark=at position 0.65 with {\draw[->] (-0.01,0) -- (0.07,0);}}
  },
  adscalar/.style={
    dashed,
    postaction={decorate},
    decoration={markings,mark=at position 0.65 with {\draw[->] (-0.01,0) -- (0.07,0);}}
  },
  line/.style={
  },
  doubleline/.style={
    double
  },
  middleArrow/.style={
    draw=white,
    decoration={markings,mark=at position 0.60 with{\arrow[scale=0.16,line width=10pt,black]{Straight Barb}}},
    decorate
  },
  dot/.style={
  	postaction={decorate},
    decoration={markings,mark=between positions 0.5 and 0.5 step 1 with {\draw[fill] (0,0) circle [radius=1.5pt];}}  
  },
  dots/.style={
  	postaction={decorate},
    decoration={markings,mark=between positions 0.33 and 0.66 step 0.33 with {\draw[fill] (0,0) circle [radius=1.5pt];}}  
  },
  dotss/.style={
  	postaction={decorate},
    decoration={markings,mark=between positions 0.25 and 0.75 step 0.25 with {\draw[fill] (0,0) circle [radius=1.5pt];}}  
  },
  blob/.style={
  	pattern=north west lines,
    draw=black
  },
  blobBlack/.style={
    draw=black,
    fill=black
  },
  blobWhite/.style={
    draw=black,
    fill=white
  },
  xSource/.style={
    circle, minimum width=6pt, draw, inner sep=0pt, path picture={\draw (path picture bounding box.south east) -- (path picture bounding box.north west) (path picture bounding box.south west) -- (path picture bounding box.north east);}
  }
}

\makeatletter
\def\gSetStdExtTypes#1{\presetkeys{gTypes}{eA=#1,eB=#1,eC=#1,eD=#1,eE=#1}{}}
\def\gSetStdIntTypes#1{\presetkeys{gTypes}{iA=#1,iB=#1,iC=#1,iD=#1,iE=#1,iF=#1,iG=#1,iH=#1,iI=#1,iJ=#1}{}}
\def\gSetStdTypes#1{\gSetStdExtTypes{#1}\gSetStdIntTypes{#1}}

\define@cmdkeys{general}{scale}
\define@cmdkeys{general}{rotate}
\define@cmdkeys{general}{font}
\define@cmdkeys{general}{yshift}
\presetkeys{general}{scale=1, rotate=0, font=\scriptsize, yshift=0pt}{}

\define@cmdkeys{gTypes}{eA,eB,eC,eD,eE,iA,iB,iC,iD,iE,iF,iG,iH,iI,iJ}
\gSetStdTypes{line}
\define@key{pre}{all}{\gSetStdTypes{#1}}
\define@key{pre}{allE}{\gSetStdExtTypes{#1}}
\define@key{pre}{allI}{\gSetStdIntTypes{#1}}

\define@cmdkeys{gLabels}{eLA,eLB,eLC,eLD,eLE,iLA,iLB,iLC,iLD,iLE,iLF,iLG,iLH,iLI,iLJ}
\presetkeys{gLabels}{eLA={},eLB={},eLC={},eLD={},eLE={},iLA={},iLB={},iLC={},iLD={},iLE={},iLF={},iLG={},iLH={},iLI={},iLJ={}}{}

\define@boolkeys{dots}{lDots,rDots}
\presetkeys{dots}{lDots=false, rDots=false}{}

\define@cmdkeys{blobs}{blobA,blobB,blobC}
\presetkeys{blobs}{blobA=blob,blobB=blob,blobC=blob}{}

\def\gTreeTri{\@ifnextchar[\@gTreeTri{\@gTreeTri[]}}
\def\@gTreeTri[#1]#2{%
  \setkeys*{pre}{#1}%
  \setrmkeys{general,gTypes,gLabels}%
  \begin{tikzpicture}
    [line width=1pt,
    baseline={([yshift=-0.5ex+\cmdKV@general@yshift]current bounding box.center)},
    scale=\cmdKV@general@scale,
    rotate=\cmdKV@general@rotate,
    font=\cmdKV@general@font]
    \draw[\cmdKV@gTypes@eA] (0.7,0.3) -- (0.4,0.3) node[label=(-180+\cmdKV@general@rotate):\cmdKV@gLabels@eLA] {};
    \draw[\cmdKV@gTypes@eB] (0.7,0.3) -- (0.9,0.6) node[label=(\cmdKV@general@rotate):\cmdKV@gLabels@eLB] {};
    \draw[\cmdKV@gTypes@eC] (0.7,0.3) -- (0.9,0) node[label=(\cmdKV@general@rotate):\cmdKV@gLabels@eLC] {};
    #2
  \end{tikzpicture}
  \gSetStdTypes{line}
}

\def\gTreeS{\@ifnextchar[\@gTreeS{\@gTreeS[]}}
\def\@gTreeS[#1]#2{%
  \setkeys*{pre}{#1}%
  \setrmkeys{general,gTypes,gLabels}%
  \begin{tikzpicture}
    [line width=1pt,
    baseline={([yshift=-0.5ex+\cmdKV@general@yshift]current bounding box.center)},
    scale=\cmdKV@general@scale,
    rotate=\cmdKV@general@rotate,
    font=\cmdKV@general@font]
    \draw[\cmdKV@gTypes@eA] (0.2,0.3) -- (0,0) node[label=(-180+\cmdKV@general@rotate):\cmdKV@gLabels@eLA] {};
    \draw[\cmdKV@gTypes@eB] (0.2,0.3) -- (0,0.6) node[label=(180+\cmdKV@general@rotate):\cmdKV@gLabels@eLB] {};
    \draw[\cmdKV@gTypes@eC] (0.7,0.3) -- (0.9,0.6) node[label=(\cmdKV@general@rotate):\cmdKV@gLabels@eLC] {};
    \draw[\cmdKV@gTypes@eD] (0.7,0.3) -- (0.9,0) node[label=(\cmdKV@general@rotate):\cmdKV@gLabels@eLD] {};    
    \draw[label distance=0,\cmdKV@gTypes@iA] (0.2,0.3) -- node[label=(90+\cmdKV@general@rotate):\cmdKV@gLabels@iLA] {} (0.7,0.3);
    #2
  \end{tikzpicture}
  \gSetStdTypes{line}
}

\def\gTreeT{\@ifnextchar[\@gTreeT{\@gTreeT[]}}
\def\@gTreeT[#1]#2{%
  \setkeys*{pre}{#1}%
  \setrmkeys{general,gTypes,gLabels}%
  \begin{tikzpicture}
    [line width=1pt,
    baseline={([yshift=-0.5ex+\cmdKV@general@yshift]current bounding box.center)},
    scale=\cmdKV@general@scale,
    rotate=\cmdKV@general@rotate,
    font=\cmdKV@general@font]
    \draw[\cmdKV@gTypes@eA] (0.45,0.1) -- (0,0) node[label=(-180+\cmdKV@general@rotate):\cmdKV@gLabels@eLA] {};
    \draw[\cmdKV@gTypes@eB] (0.45,0.5) -- (0,0.6) node[label=(180+\cmdKV@general@rotate):\cmdKV@gLabels@eLB] {};
    \draw[\cmdKV@gTypes@eC] (0.45,0.5) -- (0.9,0.6) node[label=(\cmdKV@general@rotate):\cmdKV@gLabels@eLC] {};
    \draw[\cmdKV@gTypes@eD] (0.45,0.1) -- (0.9,0) node[label=(\cmdKV@general@rotate):\cmdKV@gLabels@eLD] {};
    \draw[label distance=0,\cmdKV@gTypes@iA] (0.45,0.5) -- node[label=(\cmdKV@general@rotate):\cmdKV@gLabels@iLA] {} (0.45,0.1);
    #2
  \end{tikzpicture}
  \gSetStdTypes{line}
}

\def\gTreeU{\@ifnextchar[\@gTreeU{\@gTreeU[]}}
\def\@gTreeU[#1]#2{%
  \setkeys*{pre}{#1}%
  \setrmkeys{general,gTypes,gLabels}%
  \begin{tikzpicture}
    [line width=1pt,
    baseline={([yshift=-0.5ex+\cmdKV@general@yshift]current bounding box.center)},
    scale=\cmdKV@general@scale,
    rotate=\cmdKV@general@rotate,
    font=\cmdKV@general@font]
    \draw[\cmdKV@gTypes@eA] (0.45,0.5) -- (0,0) node[label=(-180+\cmdKV@general@rotate):\cmdKV@gLabels@eLA] {};
    \foreach \x in {5,...,14} \fill[opacity=0.12,white] (0.27,0.3) circle (\x/100);
    \draw[\cmdKV@gTypes@eB] (0.45,0.1) -- (0,0.6) node[label=(180+\cmdKV@general@rotate):\cmdKV@gLabels@eLB] {};
    \draw[\cmdKV@gTypes@eC] (0.45,0.5) -- (0.9,0.6) node[label=(\cmdKV@general@rotate):\cmdKV@gLabels@eLC] {};
    \draw[\cmdKV@gTypes@eD] (0.45,0.1) -- (0.9,0) node[label=(\cmdKV@general@rotate):\cmdKV@gLabels@eLD] {};
    \draw[label distance=0,\cmdKV@gTypes@iA] (0.45,0.5) -- node[label=(\cmdKV@general@rotate):\cmdKV@gLabels@iLA] {} (0.45,0.1);
    #2
  \end{tikzpicture}
  \gSetStdTypes{line}
}

\def\gTreeFourPt{\@ifnextchar[\@gTreeFourPt{\@gTreeFourPt[]}}
\def\@gTreeFourPt[#1]#2{%
  \setkeys*{pre}{#1}%
  \setrmkeys{general,gTypes,gLabels}%
  \begin{tikzpicture}
    [line width=1pt,
    baseline={([yshift=-0.5ex+\cmdKV@general@yshift]current bounding box.center)},
    scale=\cmdKV@general@scale,
    rotate=\cmdKV@general@rotate,
    font=\cmdKV@general@font]
    \draw[\cmdKV@gTypes@eA] (0,0) -- (-0.4,-0.4) node[label=(-180+\cmdKV@general@rotate):\cmdKV@gLabels@eLA] {};
    \draw[\cmdKV@gTypes@eB] (0,0) -- (-0.4,0.4) node[label=(180+\cmdKV@general@rotate):\cmdKV@gLabels@eLB] {};
    \draw[\cmdKV@gTypes@eC] (0,0) -- (0.4,0.4) node[label=(\cmdKV@general@rotate):\cmdKV@gLabels@eLC] {};
    \draw[\cmdKV@gTypes@eD] (0,0) -- (0.4,-0.4) node[label=(\cmdKV@general@rotate):\cmdKV@gLabels@eLD] {};
    #2
  \end{tikzpicture}
  \gSetStdTypes{line}
}

\def\gTadA{\@ifnextchar[\@gTadA{\@gTadA[]}}
\def\@gTadA[#1]#2{%
  \setkeys*{pre}{#1}%
  \setrmkeys{general,gTypes,gLabels}%
  \begin{tikzpicture}
    [line width=1pt,
    baseline={([yshift=-0.5ex+\cmdKV@general@yshift]current bounding box.center)},
    scale=\cmdKV@general@scale,
    rotate=\cmdKV@general@rotate,
    font=\cmdKV@general@font]
    \draw[\cmdKV@gTypes@eA] (-0.7,0.3) -- (-0.9,0) node[label=(180+\cmdKV@general@rotate):\cmdKV@gLabels@eLA] {};
    \draw[\cmdKV@gTypes@eB] (-0.7,0.3) -- (-0.9,0.6) node[label=(180+\cmdKV@general@rotate):\cmdKV@gLabels@eLB] {};
    \draw[\cmdKV@gTypes@eC] (0.15,0.45) -- (0.5,0.6) node[label=(\cmdKV@general@rotate):\cmdKV@gLabels@eLC] {};
    \draw[\cmdKV@gTypes@eD] (-0.3,0.3) -- (-0.2,0) node[label=(\cmdKV@general@rotate):\cmdKV@gLabels@eLD] {};
    \draw[label distance=0,\cmdKV@gTypes@iA] (-0.7,0.3) -- node[label=(90+\cmdKV@general@rotate):\cmdKV@gLabels@iLA] {} (-0.3,0.3);
    \draw[label distance=0,\cmdKV@gTypes@iB] (-0.3,0.3) -- node[label=(-80+\cmdKV@general@rotate):\cmdKV@gLabels@iLB] {} (0.15,0.45);
    \draw[label distance=0,\cmdKV@gTypes@iC] (0.15,0.45) -- node[label=(45+\cmdKV@general@rotate):\cmdKV@gLabels@iLC] {} (-0.18,0.69);
    \draw[label distance=-0.5,\cmdKV@gTypes@iD] (-0.18,0.69) .. controls ($(-0.18,0.69) + (-0.12,-0.12)$) and ($(-0.35,0.87) + (-0.12,-0.12)$) .. (-0.35,0.87) node[label=(180+\cmdKV@general@rotate):\cmdKV@gLabels@iLD] {} .. controls ($(-0.35,0.87) + (0.12,0.12)$) and ($(-0.18,0.69) + (0.12,0.12)$) .. (-0.18,0.69);
    #2
  \end{tikzpicture}
  \gSetStdTypes{line}
}

\def\gTadB{\@ifnextchar[\@gTadB{\@gTadB[]}}
\def\@gTadB[#1]#2{%
  \setkeys*{pre}{#1}%
  \setrmkeys{general,gTypes,gLabels}%
  \begin{tikzpicture}
    [line width=1pt,
    baseline={([yshift=-0.5ex+\cmdKV@general@yshift]current bounding box.center)},
    scale=\cmdKV@general@scale,
    rotate=\cmdKV@general@rotate,
    font=\cmdKV@general@font]
    \draw[\cmdKV@gTypes@eA] (0,0.3) -- (-0.2,0) node[label=(-180+\cmdKV@general@rotate):\cmdKV@gLabels@eLA] {};
    \draw[\cmdKV@gTypes@eB] (0,0.3) -- (-0.2,0.6) node[label=(180+\cmdKV@general@rotate):\cmdKV@gLabels@eLB] {};
    \draw[\cmdKV@gTypes@eC] (0.7,0.3) -- (0.9,0.6) node[label=(\cmdKV@general@rotate):\cmdKV@gLabels@eLC] {};
    \draw[\cmdKV@gTypes@eD] (0.7,0.3) -- (0.9,0) node[label=(\cmdKV@general@rotate):\cmdKV@gLabels@eLD] {};
    \draw[label distance=0,\cmdKV@gTypes@iA] (0,0.3) -- node[label=(-90+\cmdKV@general@rotate):\cmdKV@gLabels@iLA] {} (0.35,0.3);
    \draw[label distance=0,\cmdKV@gTypes@iB] (0.35,0.3) -- node[label=(\cmdKV@general@rotate):\cmdKV@gLabels@iLB] {} (0.35,0.6);
    \draw[label distance=-0.5,\cmdKV@gTypes@iC] (0.35,0.6) .. controls (0.18,0.6) and (0.18,0.85) .. (0.35,0.85) .. controls (0.52,0.85) and (0.52,0.6) .. node[label=(\cmdKV@general@rotate):\cmdKV@gLabels@iLC] {} (0.35,0.6);
    \draw[label distance=0,\cmdKV@gTypes@iD] (0.35,0.3) -- node[label=(-90+\cmdKV@general@rotate):\cmdKV@gLabels@iLD] {} (0.7,0.3);
    #2
  \end{tikzpicture}
  \gSetStdTypes{line}
}
\def\gBubA{\@ifnextchar[\@gBubA{\@gBubA[]}}
\def\@gBubA[#1]#2{%
  \setkeys*{pre}{#1}%
  \setrmkeys{general,gTypes,gLabels}%
  \begin{tikzpicture}
    [line width=1pt,
    baseline={([yshift=-0.5ex+\cmdKV@general@yshift]current bounding box.center)},
    scale=\cmdKV@general@scale,
    rotate=\cmdKV@general@rotate,
    font=\cmdKV@general@font]
    \draw[\cmdKV@gTypes@eA] (0.5,0.7) -- (0.2,0.7) node[label=(180+\cmdKV@general@rotate):\cmdKV@gLabels@eLA] {};
    \draw[\cmdKV@gTypes@eB] (0.9,0.7) -- (1.2,0.7) node[label=(\cmdKV@general@rotate):\cmdKV@gLabels@eLB] {};
    \draw[label distance=0,\cmdKV@gTypes@iA] (0.5,0.7) .. controls (0.5,0.95) and (0.9,0.95) .. node[label=(90+\cmdKV@general@rotate):\cmdKV@gLabels@iLA] {} (0.9,0.7);
    \draw[label distance=0,\cmdKV@gTypes@iB] (0.9,0.7) .. controls (0.9,0.45) and (0.5,0.45) .. node[label=(-90+\cmdKV@general@rotate):\cmdKV@gLabels@iLB] {} (0.5,0.7);
    #2
  \end{tikzpicture}
  \gSetStdTypes{line}
}
\def\gBubB{\@ifnextchar[\@gBubB{\@gBubB[]}}
\def\@gBubB[#1]#2{%
  \setkeys*{pre}{#1}%
  \setrmkeys{general,gTypes,gLabels}%
  \begin{tikzpicture}
    [line width=1pt,
    baseline={([yshift=-0.5ex+\cmdKV@general@yshift]current bounding box.center)},
    scale=\cmdKV@general@scale,
    rotate=\cmdKV@general@rotate,
    font=\cmdKV@general@font]
    \draw[\cmdKV@gTypes@eA] (0.2,0.7) -- (0,0.4) node[label=(-180+\cmdKV@general@rotate):\cmdKV@gLabels@eLA] {};
    \draw[\cmdKV@gTypes@eB] (0.2,0.7) -- (0,1) node[label=(180+\cmdKV@general@rotate):\cmdKV@gLabels@eLB] {};
    \draw[\cmdKV@gTypes@eC] (1.2,0.7) -- (1.4,1) node[label=(\cmdKV@general@rotate):\cmdKV@gLabels@eLC] {};
    \draw[\cmdKV@gTypes@eD] (1.2,0.7) -- (1.4,0.4) node[label=(\cmdKV@general@rotate):\cmdKV@gLabels@eLD] {};
    \draw[label distance=0,\cmdKV@gTypes@iA] (0.2,0.7) -- node[label=(90+\cmdKV@general@rotate):\cmdKV@gLabels@iLA] {} (0.5,0.7);
    \draw[label distance=0,\cmdKV@gTypes@iB] (0.5,0.7) .. controls (0.5,0.95) and (0.9,0.95) .. node[label=(90+\cmdKV@general@rotate):\cmdKV@gLabels@iLB] {} (0.9,0.7);
    \draw[label distance=0,\cmdKV@gTypes@iC] (0.9,0.7) .. controls (0.9,0.45) and (0.5,0.45) .. node[label=(-90+\cmdKV@general@rotate):\cmdKV@gLabels@iLC] {} (0.5,0.7);
    \draw[label distance=0,\cmdKV@gTypes@iD] (0.9,0.7) -- node[label=(90+\cmdKV@general@rotate):\cmdKV@gLabels@iLD] {} (1.2,0.7);
    #2
  \end{tikzpicture}
  \gSetStdTypes{line}
}

\def\gBubC{\@ifnextchar[\@gBubC{\@gBubC[]}}
\def\@gBubC[#1]#2{%
  \setkeys*{pre}{#1}%
  \setrmkeys{general,gTypes,gLabels}%
  \begin{tikzpicture}
    [line width=1pt,
    baseline={([yshift=-0.5ex+\cmdKV@general@yshift]current bounding box.center)},
    scale=\cmdKV@general@scale,
    rotate=\cmdKV@general@rotate,
    font=\cmdKV@general@font]
    \draw[\cmdKV@gTypes@eA] (-0.7,0.3) -- (-0.9,0) node[label=(180+\cmdKV@general@rotate):\cmdKV@gLabels@eLA] {};
    \draw[\cmdKV@gTypes@eB] (-0.7,0.3) -- (-0.9,0.6) node[label=(180+\cmdKV@general@rotate):\cmdKV@gLabels@eLB] {};
    \draw[\cmdKV@gTypes@eC] (0.3,0.5) -- (0.5,0.6) node[label=(\cmdKV@general@rotate):\cmdKV@gLabels@eLC] {};
    \draw[\cmdKV@gTypes@eD] (-0.3,0.3) -- (-0.2,0) node[label=(\cmdKV@general@rotate):\cmdKV@gLabels@eLD] {};
    \draw[label distance=0,\cmdKV@gTypes@iA] (-0.7,0.3) -- node[label=(90+\cmdKV@general@rotate):\cmdKV@gLabels@iLA] {} (-0.3,0.3);
    \draw[label distance=0,\cmdKV@gTypes@iB] (-0.3,0.3) -- node[label=(90+\cmdKV@general@rotate):\cmdKV@gLabels@iLB] {} (-0,0.4);
    \draw[label distance=0,\cmdKV@gTypes@iC] (0,0.4) .. controls (-0.03,0.6) and (0.22,0.65) .. node[label=(90+\cmdKV@general@rotate):\cmdKV@gLabels@iLC] {} (0.3,0.5);
    \draw[label distance=0,\cmdKV@gTypes@iD] (0.3,0.5) .. controls (0.38,0.3) and (0.08,0.21) .. node[label=(-90+\cmdKV@general@rotate):\cmdKV@gLabels@iLD] {} (0,0.4);
    #2
  \end{tikzpicture}
  \gSetStdTypes{line}
}
\def\gTriA{\@ifnextchar[\@gTriA{\@gTriA[]}}
\def\@gTriA[#1]#2{%
  \setkeys*{pre}{#1}%
  \setrmkeys{general,gTypes,gLabels}%
  \begin{tikzpicture}
    [line width=1pt,
    baseline={([yshift=-0.5ex+\cmdKV@general@yshift]current bounding box.center)},
    scale=\cmdKV@general@scale,
    rotate=\cmdKV@general@rotate,
    font=\cmdKV@general@font]
    \draw[\cmdKV@gTypes@eC] (0:0.3) -- (0:0.55) node[label=(\cmdKV@general@rotate):\cmdKV@gLabels@eLC] {};
    \draw[\cmdKV@gTypes@eA] (-120:0.3) -- (-120:0.55) node[label=(-180+\cmdKV@general@rotate):\cmdKV@gLabels@eLA] {};
    \draw[\cmdKV@gTypes@eB] (120:0.3) -- (120:0.55) node[label=(180+\cmdKV@general@rotate):\cmdKV@gLabels@eLB] {};    
    \draw[label distance=0,\cmdKV@gTypes@iA] (-120:0.3) -- node[label=(180+\cmdKV@general@rotate):\cmdKV@gLabels@iLA] {} (120:0.3);
    \draw[label distance=0,\cmdKV@gTypes@iB] (120:0.3) -- node[label=(60+\cmdKV@general@rotate):\cmdKV@gLabels@iLB] {} (0:0.3);
    \draw[label distance=0,\cmdKV@gTypes@iC] (0:0.3) -- node[label=(-60+\cmdKV@general@rotate):\cmdKV@gLabels@iLC] {} (-120:0.3);
    #2
  \end{tikzpicture}
  \gSetStdTypes{line}
}
\def\gTriB{\@ifnextchar[\@gTriB{\@gTriB[]}}
\def\@gTriB[#1]#2{%
  \setkeys*{pre}{#1}%
  \setrmkeys{general,gTypes,gLabels}%
  \begin{tikzpicture}
    [line width=1pt,
    baseline={([yshift=-0.5ex+\cmdKV@general@yshift]current bounding box.center)},
    scale=\cmdKV@general@scale,
    rotate=\cmdKV@general@rotate,
    font=\cmdKV@general@font]
    \draw[\cmdKV@gTypes@eA] (0.3,0.45) -- (0,0.3) node[label=(-180+\cmdKV@general@rotate):\cmdKV@gLabels@eLA] {};
    \draw[\cmdKV@gTypes@eB] (0.3,0.95) -- (0,1.1) node[label=(180+\cmdKV@general@rotate):\cmdKV@gLabels@eLB] {};
    \draw[\cmdKV@gTypes@eC] (1,0.7) -- (1.3,1.1) node[label=(\cmdKV@general@rotate):\cmdKV@gLabels@eLC] {};
    \draw[\cmdKV@gTypes@eD] (1,0.7) -- (1.3,0.3) node[label=(\cmdKV@general@rotate):\cmdKV@gLabels@eLD] {};
    \draw[label distance=0,\cmdKV@gTypes@iA] (0.3,0.45) -- node[label=(180+\cmdKV@general@rotate):\cmdKV@gLabels@iLA] {} (0.3,0.95);
    \draw[label distance=0,\cmdKV@gTypes@iB] (0.3,0.95) -- node[label=(70+\cmdKV@general@rotate):\cmdKV@gLabels@iLB] {} (0.7,0.7);
    \draw[label distance=0,\cmdKV@gTypes@iC] (0.7,0.7) -- node[label=(-70+\cmdKV@general@rotate):\cmdKV@gLabels@iLC] {} (0.3,0.45);
    \draw[label distance=0,\cmdKV@gTypes@iD] (1,0.7) -- node[label=(90+\cmdKV@general@rotate):\cmdKV@gLabels@iLD] {} (0.7,0.7);
    #2
  \end{tikzpicture}
  \gSetStdTypes{line}
}
\def\gBox{\@ifnextchar[\@gBox{\@gBox[]}}
\def\@gBox[#1]#2{%
  \setkeys*{pre}{#1}%
  \setrmkeys{general,gTypes,gLabels}%
  \begin{tikzpicture}
    [line width=1pt,
    baseline={([yshift=-0.5ex+\cmdKV@general@yshift]current bounding box.center)},
    scale=\cmdKV@general@scale,
    rotate=\cmdKV@general@rotate,
    font=\cmdKV@general@font]
    \draw[\cmdKV@gTypes@eA] (0.25,0.25) -- (0,0) node[label=(-180+\cmdKV@general@rotate):\cmdKV@gLabels@eLA] {};
    \draw[\cmdKV@gTypes@eB] (0.25,0.75) -- (0,1) node[label=(180+\cmdKV@general@rotate):\cmdKV@gLabels@eLB] {};
    \draw[\cmdKV@gTypes@eC] (0.75,0.75) -- (1,1) node[label=(\cmdKV@general@rotate):\cmdKV@gLabels@eLC] {};
    \draw[\cmdKV@gTypes@eD] (0.75,0.25) -- (1,0) node[label=(\cmdKV@general@rotate):\cmdKV@gLabels@eLD] {};
    \draw[label distance=0,\cmdKV@gTypes@iA] (0.25,0.25) -- node[label=(180+\cmdKV@general@rotate):\cmdKV@gLabels@iLA] {} (0.25,0.75);
    \draw[label distance=0,\cmdKV@gTypes@iB] (0.25,0.75) -- node[label=(90+\cmdKV@general@rotate):\cmdKV@gLabels@iLB] {} (0.75,0.75);
    \draw[label distance=0,\cmdKV@gTypes@iC] (0.75,0.75) -- node[label=(\cmdKV@general@rotate):\cmdKV@gLabels@iLC] {} (0.75,0.25);
    \draw[label distance=0,\cmdKV@gTypes@iD] (0.75,0.25) -- node[label=(-90+\cmdKV@general@rotate):\cmdKV@gLabels@iLD] {} (0.25,0.25);
    #2
  \end{tikzpicture}
  \gSetStdTypes{line}
}
\def\gPenta{\@ifnextchar[\@gPenta{\@gPenta[]}}
\def\@gPenta[#1]#2{%
  \setkeys*{pre}{#1}%
  \setrmkeys{general,gTypes,gLabels}%
  \begin{tikzpicture}
    [line width=1pt,
    baseline={([yshift=-0.5ex+\cmdKV@general@yshift]current bounding box.center)},
    scale=\cmdKV@general@scale,
    rotate=\cmdKV@general@rotate,
    font=\cmdKV@general@font]
    \draw[\cmdKV@gTypes@eA] (-144:0.35) -- (-144:0.75) node[label=(-144+\cmdKV@general@rotate):\cmdKV@gLabels@eLA] {};
    \draw[\cmdKV@gTypes@eB] (144:0.35) -- (144:0.75) node[label=(144+\cmdKV@general@rotate):\cmdKV@gLabels@eLB] {};
    \draw[\cmdKV@gTypes@eC] (72:0.35) -- (72:0.75) node[label=(72+\cmdKV@general@rotate):\cmdKV@gLabels@eLC] {};
    \draw[\cmdKV@gTypes@eD] (0:0.35) -- (0:0.75) node[label=(\cmdKV@general@rotate):\cmdKV@gLabels@eLD] {};
    \draw[\cmdKV@gTypes@eE] (-72:0.35) -- (-72:0.75) node[label=(-72+\cmdKV@general@rotate):\cmdKV@gLabels@eLE] {};
    \draw[label distance=0,\cmdKV@gTypes@iA] (-144:0.35) -- node[label=(180+\cmdKV@general@rotate):\cmdKV@gLabels@iLA] {} (144:0.35);
    \draw[label distance=0,\cmdKV@gTypes@iB] (144:0.35) -- node[label=(108+\cmdKV@general@rotate):\cmdKV@gLabels@iLB] {} (72:0.35);
    \draw[label distance=0,\cmdKV@gTypes@iC] (72:0.35) -- node[label=(36+\cmdKV@general@rotate):\cmdKV@gLabels@iLC] {} (0:0.35);
    \draw[label distance=0,\cmdKV@gTypes@iD] (0:0.35) -- node[label=(-36+\cmdKV@general@rotate):\cmdKV@gLabels@iLD] {} (-72:0.35);
    \draw[label distance=0,\cmdKV@gTypes@iE] (-72:0.35) -- node[label=(-108+\cmdKV@general@rotate):\cmdKV@gLabels@iLE] {} (-144:0.35);
    #2
  \end{tikzpicture}
  \gSetStdTypes{line}
}

\def\gBoxBox{\@ifnextchar[\@gBoxBox{\@gBoxBox[]}}
\def\@gBoxBox[#1]#2{%
  \setkeys*{pre}{#1}%
  \setrmkeys{general,gTypes,gLabels}%
  \begin{tikzpicture}
    [line width=1pt,
    baseline={([yshift=-0.5ex+\cmdKV@general@yshift]current bounding box.center)},
    scale=\cmdKV@general@scale,
    rotate=\cmdKV@general@rotate,
    font=\cmdKV@general@font]
    \draw[\cmdKV@gTypes@eA] (0.3,0.3) -- (0,0) node[label=(-180+\cmdKV@general@rotate):\cmdKV@gLabels@eLA] {};
    \draw[\cmdKV@gTypes@eB] (0.3,0.9) -- (0,1.2) node[label=(180+\cmdKV@general@rotate):\cmdKV@gLabels@eLB] {};
    \draw[\cmdKV@gTypes@eC] (1.5,0.9) -- (1.8,1.2) node[label=(\cmdKV@general@rotate):\cmdKV@gLabels@eLC] {};
    \draw[\cmdKV@gTypes@eD] (1.5,0.3) -- (1.8,0) node[label=(\cmdKV@general@rotate):\cmdKV@gLabels@eLD] {};
    \draw[label distance=0,\cmdKV@gTypes@iA] (0.3,0.3) -- node[label=(180+\cmdKV@general@rotate):\cmdKV@gLabels@iLA] {} (0.3,0.9);
    \draw[label distance=0,\cmdKV@gTypes@iB] (0.3,0.9) -- node[label=(90+\cmdKV@general@rotate):\cmdKV@gLabels@iLB] {} (0.9,0.9);
    \draw[label distance=0,\cmdKV@gTypes@iC] (0.9,0.9) -- node[label=(90+\cmdKV@general@rotate):\cmdKV@gLabels@iLC] {} (1.5,0.9);
    \draw[label distance=0,\cmdKV@gTypes@iD] (1.5,0.9) -- node[label=(\cmdKV@general@rotate):\cmdKV@gLabels@iLD] {} (1.5,0.3);
    \draw[label distance=0,\cmdKV@gTypes@iE] (1.5,0.3) -- node[label=(-90+\cmdKV@general@rotate):\cmdKV@gLabels@iLE] {} (0.9,0.3);
    \draw[label distance=0,\cmdKV@gTypes@iF] (0.9,0.3) -- node[label=(-90+\cmdKV@general@rotate):\cmdKV@gLabels@iLF] {} (0.3,0.3);
    \draw[label distance=0,\cmdKV@gTypes@iG] (0.9,0.9) -- node[label=(\cmdKV@general@rotate):\cmdKV@gLabels@iLG] {} (0.9,0.3);
    #2
  \end{tikzpicture}
  \gSetStdTypes{line}
}
\def\gBoxBoxNP{\@ifnextchar[\@gBoxBoxNP{\@gBoxBoxNP[]}}
\def\@gBoxBoxNP[#1]#2{%
  \setkeys*{pre}{#1}%
  \setrmkeys{general,gTypes,gLabels}%
  \begin{tikzpicture}
    [line width=1pt,
    baseline={([yshift=-0.5ex+\cmdKV@general@yshift]current bounding box.center)},
    scale=\cmdKV@general@scale,
    rotate=\cmdKV@general@rotate,
    font=\cmdKV@general@font]
    \draw[\cmdKV@gTypes@eA] (-1.9,0.3) -- (-2.2,0) node[label=(180+\cmdKV@general@rotate):\cmdKV@gLabels@eLA] {};
    \draw[\cmdKV@gTypes@eB] (-1.9,1.1) -- (-2.2,1.4) node[label=(180+\cmdKV@general@rotate):\cmdKV@gLabels@eLB] {};
    \draw[\cmdKV@gTypes@eC] (-0.3,0.7) -- (0,0.7) node[label=(\cmdKV@general@rotate):\cmdKV@gLabels@eLC] {};
    \draw[\cmdKV@gTypes@eD] (-1.1,0.7) -- (-1.4,0.7) node[label=(180+\cmdKV@general@rotate):\cmdKV@gLabels@eLD] {};
    \draw[label distance=0,\cmdKV@gTypes@iA] (-1.9,0.3) -- node[label=(180+\cmdKV@general@rotate):\cmdKV@gLabels@iLA] {} (-1.9,1.1);
    \draw[label distance=0,\cmdKV@gTypes@iB] (-1.9,1.1) -- node[label=(90+\cmdKV@general@rotate):\cmdKV@gLabels@iLB] {} (-0.7,1.1);
    \draw[label distance=0,\cmdKV@gTypes@iC] (-0.7,1.1) -- node[label=(45+\cmdKV@general@rotate):\cmdKV@gLabels@iLC] {} (-0.3,0.7);
    \draw[label distance=0,\cmdKV@gTypes@iD] (-0.3,0.7) -- node[label=(-45+\cmdKV@general@rotate):\cmdKV@gLabels@iLD] {} (-0.7,0.3);
    \draw[label distance=0,\cmdKV@gTypes@iE] (-0.7,0.3) -- node[label=(-90+\cmdKV@general@rotate):\cmdKV@gLabels@iLE] {} (-1.9,0.3);
    \draw[label distance=0,\cmdKV@gTypes@iF] (-0.7,1.1) -- node[label=(180+\cmdKV@general@rotate):\cmdKV@gLabels@iLF] {} (-1.1,0.7);
    \draw[label distance=0,\cmdKV@gTypes@iG] (-1.1,0.7) -- node[label=(-180+\cmdKV@general@rotate):\cmdKV@gLabels@iLG] {} (-0.7,0.3);
    #2
  \end{tikzpicture}
  \gSetStdTypes{line}
}

\def\gBoxBoxNPB{\@ifnextchar[\@gBoxBoxNPB{\@gBoxBoxNPB[]}}
\def\@gBoxBoxNPB[#1]#2{%
  \setkeys*{pre}{#1}%
  \setrmkeys{general,gTypes,gLabels}%
  \begin{tikzpicture}
    [line width=1pt,
    baseline={([yshift=-0.5ex+\cmdKV@general@yshift]current bounding box.center)},
    scale=\cmdKV@general@scale,
    rotate=\cmdKV@general@rotate,
    font=\cmdKV@general@font]
    \draw[\cmdKV@gTypes@eA] (0.3,0.3) -- (0,0) node[label=(-180+\cmdKV@general@rotate):\cmdKV@gLabels@eLA] {};
    \draw[\cmdKV@gTypes@eB] (0.3,0.9) -- (0,1.2) node[label=(180+\cmdKV@general@rotate):\cmdKV@gLabels@eLB] {};
    \draw[\cmdKV@gTypes@eC] (1.5,0.9) -- (1.8,1.2) node[label=(\cmdKV@general@rotate):\cmdKV@gLabels@eLC] {};
    \draw[\cmdKV@gTypes@eD] (1.5,0.3) -- (1.8,0) node[label=(\cmdKV@general@rotate):\cmdKV@gLabels@eLD] {};
    \draw[label distance=0,\cmdKV@gTypes@iA] (0.3,0.3) -- node[label=(180+\cmdKV@general@rotate):\cmdKV@gLabels@iLA] {} (0.3,0.9);
    \draw[label distance=0,\cmdKV@gTypes@iB] (0.3,0.9) -- node[label=(90+\cmdKV@general@rotate):\cmdKV@gLabels@iLB] {} (0.9,0.9);
    \draw[label distance=0,\cmdKV@gTypes@iC] (0.9,0.9) -- node[label=(90+\cmdKV@general@rotate):\cmdKV@gLabels@iLC] {} (1.5,0.3);
    \foreach \x in {5,...,14} \fill[opacity=0.12,white] (1.2,0.6) circle (\x/100);
    \draw[label distance=0,\cmdKV@gTypes@iD] (1.5,0.3) -- node[label=(\cmdKV@general@rotate):\cmdKV@gLabels@iLD] {} (1.5,0.9);
    \draw[label distance=0,\cmdKV@gTypes@iE] (1.5,0.9) -- node[label=(-90+\cmdKV@general@rotate):\cmdKV@gLabels@iLE] {} (0.9,0.3);
    \draw[label distance=0,\cmdKV@gTypes@iF] (0.9,0.3) -- node[label=(-90+\cmdKV@general@rotate):\cmdKV@gLabels@iLF] {} (0.3,0.3);
    \draw[label distance=-1,\cmdKV@gTypes@iG] (0.9,0.9) -- node[label=(180+\cmdKV@general@rotate):\cmdKV@gLabels@iLG] {} (0.9,0.3);
    #2
  \end{tikzpicture}
  \gSetStdTypes{line}
}

\def\gBoxBoxNPC{\@ifnextchar[\@gBoxBoxNPC{\@gBoxBoxNPC[]}}
\def\@gBoxBoxNPC[#1]#2{%
  \setkeys*{pre}{#1}%
  \setrmkeys{general,gTypes,gLabels}%
  \begin{tikzpicture}
    [line width=1pt,
    baseline={([yshift=-0.5ex+\cmdKV@general@yshift]current bounding box.center)},
    scale=\cmdKV@general@scale,
    rotate=\cmdKV@general@rotate,
    font=\cmdKV@general@font]
    \draw[\cmdKV@gTypes@eA] (0.3,0.3) -- (0,0) node[label=(-180+\cmdKV@general@rotate):\cmdKV@gLabels@eLA] {};
    \draw[\cmdKV@gTypes@eB] (0.3,0.9) -- (0,1.2) node[label=(180+\cmdKV@general@rotate):\cmdKV@gLabels@eLB] {};
    \draw[\cmdKV@gTypes@eC] (1.5,0.9) -- (1.8,1.2) node[label=(\cmdKV@general@rotate):\cmdKV@gLabels@eLC] {};
    \draw[\cmdKV@gTypes@eD] (1.5,0.3) -- (1.8,0) node[label=(\cmdKV@general@rotate):\cmdKV@gLabels@eLD] {};
    \draw[label distance=0,\cmdKV@gTypes@iA] (0.3,0.3) -- node[label=(180+\cmdKV@general@rotate):\cmdKV@gLabels@iLA] {} (0.3,0.9);
    \draw[label distance=0,\cmdKV@gTypes@iB] (0.3,0.9) -- node[label=(90+\cmdKV@general@rotate):\cmdKV@gLabels@iLB] {} (0.9,0.9);
    \draw[label distance=0,\cmdKV@gTypes@iC] (0.9,0.9) -- node[label=(90+\cmdKV@general@rotate):\cmdKV@gLabels@iLC] {} (1.5,0.9);
    \draw[label distance=0,\cmdKV@gTypes@iD] (1.5,0.9) -- node[label=(\cmdKV@general@rotate):\cmdKV@gLabels@iLD] {} (0.9,0.3);
    \foreach \x in {5,...,14} \fill[opacity=0.12,white] (1.2,0.6) circle (\x/100);
    \draw[label distance=0,\cmdKV@gTypes@iE] (1.5,0.3) -- node[label=(-90+\cmdKV@general@rotate):\cmdKV@gLabels@iLE] {} (0.9,0.3);
    \draw[label distance=0,\cmdKV@gTypes@iF] (0.9,0.3) -- node[label=(-90+\cmdKV@general@rotate):\cmdKV@gLabels@iLF] {} (0.3,0.3);
    \draw[label distance=-1,\cmdKV@gTypes@iG] (0.9,0.9) -- node[label=(180+\cmdKV@general@rotate):\cmdKV@gLabels@iLG] {} (1.5,0.3);
    #2
  \end{tikzpicture}
  \gSetStdTypes{line}
}

\def\gBoxTriNPA{\@ifnextchar[\@gBoxTriNPA{\@gBoxTriNPA[]}}
\def\@gBoxTriNPA[#1]#2{%
  \setkeys*{pre}{#1}%
  \setrmkeys{general,gTypes,gLabels}%
  \begin{tikzpicture}
    [line width=1pt,
    baseline={([yshift=-0.5ex+\cmdKV@general@yshift]current bounding box.center)},
    scale=\cmdKV@general@scale,
    rotate=\cmdKV@general@rotate,
    font=\cmdKV@general@font]
    \draw[\cmdKV@gTypes@eA] (0.3,0.6) -- (0,0) node[label=(-180+\cmdKV@general@rotate):\cmdKV@gLabels@eLA] {};
    \draw[\cmdKV@gTypes@eB] (0.3,0.6) -- (0,1.2) node[label=(180+\cmdKV@general@rotate):\cmdKV@gLabels@eLB] {};
    \draw[\cmdKV@gTypes@eC] (1.5,0.9) -- (1.8,1.2) node[label=(\cmdKV@general@rotate):\cmdKV@gLabels@eLC] {};
    \draw[\cmdKV@gTypes@eD] (1.5,0.3) -- (1.8,0) node[label=(\cmdKV@general@rotate):\cmdKV@gLabels@eLD] {};
    \draw[label distance=0,\cmdKV@gTypes@iA] (0.3,0.6) -- node[label=(90+\cmdKV@general@rotate):\cmdKV@gLabels@iLA] {} (0.9,0.9);
    \draw[label distance=0,\cmdKV@gTypes@iB] (0.9,0.9) -- node[label=(90+\cmdKV@general@rotate):\cmdKV@gLabels@iLB] {} (1.5,0.9);
    \draw[label distance=-1,\cmdKV@gTypes@iC] (0.9,0.9) -- node[label=(180+\cmdKV@general@rotate):\cmdKV@gLabels@iLC] {} (1.5,0.3);
    \foreach \x in {5,...,14} \fill[opacity=0.12,white] (1.2,0.6) circle (\x/100);
    \draw[label distance=0,\cmdKV@gTypes@iD] (1.5,0.3) -- node[label=(-90+\cmdKV@general@rotate):\cmdKV@gLabels@iLD] {} (0.9,0.3);
    \draw[label distance=0,\cmdKV@gTypes@iE] (0.9,0.3) -- node[label=(-90+\cmdKV@general@rotate):\cmdKV@gLabels@iLE] {} (0.3,0.6);
    \draw[label distance=-1,\cmdKV@gTypes@iF] (1.5,0.9) -- node[label=(180+\cmdKV@general@rotate):\cmdKV@gLabels@iLF] {} (0.9,0.3);
    #2
  \end{tikzpicture}
  \gSetStdTypes{line}
}

\def\gTriPenta{\@ifnextchar[\@gTriPenta{\@gTriPenta[]}}
\def\@gTriPenta[#1]#2{%
  \setkeys*{pre}{#1}%
  \setrmkeys{general,gTypes,gLabels}%
  \begin{tikzpicture}
    [line width=1pt,
    baseline={([yshift=-0.5ex+\cmdKV@general@yshift]current bounding box.center)},
    scale=\cmdKV@general@scale,
    rotate=\cmdKV@general@rotate,
    font=\cmdKV@general@font]
    \draw[\cmdKV@gTypes@eA] (-108:0.5) -- (-135:1) node[label=(145+\cmdKV@general@rotate):\cmdKV@gLabels@eLA] {};
    \draw[\cmdKV@gTypes@eB] (180:0.5) -- (180:1) node[label=(180+\cmdKV@general@rotate):\cmdKV@gLabels@eLB] {};
    \draw[\cmdKV@gTypes@eC] (108:0.5) -- (135:1) node[label=(-145+\cmdKV@general@rotate):\cmdKV@gLabels@eLC] {};
    \draw[\cmdKV@gTypes@eD] (0:1.2) -- (0:1.5) node[label=(\cmdKV@general@rotate):\cmdKV@gLabels@eLD] {};
    \draw[label distance=0,\cmdKV@gTypes@iA] (-108:0.5) -- node[label=(-135+\cmdKV@general@rotate):\cmdKV@gLabels@iLA] {} (-180:0.5);
    \draw[label distance=0,\cmdKV@gTypes@iB] (180:0.5) -- node[label=(135+\cmdKV@general@rotate):\cmdKV@gLabels@iLB] {} (108:0.5);
    \draw[label distance=0,\cmdKV@gTypes@iC] (108:0.5) -- node[label=(72+\cmdKV@general@rotate):\cmdKV@gLabels@iLC] {} (36:0.5);
    \draw[label distance=0,\cmdKV@gTypes@iD] (36:0.5) -- node[label=(72+\cmdKV@general@rotate):\cmdKV@gLabels@iLD] {} (0:1.2);
    \draw[label distance=2,\cmdKV@gTypes@iE] (0:1.2) -- node[label=(-90+\cmdKV@general@rotate):\cmdKV@gLabels@iLE] {} (-36:0.5);
    \draw[label distance=2,\cmdKV@gTypes@iF] (-36:0.5) -- node[label=(-90+\cmdKV@general@rotate):\cmdKV@gLabels@iLF] {} (-108:0.5);
    \draw[label distance=0,\cmdKV@gTypes@iG] (36:0.5) -- node[label=(180+\cmdKV@general@rotate):\cmdKV@gLabels@iLG] {} (-36:0.5);
    #2
  \end{tikzpicture}
  \gSetStdTypes{line}
}
\def\gTriTri{\@ifnextchar[\@gTriTri{\@gTriTri[]}}
\def\@gTriTri[#1]#2{%
  \setkeys*{pre}{#1}%
  \setrmkeys{general,gTypes,gLabels}%
  \begin{tikzpicture}
    [line width=1pt,
    baseline={([yshift=-0.5ex+\cmdKV@general@yshift]current bounding box.center)},
    scale=\cmdKV@general@scale,
    rotate=\cmdKV@general@rotate,
    font=\cmdKV@general@font]
    \draw[\cmdKV@gTypes@eA] (0.3,0.3) -- (0,0) node[label=(-180+\cmdKV@general@rotate):\cmdKV@gLabels@eLA] {};
    \draw[\cmdKV@gTypes@eB] (0.3,1.1) -- (0,1.4) node[label=(180+\cmdKV@general@rotate):\cmdKV@gLabels@eLB] {};
    \draw[\cmdKV@gTypes@eC] (1.9,1.1) -- (2.2,1.4) node[label=(\cmdKV@general@rotate):\cmdKV@gLabels@eLC] {};
    \draw[\cmdKV@gTypes@eD] (1.9,0.3) -- (2.2,0) node[label=(\cmdKV@general@rotate):\cmdKV@gLabels@eLD] {};
    \draw[label distance=0,\cmdKV@gTypes@iA] (0.3,0.3) -- node[label=(180+\cmdKV@general@rotate):\cmdKV@gLabels@iLA] {} (0.3,1.1);
    \draw[label distance=0,\cmdKV@gTypes@iB] (0.3,1.1) -- node[label=(45+\cmdKV@general@rotate):\cmdKV@gLabels@iLB] {} (0.9,0.7);
    \draw[label distance=0,\cmdKV@gTypes@iC] (0.9,0.7) -- node[label=(-45+\cmdKV@general@rotate):\cmdKV@gLabels@iLC] {} (0.3,0.3);
    \draw[label distance=0,\cmdKV@gTypes@iD] (0.9,0.7) -- node[label=(90+\cmdKV@general@rotate):\cmdKV@gLabels@iLD] {} (1.3,0.7);
    \draw[label distance=0,\cmdKV@gTypes@iE] (1.3,0.7) -- node[label=(135+\cmdKV@general@rotate):\cmdKV@gLabels@iLE] {} (1.9,1.1);
    \draw[label distance=0,\cmdKV@gTypes@iF] (1.9,1.1) -- node[label=(\cmdKV@general@rotate):\cmdKV@gLabels@iLF] {} (1.9,0.3);
    \draw[label distance=0,\cmdKV@gTypes@iG] (1.9,0.3) -- node[label=(-135+\cmdKV@general@rotate):\cmdKV@gLabels@iLG] {} (1.3,0.7);
    #2
  \end{tikzpicture}
  \gSetStdTypes{line}
}

\def\gBoxBubA{\@ifnextchar[\@gBoxBubA{\@gBoxBubA[]}}
\def\@gBoxBubA[#1]#2{%
  \setkeys*{pre}{#1}%
  \setrmkeys{general,gTypes,gLabels}%
  \begin{tikzpicture}
    [line width=1pt,
    baseline={([yshift=-0.5ex+\cmdKV@general@yshift]current bounding box.center)},
    scale=\cmdKV@general@scale,
    rotate=\cmdKV@general@rotate,
    font=\cmdKV@general@font]
    \draw[\cmdKV@gTypes@eA] (0.1,0.1) -- (-0.1,-0.1) node[label=(-180+\cmdKV@general@rotate):\cmdKV@gLabels@eLA] {};
    \draw[\cmdKV@gTypes@eB] (0.1,0.9) -- (-0.1,1.1) node[label=(180+\cmdKV@general@rotate):\cmdKV@gLabels@eLB] {};
    \draw[\cmdKV@gTypes@eC] (0.9,0.9) -- (1.1,1.1) node[label=(\cmdKV@general@rotate):\cmdKV@gLabels@eLC] {};
    \draw[\cmdKV@gTypes@eD] (0.9,0.1) -- (1.1,-0.1) node[label=(\cmdKV@general@rotate):\cmdKV@gLabels@eLD] {};
    \draw[label distance=0,\cmdKV@gTypes@iA] (0.1,0.1) -- node[label=(180+\cmdKV@general@rotate):\cmdKV@gLabels@iLA] {} (0.1,0.9);
    \draw[label distance=0,\cmdKV@gTypes@iB] (0.1,0.9) -- node[label=(90+\cmdKV@general@rotate):\cmdKV@gLabels@iLB] {} (0.32,0.9);
    \draw[label distance=0,\cmdKV@gTypes@iC] (0.32,0.9) .. controls (0.32,1.15) and (0.68,1.15) .. node[label=(90+\cmdKV@general@rotate):\cmdKV@gLabels@iLC] {} (0.68,0.9);
    \draw[label distance=0,\cmdKV@gTypes@iD] (0.68,0.9) .. controls (0.68,0.65) and (0.32,0.65) .. node[label=(-90+\cmdKV@general@rotate):\cmdKV@gLabels@iLD] {} (0.32,0.9);
    \draw[label distance=0,\cmdKV@gTypes@iE] (0.68,0.9) -- node[label=(90+\cmdKV@general@rotate):\cmdKV@gLabels@iLE] {} (0.9,0.9);
    \draw[label distance=0,\cmdKV@gTypes@iF] (0.9,0.9) -- node[label=(\cmdKV@general@rotate):\cmdKV@gLabels@iLF] {} (0.9,0.1);
    \draw[label distance=0,\cmdKV@gTypes@iG] (0.9,0.1) -- node[label=(-90+\cmdKV@general@rotate):\cmdKV@gLabels@iLG] {} (0.1,0.1);
    #2
  \end{tikzpicture}
  \gSetStdTypes{line}
}

\def\gBoxBubB{\@ifnextchar[\@gBoxBubB{\@gBoxBubB[]}}
\def\@gBoxBubB[#1]#2{%
  \setkeys*{pre}{#1}%
  \setrmkeys{general,gTypes,gLabels}%
  \begin{tikzpicture}
    [line width=1pt,
    baseline={([yshift=-0.5ex+\cmdKV@general@yshift]current bounding box.center)},
    scale=\cmdKV@general@scale,
    rotate=\cmdKV@general@rotate,
    font=\cmdKV@general@font]
    \draw[label distance=-1,\cmdKV@gTypes@eA] (0,-0.4) -- (0,-0.7) node[label=(180+\cmdKV@general@rotate):\cmdKV@gLabels@eLA] {};
    \draw[\cmdKV@gTypes@eB] (-0.4,0) -- (-0.7,0) node[label=(180+\cmdKV@general@rotate):\cmdKV@gLabels@eLB] {};
    \draw[label distance=-1,\cmdKV@gTypes@eC] (0,0.4) -- (0,0.7) node[label=(180+\cmdKV@general@rotate):\cmdKV@gLabels@eLC] {};
    \draw[\cmdKV@gTypes@eD] (1.1,0) -- (1.4,0) node[label=(\cmdKV@general@rotate):\cmdKV@gLabels@eLD] {};
    \draw[label distance=0,\cmdKV@gTypes@iA] (0,-0.4) -- node[label=(-135+\cmdKV@general@rotate):\cmdKV@gLabels@iLA] {} (-0.4,0);
    \draw[label distance=0,\cmdKV@gTypes@iB] (-0.4,0) -- node[label=(135+\cmdKV@general@rotate):\cmdKV@gLabels@iLB] {} (0,0.4);
    \draw[label distance=0,\cmdKV@gTypes@iC] (0,0.4) -- node[label=(45+\cmdKV@general@rotate):\cmdKV@gLabels@iLC] {} (0.4,0);
    \draw[label distance=0,\cmdKV@gTypes@iD] (0.4,0) -- node[label=(-45+\cmdKV@general@rotate):\cmdKV@gLabels@iLD] {} (0,-0.4);
    \draw[label distance=0,\cmdKV@gTypes@iE] (0.4,0) --  node[label=(90+\cmdKV@general@rotate):\cmdKV@gLabels@iLE] {} (0.7,0);
    \draw[label distance=0,\cmdKV@gTypes@iF] (0.7,0) .. controls (0.7,0.26) and (1.1,0.26) .. node[label=(90+\cmdKV@general@rotate):\cmdKV@gLabels@iLF] {} (1.1,0);
    \draw[label distance=0,\cmdKV@gTypes@iG] (1.1,0) .. controls (1.1,-0.26) and (0.7,-0.26) ..  node[label=(-90+\cmdKV@general@rotate):\cmdKV@gLabels@iLG] {} (0.7,0);
    #2
  \end{tikzpicture}
  \gSetStdTypes{line}
}

\def\gPentaTad{\@ifnextchar[\@gPentaTad{\@gPentaTad[]}}
\def\@gPentaTad[#1]#2{%
  \setkeys*{pre}{#1}%
  \setrmkeys{general,gTypes,gLabels}%
  \begin{tikzpicture}
    [line width=1pt,
    baseline={([yshift=-0.5ex+\cmdKV@general@yshift]current bounding box.center)},
    scale=\cmdKV@general@scale,
    rotate=\cmdKV@general@rotate,
    font=\cmdKV@general@font]
    \draw[label distance=-0.5,\cmdKV@gTypes@eA] (108:0.35) -- (108:0.75) node[label=(180+\cmdKV@general@rotate):\cmdKV@gLabels@eLA] {};
    \draw[\cmdKV@gTypes@eB] (36:0.35) -- (36:0.75) node[label=(36+\cmdKV@general@rotate):\cmdKV@gLabels@eLB] {};
    \draw[\cmdKV@gTypes@eC] (-36:0.35) -- (-36:0.75) node[label=(-36+\cmdKV@general@rotate):\cmdKV@gLabels@eLC] {};
    \draw[label distance=-0.5,\cmdKV@gTypes@eD] (-108:0.35) -- (-108:0.75) node[label=(-180+\cmdKV@general@rotate):\cmdKV@gLabels@eLD] {};
    \draw[label distance=0,\cmdKV@gTypes@iA] (108:0.35) -- node[label=(72+\cmdKV@general@rotate):\cmdKV@gLabels@iLA] {} (36:0.35);
    \draw[label distance=0,\cmdKV@gTypes@iB] (36:0.35) -- node[label=(\cmdKV@general@rotate):\cmdKV@gLabels@iLB] {} (-36:0.35);
    \draw[label distance=0,\cmdKV@gTypes@iC] (-36:0.35) -- node[label=(-72+\cmdKV@general@rotate):\cmdKV@gLabels@iLC] {} (-108:0.35);
    \draw[label distance=0,\cmdKV@gTypes@iD] (-108:0.35) -- node[label=(-144+\cmdKV@general@rotate):\cmdKV@gLabels@iLD] {} (-180:0.35);
    \draw[label distance=0,\cmdKV@gTypes@iE] (180:0.35) -- node[label=(144+\cmdKV@general@rotate):\cmdKV@gLabels@iLE] {} (108:0.35);
    \draw[label distance=0,\cmdKV@gTypes@iF] (180:0.35) -- node[label=(90+\cmdKV@general@rotate):\cmdKV@gLabels@iLF] {} (180:0.75);
    \draw[label distance=-0.5,\cmdKV@gTypes@iG] (180:0.75) .. controls ($(180:0.75)+(0,-0.18)$) and ($(180:1)+(0,-0.18)$) .. (180:1) node[label=(180+\cmdKV@general@rotate):\cmdKV@gLabels@iLG] {} .. controls ($(180:1)+(0,0.18)$) and ($(180:0.75)+(0,0.18)$) .. (180:0.75);
    #2
  \end{tikzpicture}
  \gSetStdTypes{line}
}

\def\gBoxTadA{\@ifnextchar[\@gBoxTadA{\@gBoxTadA[]}}
\def\@gBoxTadA[#1]#2{%
  \setkeys*{pre}{#1}%
  \setrmkeys{general,gTypes,gLabels}%
  \begin{tikzpicture}
    [line width=1pt,
    baseline={([yshift=-0.5ex+\cmdKV@general@yshift]current bounding box.center)},
    scale=\cmdKV@general@scale,
    rotate=\cmdKV@general@rotate,
    font=\cmdKV@general@font]
    \draw[label distance=-0.5,\cmdKV@gTypes@eA] (0,0.4) -- (0,0.7) node[label=(\cmdKV@general@rotate):\cmdKV@gLabels@eLA] {};
    \draw[\cmdKV@gTypes@eB] (0.4,0) -- (0.7,0) node[label=(\cmdKV@general@rotate):\cmdKV@gLabels@eLB] {};
    \draw[label distance=-0.5,\cmdKV@gTypes@eC] (0,-0.4) -- (0,-0.7) node[label=(\cmdKV@general@rotate):\cmdKV@gLabels@eLC] {};
    \draw[\cmdKV@gTypes@eD] (-0.7,0) -- (-0.7,-0.3) node[label=(-90+\cmdKV@general@rotate):\cmdKV@gLabels@eLD] {};
    \draw[label distance=0,\cmdKV@gTypes@iA] (0,0.4) -- node[label=(45+\cmdKV@general@rotate):\cmdKV@gLabels@iLA] {} (0.4,0);
    \draw[label distance=0,\cmdKV@gTypes@iB] (0.4,0) -- node[label=(-45+\cmdKV@general@rotate):\cmdKV@gLabels@iLB] {} (0,-0.4);
    \draw[label distance=0,\cmdKV@gTypes@iC] (0,-0.4) -- node[label=(-135+\cmdKV@general@rotate):\cmdKV@gLabels@iLC] {} (-0.4,0);
    \draw[label distance=0,\cmdKV@gTypes@iD] (-0.4,0) -- node[label=(135+\cmdKV@general@rotate):\cmdKV@gLabels@iLD] {} (0,0.4);
    \draw[label distance=0,\cmdKV@gTypes@iE] (-0.4,0) --  node[label=(90+\cmdKV@general@rotate):\cmdKV@gLabels@iLE] {} (-0.7,0);
    \draw[label distance=0,\cmdKV@gTypes@iF] (-0.7,0) -- node[label=(90+\cmdKV@general@rotate):\cmdKV@gLabels@iLF] {} (-1.1,0);
    \draw[label distance=-0.5,\cmdKV@gTypes@iG] (-1.1,0) .. controls (-1.1,-0.18) and (-1.35,-0.18) .. (-1.35,0) node[label=(180+\cmdKV@general@rotate):\cmdKV@gLabels@iLG] {} .. controls (-1.35,0.18) and (-1.1,0.18) .. (-1.1,0);
    #2
  \end{tikzpicture}
  \gSetStdTypes{line}
}

\def\gSunsetA{\@ifnextchar[\@gSunsetA{\@gSunsetA[]}}
\def\@gSunsetA[#1]#2{%
  \setkeys*{pre}{#1}%
  \setrmkeys{general,gTypes,gLabels}%
  \begin{tikzpicture}
    [line width=1pt,
    baseline={([yshift=-0.5ex+\cmdKV@general@yshift]current bounding box.center)},
    scale=\cmdKV@general@scale,
    rotate=\cmdKV@general@rotate,
    font=\cmdKV@general@font]
    \draw[\cmdKV@gTypes@eA] (0.3,0.6) -- (0,0) node[label=(-180+\cmdKV@general@rotate):\cmdKV@gLabels@eLA] {};
    \draw[\cmdKV@gTypes@eB] (0.3,0.6) -- (0,1.2) node[label=(180+\cmdKV@general@rotate):\cmdKV@gLabels@eLB] {};
    \draw[\cmdKV@gTypes@eC] (1.5,0.6) -- (1.8,1.2) node[label=(\cmdKV@general@rotate):\cmdKV@gLabels@eLC] {};
    \draw[\cmdKV@gTypes@eD] (1.5,0.6) -- (1.8,0) node[label=(\cmdKV@general@rotate):\cmdKV@gLabels@eLD] {};
    \draw[label distance=0,\cmdKV@gTypes@iA] (0.3,0.6) to[bend left=90] node[label=(90+\cmdKV@general@rotate):\cmdKV@gLabels@iLA] {} (1.5,0.6);
    \draw[label distance=0,\cmdKV@gTypes@iB] (1.5,0.6) to[bend left=90] node[label=(-90+\cmdKV@general@rotate):\cmdKV@gLabels@iLB] {} (0.3,0.6);
    \draw[label distance=-2,\cmdKV@gTypes@iC] (0.3,0.6) -- node[label=(90+\cmdKV@general@rotate):\cmdKV@gLabels@iLC] {} (1.5,0.6);
    #2
  \end{tikzpicture}
  \gSetStdTypes{line}
}

\def\gTriBubA{\@ifnextchar[\@gTriBubA{\@gTriBubA[]}}
\def\@gTriBubA[#1]#2{%
  \setkeys*{pre}{#1}%
  \setrmkeys{general,gTypes,gLabels}%
  \begin{tikzpicture}
    [line width=1pt,
    baseline={([yshift=-0.5ex+\cmdKV@general@yshift]current bounding box.center)},
    scale=\cmdKV@general@scale,
    rotate=\cmdKV@general@rotate,
    font=\cmdKV@general@font]
    \draw[\cmdKV@gTypes@eA] (0.4,0.3) -- (0,0) node[label=(-180+\cmdKV@general@rotate):\cmdKV@gLabels@eLA] {};
    \draw[\cmdKV@gTypes@eB] (0.4,0.9) -- (0,1.2) node[label=(180+\cmdKV@general@rotate):\cmdKV@gLabels@eLB] {};
    \draw[\cmdKV@gTypes@eC] (1.4,0.6) -- (1.8,1.2) node[label=(\cmdKV@general@rotate):\cmdKV@gLabels@eLC] {};
    \draw[\cmdKV@gTypes@eD] (1.4,0.6) -- (1.8,0) node[label=(\cmdKV@general@rotate):\cmdKV@gLabels@eLD] {};
    \draw[label distance=0,\cmdKV@gTypes@iA] (0.4,0.3) to[bend left=40] node[label=(180+\cmdKV@general@rotate):\cmdKV@gLabels@iLA] {} (0.4,0.9);
    \draw[label distance=0,\cmdKV@gTypes@iB] (0.4,0.9) -- node[label=(90+\cmdKV@general@rotate):\cmdKV@gLabels@iLB] {} (1.4,0.6);
    \draw[label distance=0,\cmdKV@gTypes@iC] (1.4,0.6) -- node[label=(-90+\cmdKV@general@rotate):\cmdKV@gLabels@iLC] {} (0.4,0.3);
    \draw[label distance=-1,\cmdKV@gTypes@iD] (0.4,0.9) to[bend left=40] node[label=(\cmdKV@general@rotate):\cmdKV@gLabels@iLD] {} (0.4,0.3);
    #2
  \end{tikzpicture}
  \gSetStdTypes{line}
}

\def\gBubBubA{\@ifnextchar[\@gBubBubA{\@gBubBubA[]}}
\def\@gBubBubA[#1]#2{%
  \setkeys*{pre}{#1}%
  \setrmkeys{general,gTypes,gLabels}%
  \begin{tikzpicture}
    [line width=1pt,
    baseline={([yshift=-0.5ex+\cmdKV@general@yshift]current bounding box.center)},
    scale=\cmdKV@general@scale,
    rotate=\cmdKV@general@rotate,
    font=\cmdKV@general@font]
    \draw[\cmdKV@gTypes@eA] (0.3,0.6) -- (0,0) node[label=(-180+\cmdKV@general@rotate):\cmdKV@gLabels@eLA] {};
    \draw[\cmdKV@gTypes@eB] (0.3,0.6) -- (0,1.2) node[label=(180+\cmdKV@general@rotate):\cmdKV@gLabels@eLB] {};
    \draw[\cmdKV@gTypes@eC] (1.5,0.6) -- (1.8,1.2) node[label=(\cmdKV@general@rotate):\cmdKV@gLabels@eLC] {};
    \draw[\cmdKV@gTypes@eD] (1.5,0.6) -- (1.8,0) node[label=(\cmdKV@general@rotate):\cmdKV@gLabels@eLD] {};
    \draw[label distance=0,\cmdKV@gTypes@iA] (0.3,0.6) to[bend left=90] node[label=(90+\cmdKV@general@rotate):\cmdKV@gLabels@iLA] {} (0.9,0.6);
    \draw[label distance=0,\cmdKV@gTypes@iB] (0.9,0.6) to[bend left=90] node[label=(90+\cmdKV@general@rotate):\cmdKV@gLabels@iLB] {} (1.5,0.6);
    \draw[label distance=0,\cmdKV@gTypes@iC] (1.5,0.6) to[bend left=90] node[label=(-90+\cmdKV@general@rotate):\cmdKV@gLabels@iLC] {} (0.9,0.6);
    \draw[label distance=0,\cmdKV@gTypes@iD] (0.9,0.6) to[bend left=90] node[label=(-90+\cmdKV@general@rotate):\cmdKV@gLabels@iLD] {} (0.3,0.6);
    #2
  \end{tikzpicture}
  \gSetStdTypes{line}
}

\def\gBoxBubC{\@ifnextchar[\@gBoxBubC{\@gBoxBubC[]}}
\def\@gBoxBubC[#1]#2{%
  \setkeys*{pre}{#1}%
  \setrmkeys{general,gTypes,gLabels}%
  \begin{tikzpicture}
    [line width=1pt,
    baseline={([yshift=-0.5ex+\cmdKV@general@yshift]current bounding box.center)},
    scale=\cmdKV@general@scale,
    rotate=\cmdKV@general@rotate,
    font=\cmdKV@general@font]
    \draw[\cmdKV@gTypes@eA] (0.1,0.1) -- (-0.1,-0.1) node[label=(-180+\cmdKV@general@rotate):\cmdKV@gLabels@eLA] {};
    \draw[\cmdKV@gTypes@eB] (0.1,0.9) -- (-0.1,1.1) node[label=(180+\cmdKV@general@rotate):\cmdKV@gLabels@eLB] {};
    \draw[\cmdKV@gTypes@eC] (0.9,0.9) -- (1.1,1.1) node[label=(\cmdKV@general@rotate):\cmdKV@gLabels@eLC] {};
    \draw[\cmdKV@gTypes@eD] (0.9,0.1) -- (1.1,-0.1) node[label=(\cmdKV@general@rotate):\cmdKV@gLabels@eLD] {};
    \draw[label distance=0,\cmdKV@gTypes@iA] (0.1,0.1) -- node[label=(180+\cmdKV@general@rotate):\cmdKV@gLabels@iLA] {} (0.1,0.9);
    \draw[label distance=0,\cmdKV@gTypes@iB] (0.1,0.9) to[bend left] node[label=(90+\cmdKV@general@rotate):\cmdKV@gLabels@iLB] {} (0.9,0.9);
    \draw[label distance=0,\cmdKV@gTypes@iC] (0.9,0.9) to[bend left] node[label=(-90+\cmdKV@general@rotate):\cmdKV@gLabels@iLC] {} (0.1,0.9);
    \draw[label distance=0,\cmdKV@gTypes@iD] (0.9,0.9) -- node[label=(\cmdKV@general@rotate):\cmdKV@gLabels@iLD] {} (0.9,0.1);
    \draw[label distance=0,\cmdKV@gTypes@iE] (0.9,0.1) -- node[label=(-90+\cmdKV@general@rotate):\cmdKV@gLabels@iLE] {} (0.1,0.1);
    #2
  \end{tikzpicture}
  \gSetStdTypes{line}
}

\def\gBoxZ{\@ifnextchar[\@gBoxZ{\@gBoxZ[]}}
\def\@gBoxZ[#1]#2{%
  \setkeys*{pre}{#1}%
  \setrmkeys{general,gTypes,gLabels}%
  \begin{tikzpicture}
    [line width=1pt,
    baseline={([yshift=-0.5ex+\cmdKV@general@yshift]current bounding box.center)},
    scale=\cmdKV@general@scale,
    rotate=\cmdKV@general@rotate,
    font=\cmdKV@general@font]
    \draw[\cmdKV@gTypes@eA] (0.15,0.15) -- (-0.1,-0.1) node[label=(-180+\cmdKV@general@rotate):\cmdKV@gLabels@eLA] {};
    \draw[\cmdKV@gTypes@eB] (0.15,0.85) -- (-0.1,1.1) node[label=(180+\cmdKV@general@rotate):\cmdKV@gLabels@eLB] {};
    \draw[\cmdKV@gTypes@eC] (0.85,0.85) -- (1.1,1.1) node[label=(\cmdKV@general@rotate):\cmdKV@gLabels@eLC] {};
    \draw[\cmdKV@gTypes@eD] (0.85,0.15) -- (1.1,-0.1) node[label=(\cmdKV@general@rotate):\cmdKV@gLabels@eLD] {};
    \draw[label distance=0,\cmdKV@gTypes@iA] (0.15,0.15) -- node[label=(180+\cmdKV@general@rotate):\cmdKV@gLabels@iLA] {} (0.15,0.85);
    \draw[label distance=0,\cmdKV@gTypes@iB] (0.15,0.85) -- node[label=(90+\cmdKV@general@rotate):\cmdKV@gLabels@iLB] {} (0.85,0.85);
    \draw[label distance=0,\cmdKV@gTypes@iC] (0.85,0.85) -- node[label=(\cmdKV@general@rotate):\cmdKV@gLabels@iLC] {} (0.85,0.15);
    \draw[label distance=0,\cmdKV@gTypes@iD] (0.85,0.15) -- node[label=(-90+\cmdKV@general@rotate):\cmdKV@gLabels@iLD] {} (0.15,0.15);
    \draw[label distance=-4,\cmdKV@gTypes@iE] (0.15,0.15) -- node[label=(135+\cmdKV@general@rotate):\cmdKV@gLabels@iLE] {} (0.85,0.85);
    #2
  \end{tikzpicture}
  \gSetStdTypes{line}
}

\def\gBoxBoxBoxA{\@ifnextchar[\@gBoxBoxBoxA{\@gBoxBoxBoxA[]}}
\def\@gBoxBoxBoxA[#1]#2{%
  \setkeys*{pre}{#1}%
  \setrmkeys{general,gTypes,gLabels}%
  \begin{tikzpicture}
    [line width=1pt,
    baseline={([yshift=-0.5ex+\cmdKV@general@yshift]current bounding box.center)},
    scale=\cmdKV@general@scale,
    rotate=\cmdKV@general@rotate,
    font=\cmdKV@general@font]
    \draw[\cmdKV@gTypes@eA] (-0.3,0.3) -- (-0.6,0) node[label=(-180+\cmdKV@general@rotate):\cmdKV@gLabels@eLA] {};
    \draw[\cmdKV@gTypes@eB] (-0.3,0.9) -- (-0.6,1.2) node[label=(180+\cmdKV@general@rotate):\cmdKV@gLabels@eLB] {};
    \draw[\cmdKV@gTypes@eC] (1.5,0.9) -- (1.8,1.2) node[label=(\cmdKV@general@rotate):\cmdKV@gLabels@eLC] {};
    \draw[\cmdKV@gTypes@eD] (1.5,0.3) -- (1.8,0) node[label=(\cmdKV@general@rotate):\cmdKV@gLabels@eLD] {};
    \draw[label distance=0,\cmdKV@gTypes@iA] (-0.3,0.3) -- node[label=(180+\cmdKV@general@rotate):\cmdKV@gLabels@iLA] {} (-0.3,0.9);
    \draw[label distance=0,\cmdKV@gTypes@iB] (-0.3,0.9) -- node[label=(90+\cmdKV@general@rotate):\cmdKV@gLabels@iLB] {} (0.3,0.9);
    \draw[label distance=0,\cmdKV@gTypes@iC] (0.3,0.9) -- node[label=(90+\cmdKV@general@rotate):\cmdKV@gLabels@iLC] {} (0.9,0.9);
    \draw[label distance=0,\cmdKV@gTypes@iD] (0.9,0.9) -- node[label=(90+\cmdKV@general@rotate):\cmdKV@gLabels@iLD] {} (1.5,0.9);
    \draw[label distance=0,\cmdKV@gTypes@iE] (1.5,0.9) -- node[label=(\cmdKV@general@rotate):\cmdKV@gLabels@iLE] {} (1.5,0.3);
    \draw[label distance=0,\cmdKV@gTypes@iF] (1.5,0.3) -- node[label=(-90+\cmdKV@general@rotate):\cmdKV@gLabels@iLF] {} (0.9,0.3);
    \draw[label distance=0,\cmdKV@gTypes@iG] (0.9,0.3) -- node[label=(-90+\cmdKV@general@rotate):\cmdKV@gLabels@iLG] {} (0.3,0.3);
    \draw[label distance=0,\cmdKV@gTypes@iH] (0.3,0.3) -- node[label=(-90+\cmdKV@general@rotate):\cmdKV@gLabels@iLH] {} (-0.3,0.3);
    \draw[label distance=0,\cmdKV@gTypes@iI] (0.3,0.9) -- node[label=(\cmdKV@general@rotate):\cmdKV@gLabels@iLI] {} (0.3,0.3);
    \draw[label distance=0,\cmdKV@gTypes@iJ] (0.9,0.9) -- node[label=(\cmdKV@general@rotate):\cmdKV@gLabels@iLJ] {} (0.9,0.3);
    #2
  \end{tikzpicture}
  \gSetStdTypes{line}
}

\def\gBoxBoxBoxB{\@ifnextchar[\@gBoxBoxBoxB{\@gBoxBoxBoxB[]}}
\def\@gBoxBoxBoxB[#1]#2{%
  \setkeys*{pre}{#1}%
  \setrmkeys{general,gTypes,gLabels}%
  \begin{tikzpicture}
    [line width=1pt,
    baseline={([yshift=-0.5ex+\cmdKV@general@yshift]current bounding box.center)},
    scale=\cmdKV@general@scale,
    rotate=\cmdKV@general@rotate,
    font=\cmdKV@general@font]
    \draw[\cmdKV@gTypes@eA] (0.3,0.3) -- (0,0) node[label=(-180+\cmdKV@general@rotate):\cmdKV@gLabels@eLA] {};
    \draw[\cmdKV@gTypes@eB] (0.3,1.5) -- (0,1.8) node[label=(180+\cmdKV@general@rotate):\cmdKV@gLabels@eLB] {};
    \draw[\cmdKV@gTypes@eC] (1.5,1.5) -- (1.8,1.8) node[label=(\cmdKV@general@rotate):\cmdKV@gLabels@eLC] {};
    \draw[\cmdKV@gTypes@eD] (1.5,0.3) -- (1.8,0) node[label=(\cmdKV@general@rotate):\cmdKV@gLabels@eLD] {};
    \draw[label distance=0,\cmdKV@gTypes@iA] (0.3,0.3) -- node[label=(180+\cmdKV@general@rotate):\cmdKV@gLabels@iLA] {} (0.3,0.9);
    \draw[label distance=0,\cmdKV@gTypes@iB] (0.3,0.9) -- node[label=(180+\cmdKV@general@rotate):\cmdKV@gLabels@iLB] {} (0.3,1.5);
    \draw[label distance=0,\cmdKV@gTypes@iC] (0.3,1.5) -- node[label=(90+\cmdKV@general@rotate):\cmdKV@gLabels@iLC] {} (0.9,1.5);
    \draw[label distance=0,\cmdKV@gTypes@iD] (0.9,1.5) -- node[label=(90+\cmdKV@general@rotate):\cmdKV@gLabels@iLD] {} (1.5,1.5);
    \draw[label distance=0,\cmdKV@gTypes@iE] (1.5,1.5) -- node[label=(\cmdKV@general@rotate):\cmdKV@gLabels@iLE] {} (1.5,0.3);
    \draw[label distance=0,\cmdKV@gTypes@iF] (1.5,0.3) -- node[label=(-90+\cmdKV@general@rotate):\cmdKV@gLabels@iLF] {} (0.9,0.3);
    \draw[label distance=0,\cmdKV@gTypes@iG] (0.9,0.3) -- node[label=(-90+\cmdKV@general@rotate):\cmdKV@gLabels@iLG] {} (0.3,0.3);
    \draw[label distance=0,\cmdKV@gTypes@iH] (0.9,1.5) -- node[label=(\cmdKV@general@rotate):\cmdKV@gLabels@iLH] {} (0.9,0.9);
    \draw[label distance=0,\cmdKV@gTypes@iI] (0.9,0.9) -- node[label=(\cmdKV@general@rotate):\cmdKV@gLabels@iLI] {} (0.9,0.3);
    \draw[label distance=0,\cmdKV@gTypes@iJ] (0.9,0.9) -- node[label=(90+\cmdKV@general@rotate):\cmdKV@gLabels@iLJ] {} (0.3,0.9);
    #2
  \end{tikzpicture}
  \gSetStdTypes{line}
}

\def\gEFTTree{\@ifnextchar[\@gEFTTree{\@gEFTTree[]}}
\def\@gEFTTree[#1]#2{%
  \setkeys*{pre}{#1}%
  \setrmkeys{general,gTypes,gLabels}%
  \begin{tikzpicture}
    [line width=1pt,
    baseline={([yshift=-0.5ex+\cmdKV@general@yshift]current bounding box.center)},
    scale=\cmdKV@general@scale,
    rotate=\cmdKV@general@rotate,
    font=\cmdKV@general@font]
    \draw[\cmdKV@gTypes@iA] (0,0.5) node[label=(90+\cmdKV@general@rotate):\cmdKV@gLabels@eLA] {} -- node[label=(\cmdKV@general@rotate):\cmdKV@gLabels@iLA] {} (0,-0.5) node[label=(-90+\cmdKV@general@rotate):\cmdKV@gLabels@eLB] {};
    \draw[fill] (0,0.5) circle (0.02);
    \draw[fill] (0,-0.5) circle (0.02);
    #2
  \end{tikzpicture}
  \gSetStdTypes{line}
}

\def\gEFTV{\@ifnextchar[\@gEFTV{\@gEFTV[]}}
\def\@gEFTV[#1]#2{%
  \setkeys*{pre}{#1}%
  \setrmkeys{general,gTypes,gLabels}%
  \begin{tikzpicture}
    [line width=1pt,
    baseline={([yshift=-0.5ex+\cmdKV@general@yshift]current bounding box.center)},
    scale=\cmdKV@general@scale,
    rotate=\cmdKV@general@rotate,
    font=\cmdKV@general@font]
    \draw[\cmdKV@gTypes@iA] (0,0.5) node[label=(90+\cmdKV@general@rotate):\cmdKV@gLabels@eLA] {} -- node[label=(200+\cmdKV@general@rotate):\cmdKV@gLabels@iLA] {} (0.3,-0.5) node[label=(-90+\cmdKV@general@rotate):\cmdKV@gLabels@eLC] {};
    \draw[\cmdKV@gTypes@iB] (0.6,0.5) node[label=(90+\cmdKV@general@rotate):\cmdKV@gLabels@eLB] {} -- node[label=(-20+\cmdKV@general@rotate):\cmdKV@gLabels@iLB] {} (0.3,-0.5);
    \draw[fill] (0,0.5) circle (0.02);
    \draw[fill] (0.3,-0.5) circle (0.02);
    \draw[fill] (0.6,0.5) circle (0.02);
    #2
  \end{tikzpicture}
  \gSetStdTypes{line}
}

\def\gEFTY{\@ifnextchar[\@gEFTY{\@gEFTY[]}}
\def\@gEFTY[#1]#2{%
  \setkeys*{pre}{#1}%
  \setrmkeys{general,gTypes,gLabels}%
  \begin{tikzpicture}
    [line width=1pt,
    baseline={([yshift=-0.5ex+\cmdKV@general@yshift]current bounding box.center)},
    scale=\cmdKV@general@scale,
    rotate=\cmdKV@general@rotate,
    font=\cmdKV@general@font]
    \draw[\cmdKV@gTypes@iA] (0,0.5) node[label=(90+\cmdKV@general@rotate):\cmdKV@gLabels@eLA] {} -- node[label=(225+\cmdKV@general@rotate):\cmdKV@gLabels@iLA] {} (0.3,0);
    \draw[\cmdKV@gTypes@iB] (0.6,0.5) node[label=(90+\cmdKV@general@rotate):\cmdKV@gLabels@eLB] {} -- node[label=(-45+\cmdKV@general@rotate):\cmdKV@gLabels@iLB] {} (0.3,0);
    \draw[\cmdKV@gTypes@iC] (0.3,0) -- node[label=(\cmdKV@general@rotate):\cmdKV@gLabels@iLC] {} (0.3,-0.5) node[label=(-90+\cmdKV@general@rotate):\cmdKV@gLabels@eLC] {};
    \draw[fill] (0,0.5) circle (0.02);
    \draw[fill] (0.6,0.5) circle (0.02);
    \draw[fill] (0.3,-0.5) circle (0.02);
    #2
  \end{tikzpicture}
  \gSetStdTypes{line}
}

\def\gEFTPsi{\@ifnextchar[\@gEFTPsi{\@gEFTPsi[]}}
\def\@gEFTPsi[#1]#2{%
  \setkeys*{pre}{#1}%
  \setrmkeys{general,gTypes,gLabels}%
  \begin{tikzpicture}
    [line width=1pt,
    baseline={([yshift=-0.5ex+\cmdKV@general@yshift]current bounding box.center)},
    scale=\cmdKV@general@scale,
    rotate=\cmdKV@general@rotate,
    font=\cmdKV@general@font]
    \draw[\cmdKV@gTypes@iA] (0,0.5) node[label=(90+\cmdKV@general@rotate):\cmdKV@gLabels@eLA] {} -- node[label=(225+\cmdKV@general@rotate):\cmdKV@gLabels@iLA] {} (0.5,0);
    \draw[\cmdKV@gTypes@iB] (0.5,0.5) node[label=(90+\cmdKV@general@rotate):\cmdKV@gLabels@eLB] {} -- node[label=(\cmdKV@general@rotate):\cmdKV@gLabels@iLB] {} (0.5,0);
    \draw[\cmdKV@gTypes@iC] (1,0.5) node[label=(90+\cmdKV@general@rotate):\cmdKV@gLabels@eLC] {} -- node[label=(-45+\cmdKV@general@rotate):\cmdKV@gLabels@iLC] {} (0.5,0);
    \draw[\cmdKV@gTypes@iD] (0.5,0) -- node[label=(\cmdKV@general@rotate):\cmdKV@gLabels@iLD] {} (0.5,-0.5) node[label=(-90+\cmdKV@general@rotate):\cmdKV@gLabels@eLD] {};
    \draw[fill] (0,0.5) circle (0.02);
    \draw[fill] (0.5,0.5) circle (0.02);
    \draw[fill] (1,0.5) circle (0.02);
    \draw[fill] (0.5,-0.5) circle (0.02);
    #2
  \end{tikzpicture}
  \gSetStdTypes{line}
}

\def\gEFTVinY{\@ifnextchar[\@gEFTVinY{\@gEFTVinY[]}}
\def\@gEFTVinY[#1]#2{%
  \setkeys*{pre}{#1}%
  \setrmkeys{general,gTypes,gLabels}%
  \begin{tikzpicture}
    [line width=1pt,
    baseline={([yshift=-0.5ex+\cmdKV@general@yshift]current bounding box.center)},
    scale=\cmdKV@general@scale,
    rotate=\cmdKV@general@rotate,
    font=\cmdKV@general@font]
    \draw[\cmdKV@gTypes@iA] (0,0.5) node[label=(90+\cmdKV@general@rotate):\cmdKV@gLabels@eLA] {} -- node[label=(225+\cmdKV@general@rotate):\cmdKV@gLabels@iLA] {} (0.25,0.25);
    \draw[\cmdKV@gTypes@iE] (0.25,0.25) -- node[label=(225+\cmdKV@general@rotate):\cmdKV@gLabels@iLE] {} (0.5,0);
    \draw[\cmdKV@gTypes@iB] (0.5,0.5) node[label=(90+\cmdKV@general@rotate):\cmdKV@gLabels@eLB] {} -- node[label=(\cmdKV@general@rotate):\cmdKV@gLabels@iLB] {} (0.25,0.25);
    \draw[\cmdKV@gTypes@iC] (1,0.5) node[label=(90+\cmdKV@general@rotate):\cmdKV@gLabels@eLC] {} -- node[label=(-45+\cmdKV@general@rotate):\cmdKV@gLabels@iLC] {} (0.5,0);
    \draw[\cmdKV@gTypes@iD] (0.5,0) -- node[label=(\cmdKV@general@rotate):\cmdKV@gLabels@iLD] {} (0.5,-0.5) node[label=(-90+\cmdKV@general@rotate):\cmdKV@gLabels@eLD] {};
    \draw[fill] (0,0.5) circle (0.02);
    \draw[fill] (0.5,0.5) circle (0.02);
    \draw[fill] (1,0.5) circle (0.02);
    \draw[fill] (0.5,-0.5) circle (0.02);
    #2
  \end{tikzpicture}
  \gSetStdTypes{line}
}

\def\gEFTYinvY{\@ifnextchar[\@gEFTYinvY{\@gEFTYinvY[]}}
\def\@gEFTYinvY[#1]#2{%
  \setkeys*{pre}{#1}%
  \setrmkeys{general,gTypes,gLabels}%
  \begin{tikzpicture}
    [line width=1pt,
    baseline={([yshift=-0.5ex+\cmdKV@general@yshift]current bounding box.center)},
    scale=\cmdKV@general@scale,
    rotate=\cmdKV@general@rotate,
    font=\cmdKV@general@font]
    \draw[\cmdKV@gTypes@iA] (0,0.5) node[label=(90+\cmdKV@general@rotate):\cmdKV@gLabels@eLA] {} -- node[label=(225+\cmdKV@general@rotate):\cmdKV@gLabels@iLA] {} (0.3,0.166);
    \draw[\cmdKV@gTypes@iB] (0.6,0.5) node[label=(90+\cmdKV@general@rotate):\cmdKV@gLabels@eLB] {} -- node[label=(-45+\cmdKV@general@rotate):\cmdKV@gLabels@iLB] {} (0.3,0.166);
    \draw[\cmdKV@gTypes@iC] (0.3,-0.166) -- node[label=(135+\cmdKV@general@rotate):\cmdKV@gLabels@iLC] {} (0,-0.5) node[label=(-90+\cmdKV@general@rotate):\cmdKV@gLabels@eLC] {}; 
    \draw[\cmdKV@gTypes@iD] (0.3,-0.166) -- node[label=(45+\cmdKV@general@rotate):\cmdKV@gLabels@iLD] {} (0.6,-0.5) node[label=(-90+\cmdKV@general@rotate):\cmdKV@gLabels@eLD] {};
    \draw[\cmdKV@gTypes@iE] (0.3,0.166) -- node[label=(\cmdKV@general@rotate):\cmdKV@gLabels@iLE] {} (0.3,-0.166);
    \draw[fill] (0,0.5) circle (0.02);
    \draw[fill] (0.6,0.5) circle (0.02);
    \draw[fill] (0,-0.5) circle (0.02);
    \draw[fill] (0.6,-0.5) circle (0.02);
    #2
  \end{tikzpicture}
  \gSetStdTypes{line}
}

\def\gEFTX{\@ifnextchar[\@gEFTX{\@gEFTX[]}}
\def\@gEFTX[#1]#2{%
  \setkeys*{pre}{#1}%
  \setrmkeys{general,gTypes,gLabels}%
  \begin{tikzpicture}
    [line width=1pt,
    baseline={([yshift=-0.5ex+\cmdKV@general@yshift]current bounding box.center)},
    scale=\cmdKV@general@scale,
    rotate=\cmdKV@general@rotate,
    font=\cmdKV@general@font]
    \draw[\cmdKV@gTypes@iA] (0,0.5) node[label=(90+\cmdKV@general@rotate):\cmdKV@gLabels@eLA] {} -- node[label=(225+\cmdKV@general@rotate):\cmdKV@gLabels@iLA] {} (0.3,0);
    \draw[\cmdKV@gTypes@iB] (0.6,0.5) node[label=(90+\cmdKV@general@rotate):\cmdKV@gLabels@eLB] {} -- node[label=(-45+\cmdKV@general@rotate):\cmdKV@gLabels@iLB] {} (0.3,0);
    \draw[\cmdKV@gTypes@iC] (0.3,0) -- node[label=(135+\cmdKV@general@rotate):\cmdKV@gLabels@iLC] {} (0,-0.5) node[label=(-90+\cmdKV@general@rotate):\cmdKV@gLabels@eLC] {}; 
    \draw[\cmdKV@gTypes@iD] (0.3,0) -- node[label=(45+\cmdKV@general@rotate):\cmdKV@gLabels@iLD] {} (0.6,-0.5) node[label=(-90+\cmdKV@general@rotate):\cmdKV@gLabels@eLD] {};
    \draw[fill] (0,0.5) circle (0.02);
    \draw[fill] (0.6,0.5) circle (0.02);
    \draw[fill] (0,-0.5) circle (0.02);
    \draw[fill] (0.6,-0.5) circle (0.02);
    #2
  \end{tikzpicture}
  \gSetStdTypes{line}
}

\def\gEFTH{\@ifnextchar[\@gEFTH{\@gEFTH[]}}
\def\@gEFTH[#1]#2{%
  \setkeys*{pre}{#1}%
  \setrmkeys{general,gTypes,gLabels}%
  \begin{tikzpicture}
    [line width=1pt,
    baseline={([yshift=-0.5ex+\cmdKV@general@yshift]current bounding box.center)},
    scale=\cmdKV@general@scale,
    rotate=\cmdKV@general@rotate,
    font=\cmdKV@general@font]
    \draw[\cmdKV@gTypes@iA] (0,0.5) node[label=(90+\cmdKV@general@rotate):\cmdKV@gLabels@eLA] {} -- node[label=(180+\cmdKV@general@rotate):\cmdKV@gLabels@iLA] {} (0,0);
    \draw[\cmdKV@gTypes@iB] (0.6,0.5) node[label=(90+\cmdKV@general@rotate):\cmdKV@gLabels@eLB] {} -- node[label=(\cmdKV@general@rotate):\cmdKV@gLabels@iLB] {} (0.6,0);
    \draw[\cmdKV@gTypes@iC] (0.0,0) -- node[label=(180+\cmdKV@general@rotate):\cmdKV@gLabels@iLC] {} (0,-0.5) node[label=(-90+\cmdKV@general@rotate):\cmdKV@gLabels@eLC] {}; 
    \draw[\cmdKV@gTypes@iD] (0.6,0) -- node[label=(\cmdKV@general@rotate):\cmdKV@gLabels@iLD] {} (0.6,-0.5) node[label=(-90+\cmdKV@general@rotate):\cmdKV@gLabels@eLD] {};
    \draw[\cmdKV@gTypes@iE] (0,0) -- node[label=(90+\cmdKV@general@rotate):\cmdKV@gLabels@iLE] {} (0.6,0);
    \draw[fill] (0,0.5) circle (0.02);
    \draw[fill] (0.6,0.5) circle (0.02);
    \draw[fill] (0,-0.5) circle (0.02);
    \draw[fill] (0.6,-0.5) circle (0.02);
    #2
  \end{tikzpicture}
  \gSetStdTypes{line}
}

\def\cTree{\@ifnextchar[\@cTree{\@cTree[]}}
\def\@cTree[#1]#2{%
  \setkeys*{pre}{#1}%
  \setrmkeys{general,gTypes,gLabels,blobs}%
  \begin{tikzpicture}
    [line width=1pt,
    baseline={([yshift=-0.5ex+\cmdKV@general@yshift]current bounding box.center)},
    scale=\cmdKV@general@scale,
    rotate=\cmdKV@general@rotate,
    font=\cmdKV@general@font]
    \fill[\cmdKV@blobs@blobA] (0,0) ellipse (0.25 and 0.25);
    \draw[\cmdKV@gTypes@eA] ($ (0,0) + (-135:0.25) $) -- ($ (0,0) + (-135:0.6) $) node[label=(-180+\cmdKV@general@rotate):\cmdKV@gLabels@eLA] {};
    \draw[\cmdKV@gTypes@eB] ($ (0,0) + (135:0.25) $) -- ($ (0,0) + (135:0.6) $) node[label=(180+\cmdKV@general@rotate):\cmdKV@gLabels@eLB] {};
    \draw[\cmdKV@gTypes@eC] ($ (0,0) + (45:0.25) $) -- ($ (0,0) + (45:0.6) $) node[label=(\cmdKV@general@rotate):\cmdKV@gLabels@eLC] {};
    \draw[\cmdKV@gTypes@eD] ($ (0,0) + (-45:0.25) $) -- ($ (0,0) + (-45:0.6) $) node[label=(\cmdKV@general@rotate):\cmdKV@gLabels@eLD] {};
    \node[label=(90+\cmdKV@general@rotate):\cmdKV@gLabels@iLA] at (0,0.3) {};
    \node[label=(\cmdKV@general@rotate):\cmdKV@gLabels@iLB] at (0.3,0) {};
    #2
  \end{tikzpicture}
  \gSetStdTypes{line}
}

\def\cBoxA{\@ifnextchar[\@cBoxA{\@cBoxA[]}}
\def\@cBoxA[#1]#2{%
  \setkeys*{pre}{#1}%
  \setrmkeys{general,gTypes,gLabels,dots,blobs}%
  \begin{tikzpicture}
    [line width=1pt,
    baseline={([yshift=-0.5ex+\cmdKV@general@yshift]current bounding box.center)},
    scale=\cmdKV@general@scale,
    rotate=\cmdKV@general@rotate,
    font=\cmdKV@general@font]
    \fill[\cmdKV@blobs@blobA] (-0.4,0) ellipse (0.2 and 0.4);
    \fill[\cmdKV@blobs@blobB] (0.4,0) ellipse (0.2 and 0.4);
    \draw[\cmdKV@gTypes@eA] ($ (-0.4,0) + (-135:0.26) $) -- ($ (-0.4,0) + (-135:0.7) $) node[label=(-180+\cmdKV@general@rotate):\cmdKV@gLabels@eLA] {};
    \draw[\cmdKV@gTypes@eB] ($ (-0.4,0) + (135:0.26) $) -- ($ (-0.4,0) + (135:0.7) $) node[label=(180+\cmdKV@general@rotate):\cmdKV@gLabels@eLB] {};
    \draw[\cmdKV@gTypes@eC] ($ (0.4,0) + (45:0.26) $) -- ($ (0.4,0) + (45:0.7) $) node[label=(\cmdKV@general@rotate):\cmdKV@gLabels@eLC] {};
    \draw[\cmdKV@gTypes@eD] ($ (0.4,0) + (-45:0.26) $) -- ($ (0.4,0) + (-45:0.7) $) node[label=(\cmdKV@general@rotate):\cmdKV@gLabels@eLD] {};
    \draw[label distance=1,\cmdKV@gTypes@iA] ($ (-0.4,0) + (45:0.26) $) -- node[label=(90+\cmdKV@general@rotate):\cmdKV@gLabels@iLA] {}($ (0.4,0) + (135:0.26) $);
    \draw[label distance=1,\cmdKV@gTypes@iB] ($ (0.4,0) + (-135:0.26) $) -- node[label=(-90+\cmdKV@general@rotate):\cmdKV@gLabels@iLB] {}($ (-0.4,0) + (-45:0.26) $);
    \node[label=(180+\cmdKV@general@rotate):\cmdKV@gLabels@iLC] at (-0.6,0) {};
    \node[label=(\cmdKV@general@rotate):\cmdKV@gLabels@iLD] at (0.6,0) {};
    \ifKV@dots@lDots
      \draw[dotted] ($(-0.4,0) + (-150:0.7)$) to[bend left] ($(-0.4,0) + (150:0.7)$);%
    \fi
    \ifKV@dots@rDots
      \draw[dotted] ($(0.4,0) + (30:0.7)$) to[bend left] ($(0.4,0) + (-30:0.7)$);%
    \fi
    #2
  \end{tikzpicture}
  \gSetStdTypes{line}
}

\def\cBoxAA{\@ifnextchar[\@cBoxAA{\@cBoxAA[]}}
\def\@cBoxAA[#1]#2{%
  \setkeys*{pre}{#1}%
  \setrmkeys{general,gTypes,gLabels,dots,blobs}%
  \begin{tikzpicture}
    [line width=1pt,
    baseline={([yshift=-0.5ex+\cmdKV@general@yshift]current bounding box.center)},
    scale=\cmdKV@general@scale,
    rotate=\cmdKV@general@rotate,
    font=\cmdKV@general@font]
    \fill[\cmdKV@blobs@blobA] (-0.4,0) ellipse (0.2 and 0.4);
    \fill[\cmdKV@blobs@blobB] (0.4,0) ellipse (0.2 and 0.4);
    \draw[\cmdKV@gTypes@eA] ($ (-0.4,0) + (-135:0.26) $) -- ($ (-0.4,0) + (-135:0.7) $) node[label=(-180+\cmdKV@general@rotate):\cmdKV@gLabels@eLA] {};
    \draw[\cmdKV@gTypes@eB] ($ (-0.4,0) + (135:0.26) $) -- ($ (-0.4,0) + (135:0.7) $) node[label=(180+\cmdKV@general@rotate):\cmdKV@gLabels@eLB] {};
    \draw[\cmdKV@gTypes@eC] ($ (0.4,0) + (45:0.26) $) -- ($ (0.4,0) + (45:0.7) $) node[label=(\cmdKV@general@rotate):\cmdKV@gLabels@eLC] {};
    \draw[\cmdKV@gTypes@eD] ($ (0.4,0) + (-45:0.26) $) -- ($ (0.4,0) + (-45:0.7) $) node[label=(\cmdKV@general@rotate):\cmdKV@gLabels@eLD] {};
    \draw[label distance=1,\cmdKV@gTypes@iA] ($ (-0.4,0) + (60:0.32) $) to[bend left] node[label=(90+\cmdKV@general@rotate):\cmdKV@gLabels@iLA] {}($ (0.4,0) + (120:0.32) $);
    \draw[label distance=1,\cmdKV@gTypes@iB] ($ (0.4,0) + (-120:0.32) $) to[bend left]  node[label=(-90+\cmdKV@general@rotate):\cmdKV@gLabels@iLB] {}($ (-0.4,0) + (-60:0.32) $);
    \draw[dotted] (0,0.25) -- (0,-0.25);
    \node[label=(180+\cmdKV@general@rotate):\cmdKV@gLabels@iLC] at (-0.6,0) {};
    \node[label=(\cmdKV@general@rotate):\cmdKV@gLabels@iLD] at (0.6,0) {};
    \ifKV@dots@lDots
      \draw[dotted] ($(-0.4,0) + (-150:0.7)$) to[bend left] ($(-0.4,0) + (150:0.7)$);%
    \fi
    \ifKV@dots@rDots
      \draw[dotted] ($(0.4,0) + (30:0.7)$) to[bend left] ($(0.4,0) + (-30:0.7)$);%
    \fi
    #2
  \end{tikzpicture}
  \gSetStdTypes{line}
}

\def\cBoxB{\@ifnextchar[\@cBoxB{\@cBoxB[]}}
\def\@cBoxB[#1]#2{%
  \setkeys*{pre}{#1}%
  \setrmkeys{general,gTypes,gLabels,blobs}%
  \begin{tikzpicture}
    [line width=1pt,
    baseline={([yshift=-0.5ex+\cmdKV@general@yshift]current bounding box.center)},
    scale=\cmdKV@general@scale,
    rotate=\cmdKV@general@rotate,
    font=\cmdKV@general@font]
    \fill[\cmdKV@blobs@blobA] (0,-0.3) ellipse (0.4 and 0.2);
    \fill[\cmdKV@blobs@blobB] (0,0.3) ellipse (0.4 and 0.2);
    \draw[\cmdKV@gTypes@eA] ($ (0,-0.3) + (180:0.4) $) -- ($ (0,-0.3) + (-170:0.7) $) node[label=(180+\cmdKV@general@rotate):\cmdKV@gLabels@eLA] {};
    \draw[\cmdKV@gTypes@eB] ($ (0,0.3) + (180:0.4) $) -- ($ (0,0.3) + (170:0.7) $) node[label=(180+\cmdKV@general@rotate):\cmdKV@gLabels@eLB] {};
    \draw[\cmdKV@gTypes@eC] ($ (0,0.3) + (0:0.4) $) -- ($ (0,0.3) + (10:0.7) $) node[label=(\cmdKV@general@rotate):\cmdKV@gLabels@eLC] {};
    \draw[\cmdKV@gTypes@eD] ($ (0,-0.3) + (0:0.4) $) -- ($ (0,-0.3) + (-10:0.7) $) node[label=(\cmdKV@general@rotate):\cmdKV@gLabels@eLD] {};
    \draw[label distance=1,\cmdKV@gTypes@iA] ($ (0,-0.3) + (135:0.26) $) -- node[label=(-180+\cmdKV@general@rotate):\cmdKV@gLabels@iLA] {}($ (0,0.3) + (-135:0.26) $);
    \draw[label distance=1,\cmdKV@gTypes@iB] ($ (0,0.3) + (-45:0.26) $) -- node[label=(\cmdKV@general@rotate):\cmdKV@gLabels@iLB] {}($ (0,-0.3) + (45:0.26) $);
    \node[label=(90+\cmdKV@general@rotate):\cmdKV@gLabels@iLC] at (0,0.5) {};
    \node[label=(-90+\cmdKV@general@rotate):\cmdKV@gLabels@iLD] at (0,-0.5) {};
    #2
  \end{tikzpicture}
  \gSetStdTypes{line}
}

\def\cBoxBoxA{\@ifnextchar[\@cBoxBoxA{\@cBoxBoxA[]}}
\def\@cBoxBoxA[#1]#2{%
  \setkeys*{pre}{#1}%
  \setrmkeys{general,gTypes,gLabels,blobs}%
  \begin{tikzpicture}
    [line width=1pt,
    baseline={([yshift=-0.5ex+\cmdKV@general@yshift]current bounding box.center)},
    scale=\cmdKV@general@scale,
    rotate=\cmdKV@general@rotate,
    font=\cmdKV@general@font]
    \fill[\cmdKV@blobs@blobA] (-0.4,0) ellipse (0.2 and 0.4);
    \fill[\cmdKV@blobs@blobB] (0.4,0) ellipse (0.2 and 0.4);
    \fill[\cmdKV@blobs@blobC] (1.2,0) ellipse (0.2 and 0.4);
    \draw[\cmdKV@gTypes@eA] ($ (-0.4,0) + (-135:0.26) $) -- ($ (-0.4,0) + (-135:0.7) $) node[label=(-180+\cmdKV@general@rotate):\cmdKV@gLabels@eLA] {};
    \draw[\cmdKV@gTypes@eB] ($ (-0.4,0) + (135:0.26) $) -- ($ (-0.4,0) + (135:0.7) $) node[label=(180+\cmdKV@general@rotate):\cmdKV@gLabels@eLB] {};
    \draw[\cmdKV@gTypes@eC] ($ (1.2,0) + (45:0.26) $) -- ($ (1.2,0) + (45:0.7) $) node[label=(\cmdKV@general@rotate):\cmdKV@gLabels@eLC] {};
    \draw[\cmdKV@gTypes@eD] ($ (1.2,0) + (-45:0.26) $) -- ($ (1.2,0) + (-45:0.7) $) node[label=(\cmdKV@general@rotate):\cmdKV@gLabels@eLD] {};
    \draw[label distance=1,\cmdKV@gTypes@iA] ($ (-0.4,0) + (45:0.26) $) -- node[label=(90+\cmdKV@general@rotate):\cmdKV@gLabels@iLA] {}($ (0.4,0) + (135:0.26) $);
    \draw[label distance=1,\cmdKV@gTypes@iB] ($ (0.4,0) + (45:0.26) $) -- node[label=(90+\cmdKV@general@rotate):\cmdKV@gLabels@iLB] {}($ (1.2,0) + (135:0.26) $);
    \draw[label distance=1,\cmdKV@gTypes@iC] ($ (1.2,0) + (-135:0.26) $) -- node[label=(-90+\cmdKV@general@rotate):\cmdKV@gLabels@iLC] {}($ (0.4,0) + (-45:0.26) $);
    \draw[label distance=1,\cmdKV@gTypes@iD] ($ (0.4,0) + (-135:0.26) $) -- node[label=(-90+\cmdKV@general@rotate):\cmdKV@gLabels@iLD] {}($ (-0.4,0) + (-45:0.26) $);
    \node[label=(180+\cmdKV@general@rotate):\cmdKV@gLabels@iLE] at (-0.6,0) {};
    \node[label=(\cmdKV@general@rotate):\cmdKV@gLabels@iLF] at (0.6,0) {};
    \node[label=(\cmdKV@general@rotate):\cmdKV@gLabels@iLG] at (1.4,0) {};
    #2
  \end{tikzpicture}
  \gSetStdTypes{line}
}

\def\cBoxBoxB{\@ifnextchar[\@cBoxBoxB{\@cBoxBoxB[]}}
\def\@cBoxBoxB[#1]#2{%
  \setkeys*{pre}{#1}%
  \setrmkeys{general,gTypes,gLabels}%
  \begin{tikzpicture}
    [line width=1pt,
    baseline={([yshift=-0.5ex+\cmdKV@general@yshift]current bounding box.center)},
    scale=\cmdKV@general@scale,
    rotate=\cmdKV@general@rotate,
    font=\cmdKV@general@font]
    \fill[\cmdKV@blobs@blobA] (0,0.3) ellipse (0.4 and 0.2);
    \fill[\cmdKV@blobs@blobB] (0,-0.3) ellipse (0.4 and 0.2);
    \fill[\cmdKV@blobs@blobC] (1,0) ellipse (0.2 and 0.4);
    \draw[\cmdKV@gTypes@eA] ($ (0,-0.3) + (180:0.4) $) -- ($ (0,-0.3) + (-170:0.7) $) node[label=(180+\cmdKV@general@rotate):\cmdKV@gLabels@eLA] {};
    \draw[\cmdKV@gTypes@eB] ($ (0,0.3) + (180:0.4) $) -- ($ (0,0.3) + (170:0.7) $) node[label=(180+\cmdKV@general@rotate):\cmdKV@gLabels@eLB] {};
    \draw[\cmdKV@gTypes@eC] ($ (1,0) + (45:0.26) $) -- ($ (1,0) + (45:0.7) $) node[label=(\cmdKV@general@rotate):\cmdKV@gLabels@eLC] {};
    \draw[\cmdKV@gTypes@eD] ($ (1,0) + (-45:0.26) $) -- ($ (1,0) + (-45:0.7) $) node[label=(\cmdKV@general@rotate):\cmdKV@gLabels@eLD] {};
    \draw[label distance=1,\cmdKV@gTypes@iA] ($ (0,-0.3) + (135:0.26) $) -- node[label=(-180+\cmdKV@general@rotate):\cmdKV@gLabels@iLA] {}($ (0,0.3) + (-135:0.26) $);
    \draw[label distance=1,\cmdKV@gTypes@iB] ($ (0,0.3) + (0:0.4) $) -- node[label=(90+\cmdKV@general@rotate):\cmdKV@gLabels@iLB] {} ($ (1,0) + (135:0.26) $);
    \draw[label distance=1,\cmdKV@gTypes@iC] ($ (1,0) + (-135:0.26) $) -- node[label=(-90+\cmdKV@general@rotate):\cmdKV@gLabels@iLC] {} ($ (0,-0.3) + (0:0.4) $);
    \draw[label distance=1,\cmdKV@gTypes@iD] ($ (0,0.3) + (-45:0.26) $) -- node[label=(\cmdKV@general@rotate):\cmdKV@gLabels@iLD] {}($ (0,-0.3) + (45:0.26) $);
    \node[label=(90+\cmdKV@general@rotate):\cmdKV@gLabels@iLE] at (0,0.5) {};
    \node[label=(\cmdKV@general@rotate):\cmdKV@gLabels@iLF] at (1.2,0) {};
    \node[label=(-90+\cmdKV@general@rotate):\cmdKV@gLabels@iLG] at (0,-0.5) {};
    #2
  \end{tikzpicture}
  \gSetStdTypes{line}
}

\def\cNPBox{\@ifnextchar[\@cNPBox{\@cNPBox[]}}
\def\@cNPBox[#1]#2{%
  \setkeys*{pre}{#1}%
  \setrmkeys{general,gTypes,gLabels,blob}%
  \begin{tikzpicture}
    [line width=1pt,
    baseline={([yshift=-0.5ex+\cmdKV@general@yshift]current bounding box.center)},
    scale=\cmdKV@general@scale,
    rotate=\cmdKV@general@rotate,
    font=\cmdKV@general@font]
    \fill[\cmdKV@blobs@blobA] (-0.6,0) ellipse (0.2 and 0.4);
    \fill[\cmdKV@blobs@blobB] (0.4,0) ellipse (0.2 and 0.4);
    \draw[\cmdKV@gTypes@eA] ($ (-0.6,0) + (-180:0.2) $) -- ($ (-0.6,0) + (-180:0.6) $) node[label=(-180+\cmdKV@general@rotate):\cmdKV@gLabels@eLA] {};
    \draw[\cmdKV@gTypes@eB] ($ (-0.6,0) + (0:0.2) $) -- (-0.18,0) node[label=(\cmdKV@general@rotate):\cmdKV@gLabels@eLB] {};
    \draw[\cmdKV@gTypes@eC] ($ (0.4,0) + (45:0.26) $) -- ($ (0.4,0) + (45:0.7) $) node[label=(\cmdKV@general@rotate):\cmdKV@gLabels@eLC] {};
    \draw[\cmdKV@gTypes@eD] ($ (0.4,0) + (-45:0.26) $) -- ($ (0.4,0) + (-45:0.7) $) node[label=(\cmdKV@general@rotate):\cmdKV@gLabels@eLD] {};
    \draw[label distance=1,\cmdKV@gTypes@iA] ($ (-0.6,0) + (60:0.3) $) -- node[label=(90+\cmdKV@general@rotate):\cmdKV@gLabels@iLA] {}($ (0.4,0) + (135:0.26) $);
    \draw[label distance=1,\cmdKV@gTypes@iB] ($ (0.4,0) + (-135:0.26) $) -- node[label=(-90+\cmdKV@general@rotate):\cmdKV@gLabels@iLB] {}($ (-0.6,0) + (-60:0.3) $);
    \node[label=(\cmdKV@general@rotate):\cmdKV@gLabels@iLD] at (0.6,0) {};
    #2
  \end{tikzpicture}
  \gSetStdTypes{line}
}

\def\cFourPtRec{\@ifnextchar[\@cFourPtRec{\@cFourPtRec[]}}
\def\@cFourPtRec[#1]#2{%
  \setkeys*{pre}{#1}%
  \setrmkeys{general,gTypes,gLabels,blobs}%
  \begin{tikzpicture}
    [line width=1pt,
    baseline={([yshift=-0.5ex+\cmdKV@general@yshift]current bounding box.center)},
    scale=\cmdKV@general@scale,
    rotate=\cmdKV@general@rotate,
    font=\cmdKV@general@font]
    \fill[\cmdKV@blobs@blobA] (-0.5,0) ellipse (0.2 and 0.3);
    \fill[\cmdKV@blobs@blobB] (0.5,0) ellipse (0.3 and 0.4);
    \filldraw[fill=white,draw=black] (0.55,0.15) circle (0.1);
    \filldraw[fill=white,draw=black] (0.45,-0.1) circle (0.1);
    \draw[\cmdKV@gTypes@eA] ($(-0.5,0) + (135:0.23)$) -- ($(-0.5,0) + (135:0.5)$) node[label=(180+\cmdKV@general@rotate):\cmdKV@gLabels@eLA] {};
    \draw[\cmdKV@gTypes@eB] ($(-0.5,0) + (-135:0.23)$) -- ($(-0.5,0) + (-135:0.5)$) node[label=(180+\cmdKV@general@rotate):\cmdKV@gLabels@eLB] {};
    \draw[\cmdKV@gTypes@eC] ($(0.5,0) + (45:0.35)$) -- ($(0.5,0) + (45:0.7)$) node[label=(\cmdKV@general@rotate):\cmdKV@gLabels@eLC] {};
    \draw[\cmdKV@gTypes@eD] ($(0.5,0) + (-45:0.35)$) -- ($(0.5,0) + (-45:0.7)$) node[label=(\cmdKV@general@rotate):\cmdKV@gLabels@eLD] {};
    \draw[label distance=1,\cmdKV@gTypes@iA] ($(-0.5,0) + (30:0.23)$) -- node[label=(90+\cmdKV@general@rotate):\cmdKV@gLabels@iLA] {} ($(0.5,0) + (157:0.3)$);
    \draw[label distance=1,\cmdKV@gTypes@iB] ($(0.5,0) + (-157:0.3)$) -- node[label=(-90+\cmdKV@general@rotate):\cmdKV@gLabels@iLB] {} ($(-0.5,0) + (-30:0.23)$);
    #2
 \end{tikzpicture}
  \gSetStdTypes{line}
}

\makeatother

\numberwithin{equation}{section}

\newcommand\beq{\begin{equation}}
\newcommand\eeq{\end{equation}}
\newcommand\beal{\begin{aligned}}
\newcommand\eeal{\end{aligned}}
\newcommand\bea{\begin{eqnarray}}
\newcommand\eea{\end{eqnarray}}

\newcommand\dd{{\mathrm d}}
\usepackage{xcolor}

\newcommand{\bell}{{\boldsymbol \ell}}

\newcommand{\bk}{{\boldsymbol k}}

\newcommand{\bp}{{\boldsymbol p}}

\newcommand{\bq}{{\boldsymbol q}}

\newcommand{\bx}{{\boldsymbol x}}

\newcommand{\cD}{\mathcal{D}}

\newcommand\cO{\mathcal{O}}

\newcommand\cS{\mathcal{S}}
\newcommand\cR{\mathbb{R}}

\newcommand\Mp{M_{\rm Pl}}

\makeatletter
\newcommand{\Biggg}{\bBigg@{3.5}}
\makeatother

\begin{document}
\preprint{DESY\,22-109}
\title{\center Radiation-Reaction in the Effective Field Theory Approach to Post-Minkowskian Dynamics}
\author[a]{\large  Gregor K\"alin,}
\affiliation[a]{Deutsches Elektronen-Synchrotron DESY, Notkestr. 85, 22607 Hamburg, Germany}
\author[b]{\large Jakob Neef,}
\affiliation[b]{Humboldt-Universit\"at zu Berlin,
Zum Grossen Windkanal~2, D-12489 Berlin, Germany}
\author[a]{\large and Rafael A. Porto}
\emailAdd{gregor.kaelin@desy.de}\, \emailAdd{jakob.neef@physik.hu-berlin.de}\, \emailAdd{rafael.porto@desy.de}
\abstract{ We extend the Post-Minkowskian~(PM) effective field theory (EFT) approach to incorporate conservative and dissipative radiation-reaction effects in a unified framework. This is achieved by implementing the Schwinger-Keldysh  ``in-in" formalism and separating conservative and non-conservative terms according to the formulation in \cite{chadprl}, which we show promotes Feynman's $i0$-prescription and {\it cutting} rules to a prominent role at~the~classical~level. The~resulting integrals, involving both Feynman and retarded propagators, can be bootstrapped to all orders in the velocity via differential equations with boundary conditions including potential and radiation modes. 
As~a~paradigmatic example we provide an {\it ab initio} derivation of the classical solution to the scattering problem in general relativity to ${\cal O}(G^3)$. For the sake of completeness, we also reproduce the leading order radiation-reaction effects in classical electrodynamics.}
\maketitle
\newpage

\section{Introduction} \label{sec:introduction}

 The successes of current gravitational wave (GW) detectors~\cite{LIGO}, 
together with the expected reach of future observatories \cite{LISA,ET,buosathya,tune,music,Barausse:2020rsu,Bernitt:2022aoa}, has reinvigorated various efforts to tackle the perturbative regime of the two-body dynamics in gravity stemming off of ``traditional"\cite{Damour:2008yg,blanchet,Schafer:2018kuf} as well as those inspired by effective field theory (EFT) techniques \cite{walterLH,iragrg,foffa,review}. In a concerted effort with various other theoretical frameworks, e.g. \cite{damour1,damour2,4pndjs,4pnbla,4pnbla2,Marchand:2017pir,damour3n,binidam1,binidam2,binit,Bini:2021gat,Damour:2020tta,Marchand:2020fpt,Larrouturou:2021dma,Larrouturou:2021gqo}, EFT-based approaches---both in the Post-Newtonian (PN)~\cite{nrgr,nrgrs,dis1,dis2,prl,nrgrss,nrgrs2,nrgrso,andirad,andirad2,amps,srad,chadRR,chadbr2,tail,natalia1,natalia2,apparent,nrgr4pn1,nrgr4pn2,5pn1,5pn2,hered1,hered2,tail3,blum,blum2,Blumlein:2020pyo,Blumlein:2021txe,dis3,Levi:2020uwu,Levi:2020kvb,withchad,radnrgr,pardo,Cho:2021mqw,Cho2022} and Post-Minkowskian (PM) regimes \cite{paper1,paper2,b2b3,pmeft,3pmeft,tidaleft,pmefts,4pmeft,4pmeft2,janmogul,janmogul2,Jakobsen:2022fcj,Jakobsen:2021zvh,Mougiakakos:2021ckm,Riva:2021vnj,Mougiakakos:2022sic,Riva:2022fru}---in parallel with amplitudes-based methodologies, e.g.~\cite{Holstein:2008sx,ira1,Vaidya:2014kza,Walter,Goldberger:2016iau,cheung,bohr,Guevara:2018wpp,cristof1,donal,donalvines,zvi1,zvi2,Haddad:2020que,Aoude:2020onz,Aoude:2022thd,Bjerrum-Bohr:2021din,zvispin,soloncheung,andres2,4pmzvi,4pmzvi2,Gabriele,Gabriele2,Parra,Parra3,FebresCordero:2022jts,Manohar:2022dea}, have led to major breakthroughs in our understanding of the (classical) relativistic problem in general relativity.  However, albeit with some notable exceptions, e.g. \cite{natalia1,natalia2,pardo,Cho:2021mqw,Cho2022,Riva:2022fru}, the vast majority of the EFT-rooted results have occurred in the conservative sector of PN/PM theory, culminating in the recent derivation of the state-of-the-art knowledge for potential \cite{4pmeft,4pmzvi,5pn1,5pn2,Blumlein:2020pyo} and radiation-reaction tail effects \cite{4pmeft2,4pmzvi2,hered1,hered2,Blumlein:2021txe} to 5PN and 4PM order for non-spinning bodies, respectively. Following pioneering developments in the realm of PN expansions \cite{chadprl,chadRR,chadbr2,natalia1,natalia2,tail}, the purpose of this paper is to extend the EFT approach in order to incorporate conservative and dissipative effects in the PM regime within a unified framework.\vskip 4pt In principle, radiative observables at infinity, such as the (source) radiated energy, can be computed by matching the stress-energy tensor using the trajectories in the conservative PM EFT \cite{pmeft} and subsequently squaring the (on-shell) one-point function \cite{Riva:2021vnj,Mougiakakos:2022sic,Riva:2022fru}. Mimicking the simplifications also found in PN theory \cite{andirad,radnrgr,pardo,Cho:2021mqw,Cho2022}, this piecewise derivation---{\it one scale at a time}---thus becomes simpler than constructing the full gravitational integrand, entailing instead only  one ``radiation graviton" coupled to potential exchanges. This is even more striking if we resort to an adiabatic approach within a PN multipole expansion \cite{andirad,andirad2,srad}. In this case so-called {\it moment relations} (from Ward identities) allow us to trade the non-linearities from derivatives of the multipoles that can be computed using the conservative potential,~e.g.~\cite{radnrgr,pardo,Cho:2021mqw,Cho2022}. Instead, in this paper we develop a formalism which allows us to simultaneously derive both conservative and dissipative effects all at once.\vskip 4pt

One of the major bottlenecks in computations in the PM regime has been the existence of intricate families of relativistic integrals, e.g. \cite{4pmzvi,4pmzvi2,4pmeft,4pmeft2}. Remarkably, following key ideas in particle physics \cite{Chen:1977oja,Chetyrkin:1981qh, Tkachov:1981wb,Kotikov:1991pm,Remiddi:1997ny,Beneke:1997zp,Goncharov:2001iea,Jantzen:2012mw,Smirnov:2015mct,Lee:2012cn,Smirnov:2012gma,Henn:2013pwa,Lee:2014ioa,Meyer:2016zeb,Meyer:2016slj,Prausa:2017ltv,Adams:2018yfj,Broedel:2019kmn,Primo:2017ipr,Smirnov:2019qkx,Smirnov:2020quc,Lee:2020zfb,Hidding:2020ytt,Duhr:2014woa,Duhr:2019tlz,Dlapa:2020cwj,Smirnov:2021rhf,Lee:2019zop,Lee:2013mka}, a very powerful tool has emerged as a weapon-of-choice to tackle multi-loop relativistic integration, namely the use of differential equations \cite{Parra,3pmeft}. One of the main virtues of this methodology is that only through their boundary values the solution may be sensitive to whether potential and/or radiation modes are considered. In~other words, they allow us to tackle the  so-called (classical) {\it soft} region all at once. This feature was exploited in recent amplitude-based approaches, e.g.  \cite{Parra3,Gabriele2,Bjerrum-Bohr:2021din}, to compute the relativistic impulse at 3PM order. In these methodologies one relies on taking the classical limit of a quantum amplitude, similarly to the derivations in \cite{zvi1,zvi2,4pmzvi,4pmzvi2} for the conservative sector. However, as we demonstrate here, we can bypass these manipulations and directly compute the full relativistic impulse within a worldline EFT approach instead.\vskip 4pt
The use of differential equations has been already implemented in the EFT approach to solve the conservative dynamics to 4PM order, first using only the potential region for the boundary conditions \cite{3pmeft,4pmeft} and later on incorporating radiation-reaction effects \cite{4pmeft2}. Yet, in all cases the standard ``in-out" {\it vaccum-to-vaccum} amplitude was utilised, together with Feynman's prescription for the propagators, which captures conservative-only contributions. However, the existence of such a powerful machinery in the form of differential equations---encoding in principle both conservative and radiative contributions to the dynamics---begs us for a more general framework where all dynamical effects may be incorporated in unison: {\it feeding two birds with one~scone}. Fortunately, the Schwinger-Keldysh ``in-in" formalism \cite{Schwinger:1960qe,Keldysh:1964ud,Calzetta:1986ey,Calzetta:1986cq,Jordan:1986ug} is tailor-made for this purpose. Contrary to in-out boundary conditions, the in-in approach systematically accounts for the correct causal propagation, while
at the same time the prescription in \cite{chadprl} provides us with a natural separation between conservative and dissipative terms at the level of the effective action (already  implemented in \cite{tail,nrgr4pn2} to obtain unambiguous results at 4PN). As we shall see, the procedure promotes Feynman's $i0$-prescription and {\it cutting} rules to a prominent role also at the classical level.\vskip 4pt  The full integrand including dissipative effects turns out to be a tad more involved than matching for the stress-energy tensor; however, as we show here it is practically equivalent to building the in-out counterpart. Hence, the difference between conservative and/or dissipative terms ultimately arises only through the choice of propagators. Nevertheless, as we shall see, the type of Green's functions does have an impact in the application of  standard integration tools, notably the use of `symmetry relations', which is essentially the part of the computation which requires a careful treatment. Yet, bootstrapping the solution through differential equations yields powerful constraints, in particular in the radiation sector, which allow us to read off much of the final answer without the need to compute all of the  boundary constants \cite{4pmeft2}.  As a paradigmatic example we explicitly derive the full solution of the scattering problem to 3PM order, reproducing previous results in the literature \cite{3pmeft,zvi1,Riva:2021vnj,Parra3,Gabriele2,Bjerrum-Bohr:2021din}. For~completeness, we also reproduce the leading radiation-reaction effects in classical electrodynamics~\cite{Saketh:2021sri,Bern:2021xze}.\vskip 4pt

Throughout this paper we use the following conventions: We~use the mostly minus signature: $\eta_{\mu\nu} = {\rm diag}(+,-,-,-)$. The Minkowski product between four-vectors is denoted as $k \cdot x = \eta_{\mu\nu} k^\mu x^\nu$, and we use $\bk \cdot \bx = \delta^{ij} \bk^i \bx^j$ for the Euclidean version, with bold letters representing ${\bf 3}$-vectors. We~use the shorthand notation $\int_k \equiv \int \dd^D k/\pi^{D/2}$ and  $\hat\delta^{(n)}(x) \equiv 2\pi \delta^{(n)}(x)$ for the $n$-th derivative of Dirac-$\delta$ functions. We handle divergent integrals through dimensional regularization (dim. reg.), with $D= 4-2\epsilon$.  We~define $\Mp^{-1} \equiv \sqrt{32\pi G}$ for the Planck mass in $\hbar=c=1$ units. We~denote  $M=m_1+m_2$ the total mass, $\mu = m_1m_2/M$ the reduced mass, $\nu \equiv \mu/M$ the symmetric mass ratio and $\Delta \equiv (m_1-m_2)/M = \sqrt{1-4\nu}$ for the (dimensionless) mass difference (with $m_1  \geq m_2$).  
 
 \section{Schwinger-Keldysh  worldline theory} \label{inin}
 
 The initial value formulation of non-conservative classical dynamics at the level of Hamilton's action principle was broadly discussed~in~\cite{chadprl}, and more concretely in \cite{chadRR,chadbr2,tail,natalia1,natalia2} in the context of  the EFT formalism with multipole expanded PN sources introduced in \cite{nrgr}. In this section we adapt the same formalism to the worldline EFT approach in the PM regime and encourage the reader to consult \cite{chadprl,chadRR,chadbr2,tail,natalia1,natalia2} for further details.
 
 \subsection{Basic formalism}
 
The main idea is to compute the effective action imposing only vacuum boundary conditions for the gravitational field at initial time. This is achieved by performing a `closed-time-path' integral over the field variables, e.g. \cite{Calzetta:1986ey,Calzetta:1986cq}. To this end it is convenient to introduce auxiliary fields, both for the metric and worldline sources. In our case at hand, the effective action for the two-body problem then takes the form  (schematically)
\beq
{\rm exp}  \big(i\cS_{\rm eff}[x_{a(1)}, x_{a(2)}]\big) = \!\int\! \cD h_1 \cD h_2\, {\rm exp} \big( i S_{\rm EH}[h_1] - iS_{\rm EH}[h_2] + i S_{\rm pp}[h_1,x_{a,1}] - iS_{\rm pp}[h_2,x_{a ,2}]\big)\label{seff},
\eeq
with a doubling of degrees of freedom both for the gravitational field as well as the particles $(a=1,2)$, with the latter treated as external sources. The Einstein-Hilbert action is given by,
\beq
S_{\rm EH} = -2 \Mp^2 \int d^4x \sqrt{-g} \,R \,,\label{seh}
\eeq
while the point-particle action is of the Polyakov form \cite{pmeft}
\beq
S_{\rm pp} = -\sum_a \frac{m_a}{2} \int \dd \tau_a \left( v_a^2 + \frac{h_{\mu\nu}}{\Mp} v_a^\mu v_a^\nu\right)\,,
\eeq
with $v^\mu \equiv dx^\mu/d\tau$ the four-velocity. A suitable gauge-fixing term must be added to~\eqref{seh}~\cite{pmeft}. The  path integral in \eqref{seff} over the metric degrees of freedom is then computed via the saddle point approximation, without including (quantum) `loops' of the gravitational field(s) \cite{pmeft}. \vskip 4pt 
In these variables the matrix of propagators for the $h_{1,2}$ gravitational perturbation(s) becomes, see Appendix~\ref{appA},
\begin{equation}
 G_{AB}(x-y) = i \begin{pmatrix} \Delta_{\rm F}(x-y) & -\Delta_-(x-y) \\ -\Delta_+(x-y) & \Delta_{\rm D}(x-y) \end{pmatrix}\,,\label{gmatrix}
\end{equation}
 with  $A,B \in\{1,2\}$.\vskip 4pt The equations of motion follow from $S_{\rm eff}[x_{a(1)}, x_{a(2)}]$ via the standard Euler-Lagrange procedure, and subsequently identifying $x^\alpha_{a,1} \Leftrightarrow x^\alpha_{a,2}$.\vskip 4pt  
An alternative representation, often used in the literature, is given by the following change of variables due to Keldysh~\cite{Keldysh:1964ud}
\begin{equation}
\label{hpm}
  \begin{array}{rl}
    h^-_{\mu\nu} &= \frac{1}{2}(h_{1\mu\nu}  +h_{2\mu\nu} ) \\
    h^+_{\mu\nu} &= h_{1\mu\nu}  -h_{2\mu\nu} 
  \end{array}
  \Leftrightarrow
  \begin{array}{rl}
    h_{1\mu\nu} &= h^-_{\mu\nu}+\frac{1}{2} h^+_{\mu\nu}\\
     h_{2\mu\nu} &= h^-_{\mu\nu}-\frac{1}{2} h^+_{\mu\nu}\,.
  \end{array}
  \end{equation}
and likewise to the external worldline sources,
\begin{equation}
  \begin{array}{rl}
    x^\alpha_{a,+} &= \frac{1}{2}(x^\alpha_{a,1}+x^\alpha_{a,2}) \\
    x^\alpha_{a,-} &= x^\alpha_{a,1}-x^\alpha_{a,2}
  \end{array}
  \Leftrightarrow
  \begin{array}{rl}
    x^\alpha_{a,1} &= x^\alpha_{a,+}+\frac{1}{2}x^\alpha_{a,-} \\
    x^\alpha_{a,2} &= x^\alpha_{a,+}-\frac{1}{2}x^\alpha_{a,-}\,.
  \end{array}
  \end{equation}
 In this basis the matrix of propagators takes the most familiar form dictated by causality,\footnote{Notice the difference between upper and lower indices.}
  \begin{equation}
  K^{AB}(x-y) = i \begin{pmatrix} 0 & -\Delta_{\textnormal{adv}}(x-y) \\ -\Delta_{\textnormal{ret}}(x-y) & \frac{1}{2}\Delta_H(x-y)\end{pmatrix}\,,\label{kmatrix}
\end{equation}
where this time $A,B \in\{+,-\}$. The $\Delta_H(x)\equiv \Delta_+(x)+\Delta_-(x)$ is the Hadamard propagator, see Appendix~\ref{appA}, which does not feature in the classical regime and may be ignored in the saddle point approximation. In this basis the equations of motion follow from the condition,
 \beq \left.\frac{\delta \cS_\textrm{eff}[x_{a,+},x_{a,-}]}{\delta x_{b,-}^\mu(\tau)}\right|_{\substack{x_{a,-}\to 0\\x_{a,+}\to x_a}} =0\,.
\label{seff1}
 \eeq
 It is straightforward to show that the effective action is given by the general form
\begin{equation}
  \begin{aligned}
    \cS_\textrm{eff}[x_{a,+},x_{a,-}] &= \sum_a m_a\int\dd\tau_a \left(-\frac{1}{2}{v}_{a,1}^2(\tau_a)+\frac{1}{2}{v}_{a,2}^2(\tau_a)\right)+\cS_\textrm{int}[x_{a,1},x_{a,2}]\\
    &= - \sum_a m_a\int\dd\tau_a \, v_{a,+} \cdot v_{a,-}+\cS_\textrm{int}[x_{a,+},x_{a,-}]\,,\label{seff2}
  \end{aligned}
\end{equation}
 such that
 \begin{equation}\label{eq:inineom}
  m_b \frac{d}{d\tau} {v}_{b}^\mu(\tau) = \left. -\eta^{\mu\nu}\frac{\delta\cS_\textrm{int}[x_{a,+},x_{a,-}]}{\delta x_{b,-}^\nu(\tau)}\right|_{\substack{x_{a,-}\to 0\\x_{a,+}\to x_a}}\,,
\end{equation}
and the total impulse, e.g. for particle 1, becomes
\begin{equation}\label{eq:impulse}
  \begin{aligned}
    \Delta p_1^\mu &= m_1 \Delta {v}_1^\mu = m_1 \left({v}_1^\mu(\infty) - {v}_1^\mu(-\infty)\right)\\
   & = m_1 \int_{-\infty}^\infty \dd\tau_1 \frac{d}{d \tau_1} {v}_1^\mu(\tau_1)
   = \left.-\eta^{\mu\nu} \int_{-\infty}^\infty \dd\tau_1 \frac{\delta\cS_\textrm{int}\left[x_{a,+},x_{a,-}\right]}{\delta x_{1,-}^\nu(\tau_1)}\right|_{\substack{x_{a,-}\to 0\\x_{a,+}\to x_a}}\,.
  \end{aligned}
\end{equation}
Our task now is to compute the interaction part of the effective action, $\cS_\textrm{int}\left[x_{a,+},x_{a,-}\right]$, using the Feynman rules that descend from \eqref{seff}. Despite the doubling of degrees of freedom in the in-in effective theory, we shall demonstrate next that the construction of the full integrand turns out to be remarkably similar to the computation of the impulse using the standard in-out Feynman rules \cite{3pmeft,4pmeft,4pmeft2}.

\subsection{Feynman rules for the \textit{in-in}tegrand}

The quantity of interest, once the graviton fields have been integrated out, is the variation of the effective action which appears in the impulse \eqref{eq:impulse}. Rather than the effective action, it turns out to be more efficient in this case to construct Feynman rules directly for its variation, which features both in the equations of motion and total impulse. The reason is two-fold. First of all, the physical limit $x_{a-} \to 0$ can be explicitly taken, and moreover one can readily identify the similarities with the in-out integrand prior to performing the integration. For notational simplicity throughout this section we suppress sums over particle's ($a=1,2$) labels.\vskip 4pt
 
Let us start with the propagators. To construct the integrand it turns out to be convenient to use the Keldysh representation.  Hence, since we will be dealing with the retarded (or advanced) Green's functions in \eqref{kmatrix}, we will depict the graviton propagator as
\begin{equation}
  \begin{tikzpicture}
    [baseline={([yshift=-0.5ex]current bounding box.center)}]
    \draw[graviton] (0,0) node[left] {$\mu\nu$} to node[above=0.2] {$\underset{\rightarrow}{p}$} (1.5,0) node[right] {$\alpha\beta$};
    \draw[middleArrow] (0,0) -- (1.5,0);
  \end{tikzpicture}
  = i P_{\mu\nu;\alpha\beta} \Delta_\textrm{ret}(p)\,,\label{prop}
\end{equation}
where the arrow on the momentum is associated with the standard momentum routing whereas the arrow on the wavy line indicates the direction of the {\it time flow}. The advanced propagator, therefore, may be represented by the same arrow on the momentum $p$ but with the time flowing in the opposite direction.\vskip 4pt
Next we move onto the interacting part of the point-particle action, which reads 
\begin{equation}
  \begin{aligned}
    \cS_\textnormal{pp}[x_\pm,h_\pm] &\rightarrow
    \begin{multlined}[t]
      - \frac{m}{2\Mp}\int\dd\tau \int_k \left\{\frac{h^+_{\mu\nu}(k)}{2}\left[e^{i\, k\cdot\left(x_++\frac{x_{-}}{2}\right)}\left(v_+^\mu+\frac{v_{-}^\mu}{2}\right)\left(v_+^\nu+\frac{v_-^\nu}{2}\right) \right. \right.\\
        \left. +e^{i\,k\cdot\left(x_{+}-\frac{x_{-}}{2}\right)}\left(v_+^\mu-\frac{v_{-}^\mu}{2}\right)\left(v_+^\nu-\frac{v_-^\nu}{2}\right) \right]\\
      +h^-_{\mu\nu}(k)\left[e^{i\, k\cdot\left(x_++\frac{x_{-}}{2}\right)}\left(v_+^\mu+\frac{v_{-}^\mu}{2}\right)\left(v_+^\nu+\frac{v_-^\nu}{2}\right) \right.\\
        \left.\left.-e^{i\,k\cdot\left(x_{+}-\frac{x_{-}}{2}\right)}\left(v_+^\mu-\frac{v_{-}^\mu}{2}\right)\left(v_+^\nu-\frac{v_-^\nu}{2}\right)\right]\right\}\,.
    \end{multlined}
  \end{aligned}
\end{equation}
In general, a Feynman diagram will have several insertions of the worldline coupling to the gravitational field(s). In the computation of the variation of the effective action, however, only one such insertion at a time will be hit by the derivative in \eqref{eq:impulse}. Therefore, we can distinguish two types of worldline vertices. The first is obtained when no variation is acting on it. In this case we can set the physical condition already at the level of the effective action such that we have, in Fourier space,
\begin{equation}\label{eq:WLIn}
  \left. \cS_\textrm{pp} \right|_{\substack{x_-=0\\x_+=x}}
  = -\frac{ m}{2\Mp}\int\dd\tau \, h^+_{\mu\nu}(k) e^{i\, k\cdot x} v^\mu v^\nu\,,
\end{equation}
which we represent as
\begin{equation}
  \begin{tikzpicture}
    [baseline={([yshift=-0.5ex]current bounding box.center)}]
    \draw[graviton] (0,1) -- node[left=0.1] {$\downarrow\! k$} (0,0);
    \draw[middleArrow] (0,1) -- (0,0);
    \draw[fill] (0,1) circle (0.05);
  \end{tikzpicture}
  \,
  =-\frac{i  m}{2\Mp}\int\dd\tau \, e^{i\, k\cdot x} v^\mu v^\nu\,,
\end{equation}
with the arrows defined according to \eqref{prop}. Notice that this Feynman rule resembles the standard in-out scenario, with the Feynman propagator replaced by a retarded one. For~reasons that will become clear momentarily we will refer to these vertices as {\it sources}\vskip 4pt 

The remaining case, which occurs when the variation with respect to $x_-$ hits the vertex, turns out to be surprisingly simple. The key is to notice, 
\begin{equation}\label{eq:WLOut}
  \begin{aligned}
    \left.\frac{\delta \cS_\textrm{pp}}{\delta x^\alpha_-(\sigma)}\right|_{\substack{x_-\to 0\\x_+ \to x}}
    &= -\frac{ m}{2\Mp} \int\dd\tau \int_k \, h^-_{\mu\nu}(k) e^{i\,k\cdot x}\left[ i\, k^\alpha v^\mu v^\nu \delta(\tau - \sigma)+(\eta^{\mu\alpha}v^\nu +\eta^{\nu\alpha}v^\mu)\delta'(\tau-\sigma)\right]\\
    &= -\frac{ m}{2\Mp} \int\dd\tau\int_k \, h^-_{\mu\nu}(k) e^{i\,k\cdot x}\delta(\tau-\sigma)\left[ i\, k^\alpha v^\mu v^\nu-i\,k\cdot v\eta^{\mu\alpha}v^\nu-\right. \\ & \left. \hspace{6.7cm} \eta^{\mu\alpha}\dot{v}^\nu - i\,k\cdot v\eta^{\nu\alpha}v^\mu-\eta^{\nu\alpha}\dot{v}^\mu\right]\,,
  \end{aligned}
\end{equation}
which means only one type of time flow---opposite to the previous case---must be considered. The associated Feynman rule for this particular vertex becomes
\begin{equation}\label{eq:WLOutV}
  \begin{tikzpicture}
    [baseline={([yshift=-0.5ex]current bounding box.center)}]
    \node[xSource] (A) at (0,1) {};
    \draw[graviton] (0,0) -- node[left=0.1] {$\downarrow\! k$}  (A);
    \draw[middleArrow] (0,0) -- (A);
  \end{tikzpicture}
  \,
  = -\frac{i m}{2\Mp} \,e^{i\,k\cdot x} \left[ i\, k^\alpha v^\mu v^\nu-i\,k\cdot v\, \eta^{\mu\alpha}v^\nu-\eta^{\mu\alpha}\dot{v}^\nu-i\,k\cdot v\, \eta^{\nu\alpha}v^\mu-\eta^{\nu\alpha}\dot{v}^\mu\right]\,,
\end{equation}
where the cross, which we will refer to as a {\it sink}, indicates this is the only worldline vertex where the $\delta/\delta x^\alpha$ derivative of the wordline action has been taken. This is the only rule which differs from the standard in-out treatment.\vskip 4pt 

Adding to the mix the bulk-type interactions written in the Keldysh representation, it is straightforward now to construct the in-in integrand. In particular, it is clear that any graviton vertex that leads to more than one sink can be neglected. Hence, only non-linear couplings with sources attached to a single sink linked to e.g. particle 1, for instance
\begin{equation}
  \begin{tikzpicture}
    [baseline={([yshift=-0.5ex]current bounding box.center)}]
    \draw[graviton] (0,0) -- (0,1);
    \draw[middleArrow] (0,0) -- (0,1);
    \draw[graviton] (0,0) -- (-0.66,-0.7);
    \draw[middleArrow] (-0.66,-0.7) -- (0,0);
    \draw[graviton] (0,0) -- (0.66,-0.7);
    \draw[middleArrow] (0.66,-0.7) -- (0,0);
  \end{tikzpicture}\,,
  \quad
  \begin{tikzpicture}
    [baseline={([yshift=-0.5ex]current bounding box.center)}]
    \draw[graviton] (0,0) -- (0,1);
    \draw[middleArrow] (0,0) -- (0,1);
    \draw[graviton] (0,-0.7) -- (0,0);
    \draw[middleArrow] (0,-0.7) -- (0,0);
    \draw[graviton] (0,0) -- (-0.8,-0.7);
    \draw[middleArrow] (-0.8,-0.7) -- (0,0);
    \draw[graviton] (0,0) -- (0.8,-0.7);
    \draw[middleArrow] (0.8,-0.7) -- (0,0);
  \end{tikzpicture}\,,
  \quad
  \begin{tikzpicture}
    [baseline={([yshift=-0.5ex]current bounding box.center)}, inner sep=-10pt]
    \draw[graviton] (-0.25,0.3) -- (-0.25,1);
    \draw[middleArrow] (-0.25,0.3) -- (-0.25,1);
    \draw[graviton] (-1,-0.7) -- (-0.25,0.3);
    \draw[middleArrow] (-1,-0.7) -- (-0.25,0.3);
    \draw[graviton] (0.2,-0.3) -- (-0.25,0.3);
    \draw[middleArrow] (0.2,-0.3) -- (-0.25,0.3);
    \draw[graviton] (-0.2,-0.7) -- (0.2,-0.3);
    \draw[middleArrow] (-0.2,-0.7) -- (0.2,-0.3);
    \draw[graviton] (0.7,-0.7) -- (0.2,-0.3);
    \draw[middleArrow] (0.7,-0.7) -- (0.2,-0.3);
  \end{tikzpicture}\,,
  \quad
  \begin{tikzpicture}
    [baseline={([yshift=-0.5ex]current bounding box.center)}, inner sep=-10pt]
    \draw[graviton] (0,0.15) -- (0,1);
    \draw[middleArrow] (0,0.15) -- (0,1);
    \draw[graviton] (0,-0.7) -- (0,0.15);
    \draw[middleArrow] (0,-0.7) -- (0,0.15);
    \draw[graviton] (1,1) -- (1,0.15);
    \draw[middleArrow] (1,1) -- (1,0.15);
    \draw[graviton] (1,-0.7) -- (1,0.15);
    \draw[middleArrow] (1,-0.7) -- (1,0.15);
    \draw[graviton] (1,0.15) -- (0,0.15);
    \draw[middleArrow] (1,0.15) -- (0,0.15);
  \end{tikzpicture}\,,
\end{equation} 
must be considered to compute the total $\Delta p^\alpha$. This is also consistent with having a {\it causal} time ordering attached to each Feynman diagram, as expected from causality considerations. At the end of the day, this follows directly from the fact that we are solving classical field equations with iterated retarded Green's functions. Notice that, as long as loops of the gravitational field are neglected, this also precludes the appearance of the Hadamard propagator.\vskip 4pt

Let us conclude this part with a few useful remarks. First of all, because of the property we just discussed about sinks and sources and the fact that only tree-level diagrams are needed in the classical theory, we notice that non-linear terms involving more than one $h^+$ in the in-in expansion of the Einstein-Hilbert cannot contribute to the path integral in \eqref{seff}.  This is obviously the case for even number of $h^+$'s (including zero) which identically vanish, and is also the case for odd terms greater than one simply because there is no $K^{++}$ component~in~\eqref{kmatrix}. This is a direct consequence of our in-in Feynman rules yielding a natural causal flow for each (classical) diagram.\footnote{Notice this is no longer the case in the quantum theory once loops of the gravitational field(s) are allowed.} 

Likewise, following our conventions in \eqref{hpm} and taking into account the relative minus sign in \eqref{seff}, it is also easy to show that the coefficient for the $n$th order term (schematically) $h^+ [{h^-}]^{n-1}$ in the bulk action would become identical to the standard in-out counterpart, provided we ignore the $\pm$ labels. The only apparent difference at the level of the Feynman rules are symmetry factors. The $h^+$ and $h^-$ are indistinguishable in the in-out approach, whereas these are two separate fields in the in-in framework. This means that, at the level of the in-out and in-in effective actions, we do not expect to arrive to the same result between the two by simply replacing propagators. However, at the level of the equations of motion, it is clear that we must also consider diagrams with sinks at all instances where e.g. particle 1 appears. Moreover, since in the classical theory we only include tree-level connected Feynman diagrams, it is straightforward to show that the additional Wick contractions over the metric field in the in-out computation of the impulse are compensated by the sum over sinks in the in-in formalism.\footnote{This is also manifest at the level of PN computations of the effective action in terms of multipole moments, where different overall factors appear in the in-in and in-out approach but both coincide after the variations of the respective actions are considered, see e.g. \cite{tail}.}\vskip 4pt From here we conclude that upon replacing $\Delta_\textrm{ret}\rightarrow \Delta_F$ in the in-in integrand we would directly recover the impulse computed with in-out boundary conditions. The converse is also manifestly true. After taking into account the time ordering consistently with the in-in rules, the integrand for the impulse may be equally computed directly from the in-out approach by replacing Feynman with retarded propagators consistently with a causal ordering. Another relevant comment has to do with the expression for the vertex in~\eqref{eq:WLOutV}. Since the last four terms appear due to time derivatives in the Euler-Lagrange equations, they only matter for the equations of motion but vanish in the total impulse.
 
\subsection{Conservative vs dissipative} \label{chad}
In principle the interaction part of the effective theory, $\cS_\textrm{int}[x_{a,1},x_{a,2}]$, includes both dissipative and conservative terms alike. It is clear, however, that any term in the effective action which does not mix between the $1$ and $2$ variables, e.g. of the form
\beq
\cS_\textrm{int}[x_{a,1},x_{a,2}] \supset  \int \dd \tau_a \left(V_1[x_{a,1}] - V_2[x_{a,2}]\right)\,,\label{consV}
\eeq
can be combined with the kinetic part yielding 
\begin{equation}
    \cS_\textrm{eff}[x_{a,1},x_{a,2}] = \sum_a m_a\int\dd\tau_a \left(L_{\rm cons}[x_{a,1}]-L_{\rm cons}[x_{a,2}]\right)+ \cS_\textrm{diss}[x_{a,1},x_{a,2}]\,,\\
\end{equation}
with  (up to terms that vanish in the limit $x_{a,-}\to 0$) 
\beq
\label{Lcons}
L_{\rm cons} [x_a] \equiv -\left(\sum_a \frac{m_a}{2} v_a^2 - \frac{V_1[x_a]+V_2[x_a]}{2}\right) \,,
\eeq 
such that the physics described by this Lagrangian---which can be equally described by a conserved Hamiltonian---naturally falls into the conservative sector. See \cite{chadprl} for a more detailed discussion.\vskip 4pt Notice we do not make any assumption regarding the origin of $V_a[x_a]$, which in principle can (and will) receive contributions from both off-shell potential as well as on-shell radiation modes. Needless to say, if the Green's functions are expanded in the potential region around quasi-instantaneous interactions~\cite{nrgr}, the difference between Feynman, retarded (or advanced) propagators becomes obsolete. In such case the effective action becomes entirely of the form in \eqref{consV}, as expected. However, that is not the full story. In~fact, radiation-reaction tail terms are known to contribute to the conservative sector through a term of the form in \eqref{consV}~\cite{tail}.\vskip 4pt The obvious follow up question then becomes whether one can identify these pieces in advance. As we show next, that is indeed the case for most---although not necessarily all---of conservative terms.\vskip 4pt

Let us take the matrix in \eqref{gmatrix} and split it as
\beq
G_{AB} = G_{AB}^{\rm F} + G^{\rm cut}_{AB}\,,\label{dcut}
\eeq
with
\beq
 G_{AB}^{\rm F}=i\begin{pmatrix} \Delta_{\rm F}(x-y) & 0 \\ 0 & \Delta_{\rm D}(x-y) \end{pmatrix}\,,\quad G_{AB}^{\rm cut}=i\begin{pmatrix} 0 & -\Delta_-(x-y) \\ -\Delta_+(x-y) & 0 \end{pmatrix}\,.
\eeq
Given that  $G_{AB}^{\rm F}$ is diagonal, its contribution to both the $Dh_1$ and $Dh_2$ integrals in \eqref{seff} can be computed independently of each other using the standard in-out rules. Moreover, since they come with opposite signs, it is straightforward to show that ultimately the contribution from $G_{AB}^{\rm F}$ to the effective action takes the form
\beq
  \cS_\textrm{eff}[x_{a,1},x_{a,2}]  \supset   \int \dd \tau_a \left(V_\textrm{F}[x_{a,1}]  -  V_\textrm{D}[x_{a,2}] \right)\,,
  \eeq
   where we have $V_{\rm D}[x] = V_{\rm F}^\star[x]$ (which follows from the properties of the Feynman and Dyson propagators, see Appendix~\ref{appA}).  Hence, according to \eqref{Lcons}, the in-in effective action contains a conservative piece given by
  \beq
  \label{consF}
  L_{\rm cons}[x_a] \supset  \frac{1}{2} \int \dd \tau_a \, \left( V_F[x_a] + V_D[x_a]\right) = {\cR} \int \dd \tau_a \, V_F[x_a]\,.
  \eeq
These manipulations then explicitly demonstrate that using the standard Feynman rules with Feynman's $i0$-prescription and keeping only the real part of the resulting effective action produces a contribution to the dynamical equations which can be always included as part of a conservative sector. Notably,  we can also obtain this conservative contribution directly from the $\pm$-basis by replacing $\Delta_{\rm ret} \to \Delta_F$ in the integrand {\it for the impulse}, thus landing in the in-out result as we demonstrated before.\vskip 4pt There are, of course, also {\it dissipative} effects. These are obtained by \underline{\it at least} one insertion of $G^{\rm cut}$ in \eqref{seff}, which mixes the $x_{a,1}$ and $x_{a,2}$ variables. At the level of the $\pm$-basis, these terms appear from the mismatch between retarded/advanced and Feynman Green's functions, and therefore can be entirely written in terms of $\Delta_{\pm}$ propagators. These terms, which are proportional to  $\delta(p^2)$ in Fourier space, arise due to gravitational radiation and as expected are only present when on-shell modes are turned on.\vskip 4pt  A priori, other than those identified through \eqref{consF}, we cannot tell in advance whether the effective action contains more terms of the form in \eqref{consV}. In some paradigmatic examples, however,   the contribution from \eqref{consF} does encapsulate the entire conservative sector. For instance, this is the case for all potential interactions (for which $\Delta_\pm$ vanishes). It is also the case for tail-induced effects \cite{tail,nrgr4pn1,nrgr4pn2,4pmeft,4pmeft2}. However, this may not be true for all types of non-linear radiation-reaction contributions---such as `memory' terms---and in particular for those involving several (non-vanishing) insertions~of~$\Delta_\pm$. The~latter may lead to additional  contributions of the form in \eqref{consV} which would not be captured by \eqref{consF}.\footnote{Although the calculation in \cite{Blumlein:2021txe} did not identify conservative terms as in \eqref{consV}, their result suggests that \eqref{consF} may not be the entire conservative sector when memory terms are included.}
\vskip 4pt

None of the above issues arise at leading order in the radiation-reaction at 3PM, which involves only a single non-vanishing insertion of $\Delta_-$. In this case there is a clear separation into dissipative and conservative contributions, which we study next.  
    
 \section{Scattering to 3PM} \label{3pm}
 
\subsection{Building the integrand}

We can now use the in-in diagrammatic rules to obtain the variation of the effective action necessary to compute the total impulse. For simplicity, we only display here the arrows which dictate---according to causality---whether retarded or advanced Green's functions appear in the graviton propagators.\vskip 4pt
At 1PM order there are only two diagrams 
\begin{equation}
\label{F1pm}
  \left.\frac{\delta\cS_\textnormal{eff}[x_+,x_-]}{\delta x_-^\alpha}\right|_{\substack{x_-\to 0\\x_+\to x}}^\textrm{1PM}=
  \begin{tikzpicture}
    [baseline={([yshift=-0.5ex]current bounding box.center)}]
    \node[xSource] (A) at (0,1.5) {};
    \draw[graviton] (0,0) -- (A) node[left=0.2] {};
    \draw[middleArrow] (0,0) -- (A);
    \fill (0,0) circle (0.05);
  \end{tikzpicture}
  +
  \begin{tikzpicture}
    [baseline={([yshift=-0.5ex]current bounding box.center)}]
    \node[xSource] (A) at (0,1.5) {};
    \draw[graviton] (1,1.5) to[bend left=60] (A);
    \draw[middleArrow] (1,1.5) to[bend left=60] (A);
    \fill (1,1.5) circle (0.05);
    \fill[white] (0,0) circle (0.01);
  \end{tikzpicture}
  \,.
\end{equation}
At 2PM, on the other hand, we have
\begin{equation}
  \left.\frac{\delta\cS_\textnormal{eff}[x_+,x_-]}{\delta x_-^\alpha}\right|_{\substack{x_-\to 0\\x_+\to x}}^\textrm{2PM}=
  \begin{tikzpicture}
    [baseline={([yshift=-0.5ex]current bounding box.center)}]
    \node[xSource] (A) at (-0.5,0.75) {};
    \draw[graviton] (0,0) -- (A) node[left=0.2] {};
    \draw[middleArrow] (0,0) -- (A);
    \draw[graviton] (0.5,0.75) -- (0,0);
    \draw[middleArrow] (0.5,0.75) -- (0,0);
    \fill (0.5,0.75) circle (0.05);
    \draw[graviton] (0,-0.75) -- (0,0);
    \draw[middleArrow] (0,-0.75) -- (0,0);
    \fill (0,-0.75) circle (0.05);
  \end{tikzpicture}
  +
  \begin{tikzpicture}
    [baseline={([yshift=-0.5ex]current bounding box.center)}]
    \node[xSource] (A) at (0,0.5) {};
    \draw[graviton] (0,0) -- (A) node[left=0.2] {};
    \draw[middleArrow] (0,0) -- (A);
    \draw[graviton] (-0.5,-0.75) -- (0,0);
    \draw[middleArrow] (-0.5,-0.75) -- (0,0);
    \fill (-0.5,-0.75) circle (0.05);
    \draw[graviton] (0.5,-0.75) -- (0,0);
    \draw[middleArrow] (0.5,-0.75) -- (0,0);
    \fill (0.5,-0.75) circle (0.05);
  \end{tikzpicture}
  +
  \begin{tikzpicture}
    [baseline={([yshift=-0.5ex]current bounding box.center)}]
    \node[xSource] (A) at (0,1.5) {};
    \draw[graviton] (1,1) to[bend left=25] (A);
    \draw[middleArrow] (1,1) to[bend left=25] (A);
    \draw[graviton] (2,1.5) to[bend left=25] (1,1);
    \draw[middleArrow] (2,1.5) to[bend left=25] (1,1);
    \draw[graviton] (1,1.5) -- (1,1);
    \draw[middleArrow] (1,1.5) -- (1,1);
    \fill (2,1.5) circle (0.05);
    \fill (1,1.5) circle (0.05);
    \fill[white] (0,0) circle (0.01);
  \end{tikzpicture}
  \,.
\end{equation}
whereas at 3PM order, 
\begin{equation}
  \begin{aligned}
    \left.\frac{\delta\cS_\textnormal{eff}[x_+,x_-]}{\delta x_-^\alpha}\right|_{\substack{x_-\to 0\\x_+\to x}}^\textrm{3PM}&=
    \begin{tikzpicture}
      [baseline={([yshift=-0.5ex]current bounding box.center)}]
      \node[xSource] (A) at (0,0.75) {};
      \draw[graviton] (0,0.25) -- (A) node[left=0.2] {};
      \draw[middleArrow] (0,0.25) -- (A);
      \draw[graviton] (-1,-0.75) -- (-0.5,-0.25);
      \draw[middleArrow] (-1,-0.75) -- (-0.5,-0.25);
      \draw[graviton] (-0.5,-0.25) -- (0,0.25);
      \draw[middleArrow] (-0.5,-0.25) -- (0,0.25);
      \draw[graviton] (0,-0.75) -- (-0.5,-0.25);
      \draw[middleArrow] (0,-0.75) -- (-0.5,-0.25);
      \draw[graviton] (0.75,-0.75) -- (0,0.25);
      \draw[middleArrow] (0.75,-0.75) -- (0,0.25);
      \fill (-1,-0.75) circle (0.05);
      \fill (0,-0.75) circle (0.05);
      \fill (0.75,-0.75) circle (0.05);
    \end{tikzpicture}
    +
    \begin{tikzpicture}
      [baseline={([yshift=-0.5ex]current bounding box.center)}]
      \node[xSource] (A) at (-1,0.75) {};
      \draw[graviton] (-0.5,0.25) -- (A) node[left=0.2] {};
      \draw[middleArrow] (-0.5,0.25) -- (A);
      \draw[graviton] (0,0.75) -- (-0.5,0.25);
      \draw[middleArrow] (0,0.75) -- (-0.5,0.25);
      \draw[graviton] (0,-0.25) -- (-0.5,0.25);
      \draw[middleArrow] (0,-0.25) -- (-0.5,0.25);
      \draw[graviton] (0.75,0.75) -- (0,-0.25);
      \draw[middleArrow] (0.75,0.75) -- (0,-0.25);
      \draw[graviton] (0,-0.75) -- (0,-0.25);
      \draw[middleArrow] (0,-0.75) -- (0,-0.25);
      \fill (0,0.75) circle (0.05);
      \fill (0.75,0.75) circle (0.05);
      \fill (0,-0.75) circle (0.05);
    \end{tikzpicture}
    +
    \begin{tikzpicture}
      [baseline={([yshift=-0.5ex]current bounding box.center)}]
      \node[xSource] (A) at (0.75,0.75) {};
      \draw[graviton] (0,-0.25) -- (A) node[left=0.2] {};
      \draw[middleArrow] (0,-0.25) -- (A);
      \draw[graviton] (0,0.75) -- (-0.5,0.25);
      \draw[middleArrow] (0,0.75) -- (-0.5,0.25);
      \draw[graviton] (-0.5,0.25) -- (0,-0.25);
      \draw[middleArrow] (-0.5,0.25) -- (0,-0.25);
      \draw[graviton] (-1,0.75) -- (-0.5,0.25);
      \draw[middleArrow] (-1,0.75) -- (-0.5,0.25);
      \draw[graviton] (0,-0.75) -- (0,-0.25);
      \draw[middleArrow] (0,-0.75) -- (0,-0.25);
      \fill (0,0.75) circle (0.05);
      \fill (-1,0.75) circle (0.05);
      \fill (0,-0.75) circle (0.05);
    \end{tikzpicture}
    +
    \begin{tikzpicture}
      [baseline={([yshift=-0.5ex]current bounding box.center)}]
      \node[xSource] (A) at (-0.5,0.75) {};
      \draw[graviton] (-0.5,0) -- (A) node[left=0.2] {};
      \draw[middleArrow] (-0.5,0) -- (A);
      \draw[graviton] (0.5,0.75) -- (0.5,0);
      \draw[middleArrow] (0.5,0.75) -- (0.5,0);
      \draw[graviton] (0.5,0) -- (-0.5,0);
      \draw[middleArrow] (0.5,0) -- (-0.5,0);
      \draw[graviton] (-0.5,-0.75) -- (-0.5,0);
      \draw[middleArrow] (-0.5,-0.75) -- (-0.5,0);
      \draw[graviton] (0.5,-0.75) -- (0.5,0);
      \draw[middleArrow] (0.5,-0.75) -- (0.5,0);
      \fill (0.5,0.75) circle (0.05);
      \fill (-0.5,-0.75) circle (0.05);
      \fill (0.5,-0.75) circle (0.05);
    \end{tikzpicture}\\
    &\quad+
    \begin{tikzpicture}
      [baseline={([yshift=-0.5ex]current bounding box.center)}]
      \node[xSource] (A) at (-0.5,0.75) {};
      \draw[graviton] (0,0.25) -- (A) node[left=0.2] {};
      \draw[middleArrow] (0,0.25) -- (A);
      \draw[graviton] (0.5,0.75) -- (0,0.25);
      \draw[middleArrow] (0.5,0.75) -- (0,0.25);
      \draw[graviton] (0,-0.25) -- (0,0.25);
      \draw[middleArrow] (0,-0.25) -- (0,0.25);
      \draw[graviton] (-0.5,-0.75) -- (0,-0.25);
      \draw[middleArrow] (-0.5,-0.75) -- (0,-0.25);
      \draw[graviton] (0.5,-0.75) -- (0,-0.25);
      \draw[middleArrow] (0.5,-0.75) -- (0,-0.25);
      \fill (0.5,0.75) circle (0.05);
      \fill (-0.5,-0.75) circle (0.05);
      \fill (0.5,-0.75) circle (0.05);
    \end{tikzpicture}
    +
    \begin{tikzpicture}
      [baseline={([yshift=-0.5ex]current bounding box.center)}]
      \node[xSource] (A) at (0,0.75) {};
      \draw[graviton] (0,0) -- (A) node[left=0.2] {};
      \draw[middleArrow] (0,0) -- (A);
      \draw[graviton] (-0.75,-0.75) -- (0,0);
      \draw[middleArrow] (-0.75,-0.75) -- (0,0);
      \draw[graviton] (0,-0.75) -- (0,0);
      \draw[middleArrow] (0,-0.75) -- (0,0);
      \draw[graviton] (0.75,-0.75) -- (0,-0);
      \draw[middleArrow] (0.75,-0.75) -- (0,-0);
      \fill (-0.75,-0.75) circle (0.05);
      \fill (0,-0.75) circle (0.05);
      \fill (0.75,-0.75) circle (0.05);
    \end{tikzpicture}
    +
    \begin{tikzpicture}
      [baseline={([yshift=-0.5ex]current bounding box.center)}]
      \node[xSource] (A) at (-0.5,0.75) {};
      \draw[graviton] (0,0) -- (A) node[left=0.2] {};
      \draw[middleArrow] (0,0) -- (A);
      \draw[graviton] (0.5,0.75) -- (0,0);
      \draw[middleArrow] (0.5,0.75) -- (0,0);
      \draw[graviton] (-0.5,-0.75) -- (0,0);
      \draw[middleArrow] (-0.5,-0.75) -- (0,0);
      \draw[graviton] (0.5,-0.75) -- (0,0);
      \draw[middleArrow] (0.5,-0.75) -- (0,0);
      \fill (0.5,0.75) circle (0.05);
      \fill (-0.5,-0.75) circle (0.05);
      \fill (0.5,-0.75) circle (0.05);
    \end{tikzpicture}
    +
    \begin{tikzpicture}
      [baseline={([yshift=-0.5ex]current bounding box.center)}]
      \node[xSource] (A) at (-0.75,0.75) {};
      \draw[graviton] (0,0) -- (A) node[left=0.2] {};
      \draw[middleArrow] (0,0) -- (A);
      \draw[graviton] (0,0.75) -- (0,0);
      \draw[middleArrow] (0,0.75) -- (0,0);
      \draw[graviton] (0.75,0.75) -- (0,0);
      \draw[middleArrow] (0.75,0.75) -- (0,0);
      \draw[graviton] (0,-0.75) -- (0,0);
      \draw[middleArrow] (0,-0.75) -- (0,0);
      \fill (0,0.75) circle (0.05);
      \fill (0.75,0.75) circle (0.05);
      \fill (0,-0.75) circle (0.05);
    \end{tikzpicture}
    +\dots
    \,.
  \end{aligned}
\end{equation}
\vskip 4pt
The ellipsis in the last expression include the same type of {\it self-energy} diagrams (with all sources and sinks on the same worldline) shown above, which are in general needed to include radiation-reaction effects.
Modulo the time routing of each diagram, all of these resemble the same type in the conservative sector to 3PM \cite{pmeft,3pmeft}, except for the self-energy terms which lead to scaleless integrals in the potential region, which in dim. reg. can be set to zero. More on this below. 
\vskip 4pt

Once the integrand for the impulse is constructed via the above Feynman rules we still need to compute it on solutions to the equations of motion. This entails, as before \cite{pmeft}, adding the {\it iterations} of lower order contributions to the effective action evaluated on the trajectories \cite{pmeft} to a given $n$PM order, 
\beq
\begin{aligned}
x^\mu_{a} (\tau_a) &= b_{a}^\mu + u_{a}^\mu \tau_a + \delta^{(n)} x^\mu_{a}(\tau_a)\,,\\
v^\mu_{a} (\tau_a) &=  u_{a}^\mu  + \delta^{(n)} v^\mu_{a}(\tau_a)\,,\label{traj}
\end{aligned}
\eeq
with $u_a$ the incoming velocities and $b_1^\mu-b_2^\mu \equiv b^\mu$  the (space-like) impact parameter four-vector. These trajectories can be solved iteratively using \eqref{eq:inineom} after the limit: $x_{a,-}\to 0,\, x_{a,+}\to x_a$, has been taken. Since self-energy diagrams evaluated on unperturbed trajectories vanish in dim. reg., they can only contribute through iterations on each of the sources. Therefore, only the extra term in \eqref{F1pm} is needed to 3PM order.   

\subsection{Integration problem}\label{sec:int}

After the integrand is assembled, say for the impulse of particle 1, and the a tensor reduction performed following the same steps as in \cite{3pmeft,4pmeft,4pmeft2}, we find the following families of integrals:
\begin{equation}\label{eq:intFams}
  I^{\pm\pm;T_5\dots T_9}_{\nu_1\dots\nu_9}=\int_{\ell_1,\ell_2}\frac{\hat{\delta}^{(\nu_1-1)}(\ell_1\cdot u_a)\hat{\delta}^{(\nu_2-1)}(\ell_2\cdot u_b)}{(\pm \ell_1\cdot u_{\not{a}})^{\nu_3} (\pm \ell_2\cdot u_{\not{b}})^{\nu_4}}\prod_{j=5}^9 \frac{1}{D_{j,T_j}^{\nu_j}}\,,
\end{equation}
where we use the same notation as in \cite{3pmeft,4pmeft,4pmeft2}, except for the choice of $i0$-prescription, determined by sets of square propagators of type $T_j\in\{\textrm{ret},\textrm{adv}\}$:
\begin{equation}
  \begin{aligned}
    D_{5,\textrm{ret/adv}} &= (\ell_1^0 \pm i0)^2 - \bell_1^2 \,,& D_{6,\textrm{ret/adv}} &= (\ell_2^0 \pm i0)^2 - \bell_2^2\,,\\
    D_{7,\textrm{ret/adv}} &= (\ell_1^0 + \ell_2^0 + q^0 \pm i0)^2 - (\bell_1+\bell_2-\bq)^2 \,,\\
    D_{8,\textrm{ret/adv}} &= (\ell_1^0 -q^0 \pm i0)^2 - (\bell_1 - \bq)^2 \,,& D_{9,\textrm{ret/adv}} &= (\ell_2^0 - q^0 \pm i0)^2 - (\bell_2 - \bq)^2\,.\\
  \end{aligned}
\end{equation}
The integrand itself only contains factors with $\nu_1=\nu_2=1$, i.e. no derivative. However, because the integrals in~\eqref{eq:intFams} turn out to be not all independent, derivatives of the Dirac-$\delta$ functions may be introduced by the standard  \emph{integration-by-parts} (IBP) relations  \cite{Tkachov:1981wb,Chetyrkin:1981qh,Anastasiou:2004vj}. We have used the combination of the automated programs \texttt{LiteRed} \cite{Lee:2013mka} and \texttt{FIRE6} \cite{Smirnov:2019qkx} to relate the integrals appearing in the integrand to a set of $\cO(100)$ masters.\footnote{There is one subtlety due to the presence of self-energy diagrams leading to iterations involving linear propagators with the same internal momentum but both $\pm i0$ prescriptions, e.g. $1/(\ell_1\cdot u_2+i0)(\ell_1\cdot u_2-i0)$. (Notice this cannot occur for quadratic propagators in the classical theory.)  These are both treated on equal footing by the IBP programs, irrespectively of the signs. Hence, we would not know how to reassign the correct signs  after IBP relations. Fortunately, we were able to construct a basis of master integrals for which these type of terms are absent (or appear in the numerator), thus avoiding this problem.}
\vskip 4pt

The reader will immediately notice the sharp contrast with the number of master integrals needed for the computation in \cite{3pmeft}. The reason is one of the main advantages of Feynman propagators, and the use of the symmetry relations implemented by these programs which are not present for the case of retarded/advanced Green's functions. Nevertheless, additional simplifications arise once we restrict ourselves to the conservative and dissipative sectors, respectively. First of all, for the former we can replace retarded/advanced by Feynman Green's functions, essentially returning to the analysis in \cite{3pmeft}, which we review below. For the latter, we notice only radiation modes contribute to $\Delta_{\pm}$. Hence, after isolating the different kinematical regions of integration via e.g. the {\it asy2.m} code included in the \texttt{FIESTA} package \cite{Jantzen:2012mw,Smirnov:2015mct,Smirnov:2021rhf}, we can show that only the 7th propagator in the above families has a non-zero radiation region, with all others entering only through potential modes. This means, as expected, that diagrams with a single insertion of $\Delta_{-}$ contribute to the dissipative sector at 3PM, with the others replaced by Feynman Green's functions. In such scenario, there is a subset of symmetry relations that reduce the number of master integrals to about $\cO(50)$, which we computed via the method of differential equations~\cite{Kotikov:1991pm,Remiddi:1997ny,Henn:2013pwa,Prausa:2017ltv,Lee:2020zfb,Lee:2014ioa,Adams:2018yfj}.\vskip 4pt

Since there is only a single scale in the problem, it is convenient to choose the parameter~$x$, defined through \cite{Parra}
\beq \gamma=(x^2+1)/2x\,, \quad\quad {\rm with} \quad \gamma \equiv u_1\cdot u_2\,.\eeq The dependence on the parameter~$x$ is then extracted order by order in $\epsilon$ by iterated integrations. At this order it is straightforward to show the differential equations can be brought into canonical form \cite{Henn:2013pwa}. For the problem at hand only three independent functions appear
\beq
 \big \{\log(x), \log(1+x),  \log(1-x)\big\}\,,
 \eeq
 up to polynomials in the $x$ variable. This uniquely fixes the velocity dependence of the master integrals up to a series of appropriate boundary conditions, which we choose near the static limit $x \to 1$. Notably, the solution to the differential equations allow us to extract the scaling in $(1-x)$ for each individual integral in this limit~\cite{Lee:2019zop,4pmeft2}, which becomes extremely useful to set up a series of consistency conditions that allow us to simplify the number of boundary constants needed to read off the full answer.\vskip 4pt Additional consistency requirements may be enforced to reduce the number of independent constants. The first is the lack of absorption, $\Delta m^2_a=0$, which implies the conservation of the on-shell condition $(p_a+\Delta p_a)^2=p_a^2$. This restricts the impulse to point in the direction orthogonal to $u_a^\mu$~\cite{Parra3,Gabriele2}.
At a more the technical level, we also impose the cancelation in the final answer of intermediate $1/\epsilon^k$ poles. Furthermore, terms proportional to $\log(1-x) \propto \log v$ are uniquely fixed by the universal connection to tail terms \cite{b2b3} (and therefore
do not enter until 4PM order \cite{4pmeft2}). Finally,  as anticipated e.g.~\cite{Damour:2020tta}, we can enforce continuity of the massless/high-energy limit by removing mass singularities. Remarkably, all of these properties completely determine the radiation-reaction impulse in the direction of the impact parameter at 3PM, and only a handful of boundary integrals for the longitudinal direction are needed.

\subsection{Boundary integrals}
As we mentioned, once the solution of the differential equations are known, the last input are integration constants obtained from  master integrals in the limit $x\rightarrow 1$. We first review the procedure for the conservative pieces before discussing radiative contributions.

\subsubsection{Conservative}

As it was computed in \cite{3pmeft}, and further established here from the in-in formalism, the conservative sector is obtained from the full integrand by replacing retarded/advanced propagators by Feynman counterparts and keeping the real part of the resulting impulse. In principle, at this stage both potential and radiation modes can be present. However, as we mentioned earlier, only a single propagator has support on radiation modes and therefore only the potential region contribute to the real part with time-symmetric boundary conditions (with an imaginary part consistent with the optical theorem). This also implies that all self-energy diagrams vanish in dim. reg. for the conservative sector at 3PM.\vskip 4pt The result for the potential boundary conditions was already computed in~\cite{3pmeft}. However, in order to relate the solution to the master integrals in \cite{3pmeft} an additional set of (static) IBP relations must be used at $x=1$. The result reads,
\begin{equation}
  \begin{aligned}
  \Delta p_{1,\textrm{cons}}^\mu &= \frac{4 \pi M^4 \nu b^\mu}{3 |b|^4 x^2(x^2-1)^5}
  \begin{multlined}[t]
    \big[
      c_1(x) + \nu c_2(x) +\nu c_3(x) \log(x)
    \big]
  \end{multlined}\\
  &\quad-\frac{3 \pi ^2 M^4 \nu  c_4(x) \left((\Delta -1) \check{u}_1^\mu+(\Delta +1) \check{u}_2^\mu\right)}{8 |b|^3 x^2 \left(x^2-1\right)^2}\,,
  \end{aligned}
\end{equation}
with $|b| = \sqrt{-b^\mu b_\mu}$, and introduced 
\begin{align}
  \check{u}_1 = \frac{\gamma u_2-u_1}{\gamma^2-1}\,, \quad & \check{u}_2 = \frac{\gamma u_1 - u_2}{\gamma^2-1}\,,
\end{align}
obeying $\check{u}_a\cdot u_b = \delta_{ab}$. The $c_i$'s are polynomials given by
\begin{equation}
  \begin{aligned}
    c_1(x) &= -12 x \left(x^{12}-2 x^{10}-x^8-x^4-2 x^2+1\right)\,,\\
    c_2(x) &= -(x-1)^2 \big(5 x^{12}-14 x^{11}-88 x^{10}-114 x^9-5 x^8+128 x^7\\
    &\quad+128 x^6+128 x^5-5 x^4-114 x^3-88 x^2-14 x+5\big)\,,\\
    c_3(x) &= 6 \left(x^8-8 x^6-30 x^4-8 x^2+1\right) \left(x^2-1\right)^3\,,\\
    c_4(x) &= \left(x^4+1\right) \left(5 x^4+6 x^2+5\right)\,,
  \end{aligned}
\end{equation}
in accordance with the result in \cite{3pmeft}.
\subsubsection{Dissipative}

Non-conservative terms are obtained by systematically including the factors of $\Delta_-$ accounting for the mismatch between Feynman and retarded propagators.  Unlike Feynman Green's functions, $\Delta_-$ vanishes for off-shell modes and therefore we just need to consider the single propagator with support on the radiation region. Furthermore, as we discussed in \cite{4pmeft,4pmeft2}, each radiation mode scales with a power of $(1-x)^{-2\epsilon}$. This already allows us to individually target radiative contributions.\vskip 4pt
Surprisingly, most of the integration constants for the dissipative sector can be determined by the same type of consistency constraints  already implemented in \cite{4pmeft,4pmeft2} to study radiation-reaction effects in the conservative sector at 4PM. In particular, the contribution in direction of the impact parameter $b^\mu$ turns out to be uniquely fixed. On the other hand, for the longitudinal contribution, which must be proportional to $\check{u}_2^\mu$, the master integrals have already been computed in the literature \cite{Riva:2021vnj,Gabriele2,Parra3} via Cutkosky rules~\cite{Cutkosky:1960sp} (see e.g. Appendix C in \cite{Riva:2021vnj}), which we confirm in terms of solutions of the differential equations.\vskip 4pt

As an example, consider the H-type master integral resulting after IBP relations
\begin{equation}
  I_{\rm H} = \int_{\ell_1,\ell_2}\frac{\hat{\delta}'(\ell_1\cdot u_a)\hat{\delta}(\ell_2\cdot u_b)}{\ell_1^2 \ell_2^2 (((\ell_1+\ell_2-q)^0+i0)^2-(\bell_1+\bell_2+\bq)^2) (\ell_1-q)^2 (\ell_2-q)^2}\,,
\end{equation}
As we mentioned, only one leg may be taken as a full retarded propagator, with the other ones having support only on potential modes. After replacing $\Delta_{\rm ret} \to \Delta_{\rm F} + \Delta_{-}$ and identifying the conservative part from the real part of the Feynman piece,\footnote{Remarkably, the resulting imaginary part of the Feynman integrals entering the impulse in the $\check{u}_2$ direction vanish. Consequently, the contribution from the remaining cut integrals are real functions such that the cancelation of imaginary parts becomes trivial in this case.} the dissipative term is obtained by replacing $\Delta_{\rm ret} \to \Delta_{-}$, yielding for this integral 
\begin{equation}
  I_{\rm H}^\textrm{cut} = \int_{\ell_1,\ell_2}\frac{\hat{\delta}'(\ell_1\cdot u_a)\hat{\delta}(\ell_2\cdot u_b)\hat{\delta}((\ell_1+\ell_2-q)^2)\Theta(-\ell_1^0-\ell_2^0+q^0)}{\ell_1^2 \ell_2^2 (\ell_1-q)^2 (\ell_2-q)^2}\,,
\end{equation}
which (up to a sign) corresponds to the master integral $f_4$ in \cite{Riva:2021vnj} after performing the shift $\ell_i \leftrightarrow -\ell_i$ and $q\leftrightarrow -q$. This integral was computed by resorting to Cutkosky's rules~\cite{Cutkosky:1960sp}, including also a cut over hypothetical linear `worldline propagators,' which account for the Dirac-$\delta$ functions already present in $I_{\rm H}$, and conspicuously everywhere in the worldline formalism. We may describe the procedure diagrammatically as follows
\begin{equation}
  I_{\rm H}^\textrm{cut}=
  \begin{tikzpicture}
    [
      baseline={([yshift=-1.7ex]current bounding box.center)}
    ]
    \draw[graviton] (-0.75,0.75) -- (-0.75,0);
    \draw[graviton] (-0.75,0) -- (-0.75,-0.75);
    \draw[graviton] (0.75,0.75) -- (0.75,0);
    \draw[graviton] (0.75,0) -- (0.75,-0.75);
    \draw[graviton] (-0.75,0) -- (0.75,0);
    \draw[middleArrow] (0.75,0) -- (-0.75,0);
    \draw[dotted] (-0.75,0.75) -- (0.75,0.75);
    \draw[dotted] (-0.75,-0.75) -- (0.75,-0.75);
    \fill (-0.75,0.75) circle (0.05);
    \fill (0.75,0.75) circle (0.05);
    \fill (-0.75,-0.75) circle (0.05);
    \fill (0.75,-0.75) circle (0.05);
    \fill (0,0.75) circle (0.08);
    \draw[dashed,red] (-0.3,1) -- (0.1,-1);
    \node[red,rotate=-80] at (-0.33,1.1) {\Rightscissors};
  \end{tikzpicture}\,,
\end{equation}
where the dotted lines represent the $\delta^{(n)}$ functions as cut linear propagators, with the big dot on the upper line denoting the case $\nu_1=2$ (i.e. a derivative) for one of the $\delta$'s. The cut over the retarded propagator denotes the replacement $\Delta_{\rm ret}(p) \to 2\pi i\, \delta(p^2) \Theta(-p^0)$ in momentum space.\footnote{Because of the $\Theta(-p^0)$, we have flipped the arrow on the retarded propagator to illustrate that positive energy is flowing towards the right of the diagram, as it is standard lore.}  By Cutkosky's rules, we can relate the cut integral it to the imaginary part of its {\it uncut} version,\footnote{The only subtlety here is the cut over the linear propagator, which relies on the identity 
\begin{equation}
  \begin{aligned}
    \int\frac{\dd^4\ell}{(2\pi)^4}\frac{e^{-i\,\ell\cdot x}}{\ell\cdot u+i0} &= - 2\pi i\int\frac{\dd^4\ell}{(2\pi)^4} e^{-i\,\ell\cdot x}\delta(\ell\cdot u) \left( \Theta(u^0) \Theta(x^0)+\Theta(-u^0)\Theta(-x^0)\right)\,,
  \end{aligned}
\end{equation}
such that the cut propagator is given by $2\pi i\, \delta(\ell\cdot u)\Theta(u^0)$, as expected. (Note this differs from the version given in \cite{Parra3} in terms of $\Theta(\ell^0)$ which, however, we believe to be a typo.)  In this fashion, the $\delta(\ell\cdot u)$ from the time-integration of the worldline sources---naturally obeying $u^0>1$ in the classical theory---can then be equally described by cut linear propagators.} 
\begin{equation}
  \begin{tikzpicture}
    [
      baseline={([yshift=-1.7ex]current bounding box.center)}
    ]
    \draw[graviton] (-0.75,0.75) -- (-0.75,0);
    \draw[graviton] (-0.75,0) -- (-0.75,-0.75);
    \draw[graviton] (0.75,0.75) -- (0.75,0);
    \draw[graviton] (0.75,0) -- (0.75,-0.75);
    \draw[graviton] (-0.75,0) -- (0.75,0);
    \draw[middleArrow] (0.75,0) -- (-0.75,0);
    \draw[dotted] (-0.75,0.75) -- (0.75,0.75);
    \draw[dotted] (-0.75,-0.75) -- (0.75,-0.75);
    \fill (-0.75,0.75) circle (0.05);
    \fill (0.75,0.75) circle (0.05);
    \fill (-0.75,-0.75) circle (0.05);
    \fill (0.75,-0.75) circle (0.05);
    \fill (0,0.75) circle (0.08);
    \draw[dashed,red] (-0.3,1) -- (0.1,-1);
    \node[red,rotate=-80] at (-0.33,1.1) {\Rightscissors};
  \end{tikzpicture}
  =
  2\textrm{Im}\left(\,
  \begin{tikzpicture}
    [
      baseline={([yshift=-0.5ex]current bounding box.center)}
    ]
    \draw[graviton] (-0.75,0.75) -- (-0.75,0);
    \draw[graviton] (-0.75,0) -- (-0.75,-0.75);
    \draw[graviton] (0.75,0.75) -- (0.75,0);
    \draw[graviton] (0.75,0) -- (0.75,-0.75);
    \draw[graviton] (-0.75,0) -- (0.75,0);
    \draw (-0.75,0.75) -- (0.75,0.75);
    \draw (-0.75,-0.75) -- (0.75,-0.75);
    \fill (-0.75,0.75) circle (0.05);
    \fill (0.75,0.75) circle (0.05);
    \fill (-0.75,-0.75) circle (0.05);
    \fill (0.75,-0.75) circle (0.05);
    \fill (0,0.75) circle (0.08);
  \end{tikzpicture}
  \,\right)\,,
\end{equation}
with the dotted and straight lines representing the cut and uncut linear propagators, including higher derivatives of Dirac-$\delta$ functions corresponding to higher powers of linear propagators. The~uncut integral, which is now entirely written in terms of Feynman propagators, can then be computed in the near-static limit $x\rightarrow 1$ by a direct integration. See \cite{Riva:2021vnj} for more details.
\vskip 4pt
Combining all these results we find
\begin{equation}
  \begin{aligned}
  \Delta p_{1,\textrm{diss}}^\mu &=
  \frac{M^4 \nu^2 (1+x^4)^2b^\mu}{3 |b|^4 x^2(x^2-1)^5}\left[d_1(x) + d_2(x)\log(x) \right]\\
   &\quad + \frac{\pi M^4 \nu^2\check{u}_2^\mu}{192 |b|^3 x^3(x^2-1)^4} \left[d_3(x)+d_4(x) \log \left(\frac{x+1}{2}\right) + d_5(x)\log (x)\right]
   \end{aligned}
\end{equation}
with
\begin{equation}
  \begin{aligned}
    d_1(x) &= 4 \left(5 x^6-27 x^4+27 x^2-5\right)\,,\quad
    d_2(x) = -24 \left(x^6-3 x^4-3 x^2+1\right)\,,\\
    d_3(x) &= 105 x^{14}-552 x^{13}+1203 x^{12}-5856 x^{11}+28163 x^{10}-63432 x^9+63897 x^8\\
    &\quad-63897 x^6+63432 x^5-28163 x^4+5856 x^3-1203 x^2+552 x-105\,,\\
    d_4(x) &= -6 \left(x^2-1\right)^3 \left(35 x^8+120 x^7-460 x^6+968 x^5-1070 x^4+968 x^3-460 x^2+120 x+35\right)\,,\\
    d_5(x) &= 6 x \big(35 x^{13}+60 x^{12}-325 x^{11}+304 x^{10}+198 x^9-788 x^8\\
    &\quad+446 x^7-889 x^5+788 x^4-217 x^3-304 x^2+240 x-60\big)\,,
  \end{aligned}
\end{equation}
which reproduces the result in the literature \cite{Parra3,Gabriele2}. From here one can extract the total radiated energy in hyperbolic-like motion in the center-of-mass frame,
\beq
\Delta E_{\rm hyp} = -(\Delta p_1 + \Delta p_2)\cdot \frac{ (m_1 u_1+m_2 u_2)}{|m_1 u_1+m_2 u_2|}\,,
\eeq
which can then be transformed via the boundary-to-bound dictionary, 
\beq
\Delta E_{\rm ell}(J) =  \Delta E_{\rm hyp}(J) - \Delta E_{\rm hyp}(-J)\,,
\eeq
with $J$ the angular momentum,  to obtain the radiated energy over one period of a bound elliptic-like orbit \cite{b2b3}. 

 \section{Discussion \& Outlook}\label{conc}
 
Building upon the formalism introduced in \cite{chadprl,chadRR,chadbr2,natalia1,natalia2,tail}, we developed the in-in framework to study scattering processes within the worldline EFT approach. Despite the apparent increase in complexity, we have shown how the computation of the relativistic integrand for the total impulse can be mapped to the in-out counterpart, and vice verse, provided a consistent flow (of time) is assigned to each diagram following the in-in Feynman rules. After reducing the problem to a series of master integrals involving retarded Green's functions, these can then be computed to all orders in velocity through differential equations. In comparison with standard Feynman propagators, only the issue of symmetry relations (under $p \to -p$) must be reevaluated.\footnote{In principle, because of the $\Theta(\pm p^0)$ in advance/retarded propagators, it may seem these would introduce extra terms once they are hit by derivatives w.r.t. the momentum. However, one can also think of the $i0$-prescription as a choice of contour integration for Green's function of the {\it box} operator. Hence, we do not expect the ultimate choice to alter the form of the differential equations other than through the boundary conditions. We thank Ruth Britto for a discussion on this point.}  The final solution then follows up to a set of boundary integrals in the near-static limit. Notably, consistency conditions imposed by the full solution can be further exploited to bootstrap the final answer up to a handful of constants.\vskip 4pt   
 \vskip 4pt In order to identify the different terms we used the identity $\Delta_{\rm ret} = \Delta_{\rm F} + \Delta_{-}$ 
and split the total in-in impulse as
\beq
\Delta p^\mu_{\rm tot} = \Delta p^\mu_{\rm F} + \Delta p^\mu_{\rm cut}\,,\label{pmutot0}
\eeq
where $ \Delta p^\mu_{\rm F}$ is obtained by replacing $\Delta_{\rm ret}$ with $\Delta_{\rm F}$, which we showed agrees with the in-out result, whereas $\Delta p^\mu_{\rm cut}$ involves the remaining insertions of $\Delta_{-}$'s, where the time ordering plays an important role. Following \cite{chadprl} we demonstrated that $ \Delta p^\mu_{\rm cons} \equiv {\cR} \Delta p^\mu_{\rm F} $ can be obtained from a conservative Lagrangian/Hamiltonian, such that $\Delta p^\mu_{\rm diss} \equiv {\cR}\Delta p^\mu_{\rm cut}$ includes all the dissipative effects, yielding for the total impulse the  decomposition
\beq
\Delta p^\mu_{\rm tot} =\Delta p^\mu_{\rm cons}+\Delta p^\mu_{\rm diss} \,,\label{pmutot}
\eeq
which is independent of the nature of the interaction. As expected, the imaginary parts cancel out in the sum, which must be add up to a real value for the impulse. Diagrammatically, following our previous conventions and up to iterations from lower order equations of motion, the computation for the variation of the effective action may be represented as follows
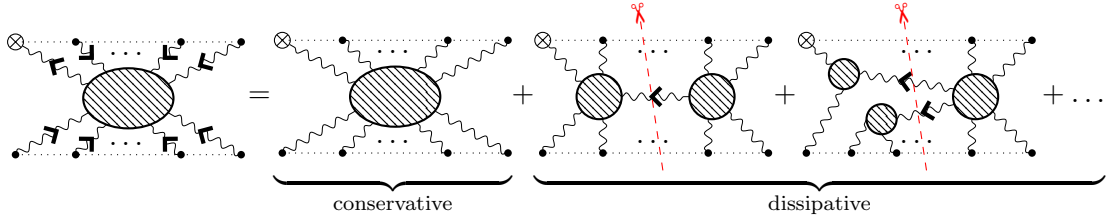
\begin{equation}
\nonumber
  \begin{tikzpicture}
    [baseline={([yshift=-0.5ex]current bounding box.center)}]
    \draw[dotted] (-1.5,0.75) -- (1.5,0.75);
    \draw[dotted] (-1.5,-0.75) -- (1.5,-0.75);
    \fill[draw=black,thick,pattern=north west lines] (0,0) ellipse (0.6 and 0.4);
    \node[xSource] (A) at (-1.5,0.75) {};
    \draw[graviton]  (160:0.57) -- (A);
    \draw[middleArrow] (160:0.57) -- (A);
    \draw[graviton] (-0.70,0.75) -- (130:0.46);
    \draw[middleArrow] (-0.70,0.75) -- (130:0.46);
    \node at (90:0.6) {\dots};
    \draw[graviton] (-1.5,-0.75) -- (200:0.57);
    \draw[middleArrow] (-1.5,-0.75) -- (200:0.57);
    \draw[graviton] (-0.7,-0.75) -- (230:0.46);
    \draw[middleArrow] (-0.7,-0.75) -- (230:0.46);
    \draw[graviton] (0.7,0.75) -- (50:0.46);
    \draw[middleArrow] (0.7,0.75) -- (50:0.46);
    \node at (-90:0.6) {\dots};
    \draw[graviton] (0.7,-0.75) -- (-50:0.46);
    \draw[middleArrow] (0.7,-0.75) -- (-50:0.46);
    \draw[graviton] (1.5,-0.75) -- (-20:0.57);
    \draw[middleArrow] (1.5,-0.75) -- (-20:0.57);
    \draw[graviton] (1.5,0.75) -- (20:0.57);
    \draw[middleArrow] (1.5,0.75) -- (20:0.57);
    \fill (1.5,0.75) circle (0.05);
    \fill (-0.7,0.75) circle (0.05);
    \fill (-1.5,-0.75) circle (0.05);
    \fill (-0.7,-0.75) circle (0.05);
    \fill (1.5,-0.75) circle (0.05);
    \fill (0.7,-0.75) circle (0.05);
    \fill (0.7,0.75) circle (0.05);
  \end{tikzpicture}
  \!=\!
  \underbrace{
  \begin{tikzpicture}
    [baseline={([yshift=-0ex]current bounding box.center)}]
    \draw[dotted] (-1.5,0.75) -- (1.5,0.75);
    \draw[dotted] (-1.5,-0.75) -- (1.5,-0.75);
    \fill[draw=black,thick,pattern=north west lines] (0,0) ellipse (0.6 and 0.4);
    \node[xSource] (A) at (-1.5,0.75) {};
    \draw[graviton] (160:0.57) -- (A);
    \draw[graviton] (-0.70,0.75) -- (130:0.46);
    \node at (90:0.6) {\dots};
    \draw[graviton] (-1.5,-0.75) -- (200:0.57);
    \draw[graviton] (-0.7,-0.75) -- (230:0.46);
    \draw[graviton] (0.7,0.75) -- (50:0.46);
    \node at (-90:0.6) {\dots};
    \draw[graviton] (0.7,-0.75) -- (-50:0.46);
    \draw[graviton] (1.5,-0.75) -- (-20:0.57);
    \draw[graviton] (1.5,0.75) -- (20:0.57);
    \fill (1.5,0.75) circle (0.05);
    \fill (-0.7,0.75) circle (0.05);
    \fill (-1.5,-0.75) circle (0.05);
    \fill (-0.7,-0.75) circle (0.05);
    \fill (1.5,-0.75) circle (0.05);
    \fill (0.7,-0.75) circle (0.05);
    \fill (0.7,0.75) circle (0.05);
    \fill[white] (0,-1) circle (0.01);
  \end{tikzpicture}
  }_{\textrm{conservative}}
  \!+\!\underbrace{
  \begin{tikzpicture}
    [baseline={([yshift=-1.4ex]current bounding box.center)}]
    \draw[dotted] (-1.5,0.75) -- (1.5,0.75);
    \draw[dotted] (-1.5,-0.75) -- (1.5,-0.75);
    \node[xSource] (A) at (-1.5,0.75) {};
    \fill[draw=black,thick,pattern=north west lines] (-0.75,0) ellipse (0.3 and 0.3);
    \fill[draw=black,thick,pattern=north west lines] (0.75,0) ellipse (0.3 and 0.3);
    \draw[graviton] (-0.98,0.2) -- (A);;
    \draw[graviton] (-0.7,0.75) -- (-0.7,0.3);
    \draw[graviton] (0.7,0.75) -- (0.7,0.3);
	\node at (90:0.6) {\dots};
    \draw[graviton] (-1.5,-0.75) -- (-0.98,-0.2);
    \draw[graviton] (-0.7,-0.75) -- (-0.7,-0.3);
    \draw[graviton] (1.5,-0.75) -- (0.98,-0.2);
    \draw[graviton] (0.7,-0.75) -- (0.7,-0.3);
    \node at (-90:0.6) {\dots};
    \draw[graviton] (1.5,0.75) -- (0.98,0.2);
    \draw[graviton] (-0.45,0) -- (0.45,0);
    \draw[middleArrow] (0.45,0) -- (-0.45,0);
    \fill (1.5,0.75) circle (0.05);
    \fill (-0.7,0.75) circle (0.05);
    \fill (-1.5,-0.75) circle (0.05);
    \fill (-0.7,-0.75) circle (0.05);
    \fill (1.5,-0.75) circle (0.05);
    \fill (0.7,-0.75) circle (0.05);
    \fill (0.7,0.75) circle (0.05);
    \draw[dashed,red] (-0.2,1) -- (0.1,-1);
    \node[red,rotate=-80] at (-0.23,1.1) {\Rightscissors};
  \end{tikzpicture}
  \!+\!
  \begin{tikzpicture}
    [baseline={([yshift=-1.4ex]current bounding box.center)}]
    \draw[dotted] (-1.5,0.75) -- (1.5,0.75);
    \draw[dotted] (-1.5,-0.75) -- (1.5,-0.75);
    \node[xSource] (A) at (-1.5,0.75) {};
    \fill[draw=black,thick,pattern=north west lines] (-1,0.3) ellipse (0.2 and 0.2);
    \fill[draw=black,thick,pattern=north west lines] (-0.5,-0.3) ellipse (0.2 and 0.2);
    \fill[draw=black,thick,pattern=north west lines] (0.75,0) ellipse (0.3 and 0.3);
    \draw[graviton] (-1.15,0.43) -- (A);
    \draw[graviton] (0.7,0.75) -- (0.7,0.3);
	\node at (90:0.6) {\dots};
    \draw[graviton] (-1.5,-0.75) -- (-1,0.1);
    \draw[graviton] (-0.9,-0.75) -- (-0.7,-0.3);
    \draw[graviton] (-0.3,-0.75) -- (-0.5,-0.5);
    \draw[graviton] (1.5,-0.75) -- (0.98,-0.2);
    \draw[graviton] (0.7,-0.75) -- (0.7,-0.3);
    \node at (-90:0.6) {\dots};
    \draw[graviton] (0.98,0.2) -- (1.5,0.75);
    \draw[graviton] (-0.3,-0.3) -- (0.48,-0.1);
    \draw[middleArrow] (0.48,-0.1) -- (-0.3,-0.3);
    \draw[graviton] (-0.79,0.3) -- (0.48,0.1);
    \draw[middleArrow] (0.48,0.1) -- (-0.79,0.3);
    \fill (1.5,0.75) circle (0.05);
    \fill (-1.5,-0.75) circle (0.05);
    \fill (-0.9,-0.75) circle (0.05);
    \fill (-0.3,-0.75) circle (0.05);
    \fill (1.5,-0.75) circle (0.05);
    \fill (0.7,-0.75) circle (0.05);
    \fill (0.7,0.75) circle (0.05);
    \draw[dashed,red] (-0.2,1) -- (0.1,-1);
    \node[red,rotate=-80] at (-0.23,1.1) {\Rightscissors};
  \end{tikzpicture}
  +\dots}_{\textrm{dissipative}}
\end{equation}
using the same notation we introduced before. The conservative piece of the impulse is thus built up entirely in terms of (unoriented) Feynman propagators, whereas the dissipative pieces are obtained by systematically replacing retarded propagators by factors of $\Delta_-$.\footnote{Let us continue to emphasize that in principle we cannot rule out specific situations where additional conservative terms may result from even-in-time products of $\Delta_{-}$'s, which would not be captured by Feynman's prescription.} After implementing the method of differential equations to compute the resulting master integrals, the remaining task is reduced to obtaining the necessary boundary conditions, both in the conservative and dissipative sectors, which may be then further decomposed using the method of regions \cite{nrgr}.\vskip 4pt  

 As we have seen, the contribution from radiation modes to $\Delta p^\mu_{\rm F}$ produces only imaginary terms at 3PM so that the conservative sector is dominated by the potential region at this order. The impulse was already computed in \cite{3pmeft} and reproduced here from the full integrand. However, radiation modes do enter at higher PM orders. As~discussed in \cite{4pmeft,4pmeft2}, the above procedure yields a conservative piece in the scattering angle which includes both potential and radiation-reaction tail effects at 4PM. Because of the presence of $\delta(p^2)$ in $\Delta_{-}$, radiation modes are necessary for a non-zero $\Delta p^\mu_{\rm diss}$. It turns out to be somewhat useful to obtain the associated (real) boundary integrals from (the imaginary part of) those involving Feynman propagators using Cutkosky's rules~\cite{Cutkosky:1960sp}. For this purpose one must generalize the standard cutting rules, such that the Dirac-$\delta$ functions for the sources in the worldline EFT approach may be described in terms of cuts over linear propagators. Once again, in addition to the incursion of Feynman's $i0$-prescription dictated by the path integral, here we find yet another standard tool from quantum field theory, i.e. unitary relations, playing a key role at the classical level. Even though all the resulting integral cuts at 3PM may be computed following the analysis in \cite{Riva:2021vnj,Parra3,Gabriele2}, we have shown a dramatic simplification occurs after applying a set of consistency conditions to the full answer \cite{4pmeft2}. This is another instance in which the use of differential equations turn out to be a powerful tool to solve the entire dynamics.\vskip 4pt

Let us comment on some important points regarding the above manipulations, and in particular with regards to the different choices of basis in the in-in formalism, namely either using \eqref{gmatrix} or \eqref{kmatrix}. On the one hand, the former provides us with a natural identification of conservative terms via Feynman Green's functions within the standard in-out prescription; whereas, on the other hand, the latter makes causality explicit in the form of retarded propagators.  Yet, as we demonstrated here, the map between the in-in and in-out results for the impulse allows us to implement the split in \eqref{pmutot0} even though, in principle, the integrand was constructed in the $\pm$-basis. In turn, the choice of basis influences the way the real part of the impulse computed with Feynman propagators arises. While, as we discussed, this is straightforward in the $1/2$-basis, the cancelation of imaginary parts in the $\pm$-basis becomes manifest only after using \eqref{eq:propRel1} and re-writing $\Delta_{\rm ret}$ as the average between Feynman and Dyson pieces. (Although this cancelation turned out to be trivial at 3PM order, it will become a useful consistency check in higher order computations.)\vskip 4pt

Let us conclude with a few remarks regarding the connection between the in-in worldline formalism and the so-called KMOC approach introduced in \cite{donal}. Following unitarity of the $S$-matrix, the KMOC prescription to compute the impulse is split into two contributions, denoted as $I_{(1)}$ and $I_{(2)}$ in \cite{donal}. The former is constructed in terms of the scattering amplitude in the standard in-out approach, whereas the latter appears in the form of a sum over cuts. As in our case, the sum of these two terms must add up to the (real) total impulse, and therefore intermediate imaginary parts must cancel out. Let us concentrate first on the conservative sector. Following momentum conservation, the impulse (in general relativity) can be uniquely determined at any PM order from $I_{\perp}$, the perpendicular part introduced in \cite{Parra3}.  Hence, after taking the classical limit and subtracting {\it super-classical} terms, the real part of the (time-symmetric) contribution linear in the scattering amplitude ought to be identified with a conservative contribution.\footnote{Notice that due to the pollution from super-classical terms additional subtractions, also in the form of cuts, may still be needed even with potential-only modes, see e.g. Eq. 6.18 in \cite{Parra3}, resembling the iterated terms in \cite{zvi1}. However, these cuts also contain physical information needed to include non-conservative effects.} 
Modulo the lack of subtractions in our formalism, this part then resembles the first term in our decomposition of the total impulse---provided massive lines are further reduced and combined into Dirac-$\delta$ functions using reversed unitarity~\cite{Parra}. 
For dissipative terms, on the other hand, both the perpendicular and longitudinal parts matter. Moreover, since the real part of the Feynman piece is invariant under $t \to -t$, only the cuts contribute to the dissipative sector.  The latter therefore must include, among other things, the remaining terms in our decomposition. However, because of super-classical pieces and the appearance of cuts over massive lines needed to account for the conservative impulse in the longitudinal direction (consistently with momentum conservation), the direct comparison between terms is less transparent.  The map also becomes less straightforward at higher orders once we include iterations of lower order deflections that include radiation-reaction effects. Hence, although the KMOC and in-in worldline formalism are clearly tightly related, the identification of terms requires further study. Yet, understanding the map between the different frameworks, and also the relation to the eikonal phase \cite{Gabriele2}, can ultimately allow us to freely incorporate powerful tools and simplifications that can help us not only fine-tune the computational machinery, but also elucidate the role of powerful quantum-based tools from the theory of scattering amplitudes in classical computations.\\

{\bf Note added:} While the results in this paper were prepared for submission we became aware of the work in \cite{rrHU} which has some overlap with the derivations in our paper. We~thank the authors for sharing a draft prior to submission.\\

\section*{Acknowledgments}

We thank 	Ruth Britto, Christoph Dlapa, Walter Goldberger, Zhengwen Liu, Massimiliano Riva and Chia-Hsien Shen for illuminating discussions, and in particular to Christoph Dlapa for invaluable help on the topic of differential equations. J.N. would like to thank Benjamin Sauer for useful exchanges. 
The work of G.K. and R.A.P. received support from the ERC-CoG {\it Precision Gravity: From the LHC to LISA}, provided by the European Research Council (ERC) under the European Union's H2020 research and innovation programme (grant No. 817791). 

\appendix
\section{Green's functions}\label{appA}

Throughout this paper we use  the following conventions for Feynman/Dyson, retarded/advanced and Wightman Green's functions:
 \begin{equation}\label{eq:props}
  \begin{aligned}
    \Delta_{\rm F}(x) &= \int \frac{\dd^4 p}{(2\pi)^4} \frac{e^{-i\,p\cdot x}}{p^2+i0}\,, \quad\quad \Delta_{\rm D}(x) =\int \frac{\dd^4 p}{(2\pi)^4} \frac{e^{-i\,p\cdot x}}{p^2-i0}\,,\\
    \Delta_{\textnormal{ret}}(x) &= \int \frac{\dd^4 p}{(2\pi)^4} \frac{e^{-i\,p\cdot x}}{(p^0+i0)^2-\bp^2}\,, \quad\quad
    \Delta_{\textnormal{adv}}(x) = \int \frac{\dd^4 p}{(2\pi)^4}\frac{e^{-i\,p\cdot x}}{(p^0-i0)^2-\bp^2}\,,\\
    \Delta_\pm(x) &= 2\pi i\,\int  \frac{\dd^4 p}{(2\pi)^4} e^{-i\,p\cdot x} \delta(p^2)\Theta(\pm p^0)\,,\\
  \end{aligned}
\end{equation}
obeying, in position space,
\begin{equation}\label{eq:propRel0}
 \begin{aligned}
  \Delta_{\rm F}(x) &= -\Delta_+(x)\Theta(x^0) - \Delta_-(x)\Theta(-x^0)\,,\\
    \Delta_{\rm D} (x) &= \Delta_-(x)\Theta(x^0)+\Delta_+(x)\Theta(-x^0)\,,\\
      \Delta_\textrm{ret}(x) &= -\big(\Delta_+(x) - \Delta_-(x)\big)\Theta(x^0),\\
  \Delta_\textrm{adv}(x) &=\big(\Delta_+(x) - \Delta_-(x)\big)\Theta(-x^0)\,,
\end{aligned}
\end{equation}
 implying the relations
 \begin{equation}\label{eq:propRel1}
  \begin{aligned}
    \Delta_{\textnormal{ret}}(x) &= \Delta_{\rm F}(x)+\Delta_-(x)=\Delta_{\rm D}(x)-\Delta_+(x)\,,\\
    \Delta_{\textnormal{adv}}(x) &= \Delta_{\rm F}(x)+\Delta_+(x) =\Delta_{\rm D}(x) - \Delta_-(x)\,,
  \end{aligned}
\end{equation}

\section{Classical electrodynamics}

It is straightforward to use the formalism developed in this paper to study the case of classical electromagnetism, controlled by the worldline action
\beq
S = -\frac{1}{4} \int \dd^4 x \, F_{\mu\nu} F^{\mu\nu} - \sum_a \int \dd \tau_a \left(\frac{m_a}{2} v_a^2  + e\, q_a v_a^\mu A_\mu(x^\alpha_a)\right) \,,
 \eeq
 where $F_{\mu\nu} = \partial_\mu A_\nu -  \partial_\nu A_\mu$, $A_\mu$ is the four-vector potential, and $q_{a=1,2}$ are the respective charges. The computation simplifies drastically compared to the gravity case. The main reason is the absence of bulk interactions, which implies only two diagram are needed for the complete, all-order variation of the effective action:
\begin{equation}
  \left.\frac{\delta\cS_\textnormal{eff}[x_+,x_-]}{\delta x_-^\alpha}\right|_{\substack{x_-\to 0\\x_+\to x}}=
  \begin{tikzpicture}
    [baseline={([yshift=-0.5ex]current bounding box.center)}]
    \node[xSource] (A) at (0,1.5) {};
    \draw[graviton] (0,0) -- (A) node[left=0.2] {};
    \draw[middleArrow] (0,0) -- (A);
    \fill (0,0) circle (0.05);
  \end{tikzpicture}
  +
  \begin{tikzpicture}
    [baseline={([yshift=-0.5ex]current bounding box.center)}]
    \node[xSource] (A) at (0,1.5) {};
    \draw[graviton] (1,1.5) to[bend left=60] (A);
    \draw[middleArrow] (1,1.5) to[bend left=60] (A);
    \fill (1,1.5) circle (0.05);
    \fill[white] (0,0) circle (0.01);
  \end{tikzpicture}
  \,.
\end{equation}
Hence, all higher order contributions to the deflection are exclusively coming from iterations via the equations of motion.
The rest of the calculation is unchanged, such that the result depends on a subset of the master integrals for the gravity case.
Expanding the impulse in powers of $\alpha=e^2/4\pi$,
\begin{equation}
  \Delta p_1^\mu = \sum_{n=1}^\infty \Delta^{(n)} p_1^\mu \alpha^n\,,
\end{equation}
we find for the conservative contribution
\begin{equation}
  \begin{aligned}
    \Delta^{(3)} p_{1,\textrm{cons}}^\mu &= \frac{q_1^3 q_2^3 b^\mu}{8 \pi^2 |b|^4 M^2\nu^2  (x^2-1)^5}\left[4 \left(x^2+1\right) x^4+\nu  \left(x^9-8 x^6+14 x^5-8 x^4+x\right)\right]\\
    &\quad+ \frac{q_1^3 q_2^3 x \left(x^2+1\right) \left((\Delta -1) \check{u}_1^\mu+(\Delta +1) \check{u}_2^\mu\right)}{4 \pi  |b|^3 M^2 (\Delta^2 -1)^2 (x^2-1)^2 }\,,
  \end{aligned}
\end{equation}
whereas the dissipative part becomes
\begin{equation}
  \begin{aligned}
    \Delta^{(3)} p_{1,\textrm{diss}}^\mu &= \frac{q_1^2 q_2^2 b^\mu}{96 \pi^3 |b|^4 M^2 (x^2-1)^2 \nu^2}
    \begin{multlined}[t]
      \bigg[\left(x^2+1\right)^2 \left(q_1^2 (\Delta +2 \nu -1)-q_2^2 (\Delta -2 \nu +1)\right)\\
        +\frac{12 \nu  q_1 q_2 x \left(x^2+1\right)^2 \left(x^4-1-4 x^2 \log (x)\right)}{\left(x^2-1\right)^3}\bigg]
    \end{multlined}\\
    &\quad + \frac{q_1^2 q_2^2 \check{u}_2^\mu}{768 \pi^2 |b|^3 M^2 }
    \begin{multlined}[t]
      \bigg[\frac{\left(x^2+3\right) \left(3 x^2+1\right) \left((\Delta -1)^2 q_1^2 \left(x^2+1\right)+2 (\Delta +1)^2 q_2^2 x\right)}{\left(\Delta ^2-1\right)^2 x^2 (x^2-1)}\\
        -\frac{3 q_1 q_2 \left(3 x^6-8 x^5+45 x^4-48 x^3+45 x^2-8 x+3\right)}{\nu  \left(x^2-1\right)^3}\\
        +\frac{12 q_1 q_2 x^2 \left(3 x^4+10 x^2+3\right) \log (x)}{\nu  \left(x^2-1\right)^4}
        \bigg]\,,
    \end{multlined}
  \end{aligned}
\end{equation}
which agrees with the result in \cite{Saketh:2021sri,Bern:2021xze}.

\bibliographystyle{JHEP}
\bibliography{Ref4PM}

\end{document}